\documentclass[]{aastex631}

\usepackage{array}
\shorttitle{ALFALFA Almost-Dark TDG Candidates}
\shortauthors{Gray et al.}
\graphicspath{{./}{figures/}}

\begin{document}

\title{Catching Tidal Dwarf Galaxies at a Later Evolutionary Stage with ALFALFA}

\author[0000-0001-6389-5639]{Laurin M. Gray}
\affiliation{Indiana University, 727 East Third Street, Bloomington, IN 47405, USA}
\email{grayla@iu.edu}

\author[0000-0001-8283-4591]{Katherine L. Rhode}
\affiliation{Indiana University, 727 East Third Street, Bloomington, IN 47405, USA}

\author[0000-0001-8849-7987]{Lukas Leisman}
\affiliation{Department of Astronomy, University of Illinois, 1002 W. Green Street, Urbana, IL 61801, USA}

\author[0000-0001-5175-939X]{Pavel E. Mancera Pi\~{n}a}
\affiliation{Leiden Observatory, Leiden University, P.O. Box 9513, 2300 RA Leiden, The Netherlands}

\author[0000-0002-1821-7019]{John M. Cannon}
\affiliation{Department of Physics \& Astronomy, Macalester College, 1600 Grand Avenue, Saint Paul, MN 55105, USA}

\author[0000-0001-8483-603X]{John J. Salzer}
\affiliation{Indiana University, 727 East Third Street, Bloomington, IN 47405, USA}

\author[0000-0002-2492-7973]{Lexi Gault}
\affiliation{Indiana University, 727 East Third Street, Bloomington, IN 47405, USA}

\author[0000-0002-8598-439X]{Jackson Fuson}
\affiliation{Department of Physics \& Astronomy, Macalester College, 1600 Grand Avenue, Saint Paul, MN 55105, USA}

\author[0000-0003-0608-6258]{Gyula I. G. J\'{o}zsa}
\affiliation{Max-Planck-Institut f\"ur Radioastronomie, Auf dem H\"ugel 69, 53121 Bonn, Germany}
\affiliation{Department of Physics and Electronics, Rhodes University, PO Box 94, Makhanda, 6140, South Africa}

\author[0000-0002-9798-5111]{Elizabeth A. K. Adams}
\affiliation{ASTRON, Netherlands Institute for Radio Astronomy, Oude Hoogeveensedijk 4, 7991 PD Dwingeloo, The Netherlands}
\affiliation{Kapteyn Astronomical Institute, Landleven 12, 9747 AD Groningen, The Netherlands}

\author[0000-0002-3222-2949]{Nicholas J. Smith}
\affiliation{Indiana University, 727 East Third Street, Bloomington, IN 47405, USA}

\author[0000-0001-5334-5166]{Martha P. Haynes}
\affiliation{Cornell Center for Astrophysics and Planetary Science, Space Sciences Building, Cornell University, Ithaca, NY 14853, USA}

\author[0000-0001-9165-8905]{Steven Janowiecki}
\affiliation{University of Texas, Hobby-Eberly Telescope, McDonald Observatory, TX 79734, USA}

\author{Hannah J. Pagel}
\affiliation{Indiana University, 727 East Third Street, Bloomington, IN 47405, USA}



\begin{abstract}

We present deep optical imaging and photometry of four objects classified as “Almost-Dark” galaxies in the ALFALFA survey because of their gas-rich nature and extremely faint or missing optical emission in existing catalogs.  They have HI masses of $10^7$-$10^9$ $M_{\odot}$ and distances of $\sim$9-100 Mpc. Observations with the WIYN 3.5m telescope and One Degree Imager reveal faint stellar components with central surface brightnesses of $\sim$24-25 $\mathrm{mag}\,\mathrm{arcsec}^{-2}$ in the $g$-band.  We also present the results of HI synthesis observations with the Westerbork Synthesis Radio Telescope.  These Almost-Dark galaxies have been identified as possible tidal dwarf galaxies (TDGs) based on their proximity to one or more massive galaxies. We demonstrate that AGC~229398 and AGC~333576 likely have the low dark matter content and large effective radii representative of TDGs.  They are located much farther from their progenitors than previously studied TDGs, suggesting they are older and more evolved.  AGC~219369 is likely dark matter dominated, while AGC~123216 has a dark matter content that is unusually high for a TDG, but low for a normal dwarf galaxy.  We consider possible mechanisms for the formation of the TDG candidates such as a traditional major merger scenario and gas ejection from a high velocity fly-by.  Blind HI surveys like ALFALFA enable the detection of gas-rich, optically faint TDGs that can be overlooked in other surveys, thereby providing a more complete census of the low-mass galaxy population and an opportunity to study TDGs at a more advanced stage of their life cycle.

\end{abstract}


\section{Introduction} \label{sec:intro}

The Arecibo Legacy Fast ALFA (ALFALFA) blind extragalactic HI survey \citep{Giovanelli2005, Haynes2018} was designed primarily to probe the low-mass end of the HI mass function, and has identified over 31,000 sources out to a redshift z $\sim$ 0.06.  ALFALFA covered nearly 7000 square degrees of sky and detected HI sources without regard to their optical properties, which makes it a powerful tool to search for tidal remnants such as tidal dwarf galaxies (TDGs) on a larger scope than previously attempted \citep{Giovanelli2005}.  

TDGs are generally composed of material removed from a parent galaxy through tidal interactions, which means they are usually very gas-rich \citep{Duc2012}.  They have masses and sizes comparable to dwarf galaxies, and over time will become dynamically stable systems independent of their parent galaxy \citep{Duc2012}.  TDGs contain clues to the interaction histories of their parent galaxies, potentially constraining the interaction type and properties of the merging systems \citep{LeeWaddell2014}.  They are also vital for determining cosmological constraints in group environments \citep{LeeWaddell2014}.  

TDGs usually have higher metallicities than normal dwarf galaxies because they are created out of pre-enriched material stripped from the outer disks of larger galaxies \citep{Duc1999, Duc2000, Weilbacher2001, Duc2012}.  However, \cite{Hunter2000} point out that very old tidal dwarfs which were formed before spirals could become more enriched could have metallicities similar to modern metal-poor dwarfs.  TDGs have a very low dark matter content, because tidal forces create gas-rich streams that contain a small fraction of pre-existing stars and almost no dark matter, and the potential gravitational well of the dwarf is too shallow to capture dark matter \citep{Bournaud2010}.  The turbulence of the tidal interaction also frequently triggers star formation, leading to a population of young blue stars \citep{Duc2000}.  

\cite{Kaviraj2012} conducted a statistical survey of the properties of 405 optically identified TDG candidates and their parent systems and found that the vast majority of TDGs are within 20 kpc of their progenitors.  However, the most easily identified TDG candidates still have a visible tidal tail connecting them to their parents.  As a result, the sample of known TDGs is biased towards objects that are near their parent galaxies and still in the early stages of their evolution, as the tail takes time to dissolve.  Simulations suggest that to become a bona fide TDG, a tidal knot must have enough mass ($\geq$ $10^8$ $M_{\odot}$) and distance from the gravitational well of the parent to become self-gravitating and avoid falling back into the parent system \citep{Bournaud2006}.  Therefore, the overall sample of objects that have been studied consists mainly of TDG candidates, with some uncertainty as to whether they will become long-lived ($>$1 Gyr; \citealt{Kaviraj2012, Bournaud2006}). It is also uncertain how long TDGs are able to remain stable without a massive dark matter halo, as simulations indicate that dark-matter-deficient, rotationally supported galaxy disks are prone to axisymmetric and non-axisymmetric instabilities on dynamical timescales \citep{Sellwood2022}. As TDGs evolve, they become optically indistinct from other satellite dwarf galaxies \citep{Duc2012}.  Identification relies on metallicity measurements and age estimations, usually from time-intensive spectroscopy.  Additionally, the initial burst of star formation will eventually end, the young stellar population will age, and the higher metallicity will contribute to a faster rate of dimming, making older TDGs even more difficult to detect through optical methods \citep{Roman2021}.  However, they will remain gas-rich, so HI observations are useful for locating objects of interest when searching for TDGs because they will be able to catch sources with extremely low optical surface brightnesses.

Around 99\% of the objects observed in ALFALFA could be matched to objects in previously published optical catalogs such as the Sloan Digital Sky Survey (SDSS; \citealt{SDSS-III}), though many had never been observed in HI before.  The remaining 1\% is a set of gas-rich objects that are missing a clear counterpart in optical catalogs.  A subset of these are classified as Ultra-Compact High-Velocity Clouds (UCHVCs; \citealp{Giovanelli2010, Adams2013}), which are compact HI clouds with velocities consistent with a location in the vicinity of the Local Group.  Follow-up optical imaging of the UCHVCs has in some cases revealed a dwarf galaxy or other possible stellar counterpart \citep[see][]{Rhode2013, Sand2015, Janesh2019}.  The remaining objects without clear optical counterparts are classified as “Almost-Dark" galaxies (ADs); these objects have velocities that place them outside of the Local Group and HI masses ranging from $10^7$ - $10^9$ $M_{\odot}$.  Deep optical imaging of several ADs has revealed low surface brightness stellar components that were below the sensitivity limits of previous surveys \citep{Cannon2015, Leisman2021}.  As explained in \cite{Cannon2015}, while the majority were either OH megamasers or tidal debris from a nearby galaxy interaction, a small subset appeared to be more isolated.  Some of the ADs have properties that make them comparable to ultra-diffuse galaxies (UDGs;  \citealt{vanDokkum2015, Leisman2017, Gault2021}).  Other ADs have been identified as likely TDGs that have moved farther away from their parent sources than the majority of previously studied TDGs \citep{LeeWaddell2016}.  As we will discuss in this paper, TDGs (particularly evolved TDGs) have low central surface brightnesses and large effective radii, which means that some TDGs may fit the criteria used to define UDGs, so there may be overlap between the two categories.

In this paper, we present HI and optical data of four objects that have been classified as potential Almost-Dark Tidal Dwarf Galaxies (AD-TDGs). The four objects, AGC~123216, AGC~219369, AGC~229398, and AGC~333576, were selected from the AD sample and identified as potential TDGs because of the presence of at least one HI-rich source that could be classified as a potential parent galaxy based on that galaxy's proximity and velocity relative to the AD.  Specifically, the potential parent galaxies for each AD-TDG candidate are located within 500 kpc on the sky at the reported distance for the AD and have a measured heliocentric velocity that is within $\pm$500 $\mathrm{km}\,\mathrm{s}^{-1}$ of the AD as measured by \cite{Haynes2011}.  The four AD-TDG candidates we analyzed each have between one and three potential parent galaxies meeting these criteria.

The paper is organized as follows.  Section \ref{sec:obs} covers the HI and optical observations and data processing.  Section \ref{sec:analysis} describes the procedures used to carry out our measurements of quantities such as optical surface brightness, and stellar, baryonic, and dynamical masses.  A discussion of the properties and environment of each AD is given in Section \ref{sec:results}.  Section \ref{sec:discussion} discusses the ADs in the context of our current understanding of TDGs, including comments on their possible origins.  The last section includes a brief summary of our main conclusions.

\section{Observations and Data Processing} \label{sec:obs}

\subsection{HI Data}
The AD ALFALFA observations and data reduction processes are described in detail in other papers \citep{Giovanelli2005, Saintonge2007, Haynes2011}, and the relevant ALFALFA catalog \citep{Haynes2018} measurements for these objects and their possible parents are reproduced in Table \ref{tab:ALFALFAdata}, along with photometric data from SDSS for the possible parents for reference \citep{SDSSDR16}.  Generally, the distance for objects with $v_{helio} > 6000$ $\mathrm{km}\,\mathrm{s}^{-1}$ was estimated with the Hubble Law (with $H_0$ = 70 $\mathrm{km}\,\mathrm{s}^{-1}\,\mathrm{Mpc}^{-1}$), and the local peculiar velocity model of \cite{Masters2005} was used for distances of objects with $v_{helio} < 6000$ $\mathrm{km}\,\mathrm{s}^{-1}$.

In addition to the original ALFALFA observations, the ADs presented in this paper and their prospective progenitors were also observed with the Westerbork Synthesis Radio Telescope (WSRT) in an exploratory observational program of a larger subset of Almost-Dark galaxies (program R13B/001; PI Adams).  Observations consisted of two 12 hour pointings centered on the central HI velocity from ALFALFA, using a 10 MHz bandpass with 1024 channels and 2 polarizations.  Data reduction is detailed in \cite{Janowiecki2015} and \cite{Leisman2016}.  The reduction used an automated pipeline based on the MIRIAD \citep{Sault1995} data software \citep[see][]{Serra2012, Wang2013}.  The pipeline automates radio frequency interference flagging and implements primary bandpass calibration, and iterative deconvolution of the data with the CLEAN algorithm to apply a self-calibration.  We used a robustness weighting r=0.4 and a velocity resolution binning of 6.0 $\mathrm{km}\,\mathrm{s}^{-1}$ after Hanning smoothing.  The resulting images had synthesized beams with major (north-south) axes ranging from 32-38$\arcsec$ and minor axes from 13-17$\arcsec$.  We also performed a primary beam correction to our measured column densities; note that the FWHM of the WSRT primary beam is $\sim$35$\arcmin$ at 1.4 GHz, so the column density sensitivity for potential parent sources near the edges of our images is low.  Total intensity and velocity field maps were created as detailed in \cite{Gault2021}.  We were able to retrieve the peak HI column density from the total intensity maps, under the assumption that the flux is spread out over the beam, so this measurement is dependent on the resolution of the beam. The peak HI column density is reported in the Results section for each source.

\movetabledown=30mm
\begin{rotatetable}
\begin{deluxetable}{ccccccccccc}
    \centering
    \tablecaption{\protect\label{tab:ALFALFAdata} HI Measurements and Other Properties of Candidate AD-TDGs and Potential Parent Galaxies}
    \tablehead{ \colhead{ID} & \colhead{Classification} & \colhead{RA} & \colhead{Dec} &
    \colhead{$v_{helio}$} & \colhead{W20} &
    \colhead{Distance} & \colhead{Angular} & \colhead{log($M_{HI}$)} & \colhead{g} & \colhead{g-r} \\ 
    & & [deg] & [deg] & [$\mathrm{km}\,\mathrm{s}^{-1}$] & [$\mathrm{km}\,\mathrm{s}^{-1}$] & [Mpc] & Separation [kpc] & & [mag] & [mag] }
    \startdata
        AGC~123216 & Almost-Dark & 31.17792 & 28.80583 & 5111 $\pm$ 14 & 51 $\pm$ 7 & 70.5 $\pm$ 2.3 & -- & 8.65 $\pm$ 0.08 & -- & -- \\
        MRK~365\tablenotemark{a} & Spiral & 31.07729 & 28.65592 & 5151 $\pm$ 5 & -- & 72.2 $\pm$ 5.1 & 214 $\pm$ 7 & -- & 14.33 & 0.67 \\
        NGC~807 & Elliptical\tablenotemark{b} & 31.22917 & 28.98889 & 4750 $\pm$ 218 & 492 $\pm$ 12 & 67.7 $\pm$ 4.4 & 231 $\pm$ 8 & 9.79 $\pm$ 0.07 & 14.20 & 0.87 \\
        MCG~05-06-003 & Spiral & 31.38208 & 28.61083 & 5312 $\pm$ 92 & 202 $\pm$ 5 & 73.4 $\pm$ 2.3 & 331 $\pm$ 11 & 9.7 $\pm$ 0.05 & 14.47 & 0.39\\
        \hline
        AGC~219369\tablenotemark{c} & Almost-Dark & 165.96333 & 28.68583 & 667 $\pm$ 11 & 45 $\pm$ 7 & 9.2 $\pm$ 2.2 & -- & 7.25 $\pm$ 0.21 & -- & -- \\
        NGC~3510 & Spiral & 165.93668 & 28.88806 & 704 $\pm$ 92 & 199 $\pm$ 2 & 16.7 $\pm$ 3.3 & 33 $\pm$ 8 & 9.45 $\pm$ 0.18 & 13.33 & 0.28 \\
        \hline
        AGC~229398 & Almost-Dark & 180.16251 & 21.41556 & 6965 $\pm$ 14 & 49 $\pm$ 3 & 104 $\pm$ 2.3 & -- & 9.28 $\pm$ 0.05 & -- & -- \\
        KUG~1158+216 & Spiral & 180.16458 & 21.3525 & 7388 $\pm$ 125 & 268 $\pm$ 42 & 110.1 $\pm$ 2.3 & 102 $\pm$ 2 & 9.47 $\pm$ 0.06 & 16.36 & 0.59 \\
        UGC~6989 & Spiral & 180.02415 & 21.64528 & 6404 $\pm$ 99 & 220 $\pm$ 3 & 96 $\pm$ 2.4 & 482 $\pm$ 11 & 9.87 $\pm$ 0.05 & 15.80 & 0.38 \\
        \hline
        AGC~333576 & Almost-Dark & 358.185 & 28.74667 & 7031 $\pm$ 13 & 37 $\pm$ 3 & 93.9 $\pm$ 4.3 & -- & 9.1 $\pm$ 0.07 & -- & -- \\
        NGC~7775 & Spiral & 358.09833 & 28.77083 & 6760 $\pm$ 65 & 182 $\pm$ 8 & 93.9 $\pm$ 4.3 & 127 $\pm$ 6 & 9.88 $\pm$ 0.06 & 13.74 & 0.54 \\
    \enddata
    \tablecomments{Data are taken from \cite{Haynes2018} except when otherwise noted.  Almost-Dark galaxies are listed at the top of each subsection and followed by their potential progenitors.  Classification for the potential parent galaxies is the galaxy type listed on the NASA/IPAC Extragalactic Database (NED; \citeyear{NED}). RA and Dec are the coordinates for the center of the HI source from ALFALFA, which is not necessarily the center of the optical component.  Angular separation in kpc is calculated at the ALFALFA distance of the relevant AD.  Listed magnitudes are the CModelMags from SDSS DR16 \citep{SDSSDR16}, corrected for extinction using the \cite{Schlegel1998} extinction maps.}
    \tablenotetext{a}{MRK~365 is not included in the ALFALFA catalog, so the distance listed here is the NED Hubble flow distance for the heliocentric velocity measured by \cite{Thuan1999}, corrected for a 3K CMB according to \cite{Fixsen1996}.}
    \tablenotetext{b}{See discussion of classification for NGC~807 in Section \ref{sec:A123216}}
    \tablenotetext{c}{The HI mass and angular separation in kpc were calculated using the \cite{Haynes2018} distance of 9.2 Mpc, see Section \ref{sec:A219369} for a discussion on the ambiguity in this measurement.}
\end{deluxetable}
\end{rotatetable}
\newpage

\subsection{Optical Data}

Optical observations of the ADs and their surrounding areas were carried out with the WIYN 3.5-m telescope.  One object (AGC~229398) was observed in March 2014 with the One Degree Imager with a partially-populated focal plane (pODI).  Before a larger detector array was installed in 2015, pODI had a 3$\times$3 grid of orthogonal transfer arrays (OTAs), with a field of view of 24$\arcmin\times$24$\arcmin$.  There were also four more OTAs in the outer corners that were used to image guide stars.  Each OTA is an 8$\times$8 arrangement of 480$\times$496 pixel CCDs, where each pixel has a pixel scale of 0.11$\arcsec$/pixel.  The other three objects were observed on multiple occasions between 2017 and 2019 with the One Degree Imager (ODI; \citealt{Harbeck2014}).  The ODI camera has 17 additional OTAs compared to pODI, arranged in a 5$\times$6 grid for a field of view of 40$\arcmin \times$48$\arcmin$.  During each observing run, each source was observed in the $g'$ and $r'$ bands, using a 9-point dither pattern of 300 seconds per exposure to avoid gaps in the final image resulting from gaps in the CCD array.

We reduced our observations using the One Degree Imager Pipeline, Portal, and Archive (ODI-PPA; \citealt{Gopu2014}) QuickReduce \citep{Kotulla2014} data reduction pipeline.  The pipeline consists of the following: saturated pixel masking; corrections for crosstalk, persistence, and nonlinearity; overscan subtraction; bias, dark, and flat-field corrections; pupil ghost corrections; and cosmic ray removal.  We then stacked our images using the SWarp program \citep{Bertin2002} integrated with ODI-PPA's Stack interface.  Due to the faintness of these objects, we created the deepest combined images possible without sacrificing image quality by using the best exposure at each dither point and then including any additional exposures with good focus and seeing below 1.5$\arcsec$.  The total integration times of the images vary from 45 to 90 minutes based on the number of exposures included in the stack. The background of each image was mapped and subtracted, and then the images were resampled, combined with a weighted average, and reprojected to a pixel scale of 0.125 $\arcsec$/pixel.  An average background value was then added back to each image.  The average FWHM values of the point spread function (PSF) in the stacked images ranges from 0.8$\arcsec$ to 1.1$\arcsec$ in $g'$, and 0.6$\arcsec$ to 1.0$\arcsec$ in $r'$.

\section{Data Analysis and Measurement Procedures} \label{sec:analysis}

\subsection{Gas Mass} \label{sec:gasmass}

In this paper, we make use of two measurements of the HI mass, which are derived independently from the ALFALFA observations and the WSRT observations.  Measurements derived from the ALFALFA HI measurements are denoted with a prime ($'$).  The ALFALFA HI mass was taken from \cite{Haynes2018}, which calculated it as a function of the flux $S_{21}'$ and the distance $D'$ in Mpc using 
\begin{equation} \label{eq:HImass}
    M_{HI}' [M_{\odot}] = 2.356 \times 10^5 D'^2 S_{21}',
\end{equation}
with an uncertainty calculated with
\begin{equation}
    \sigma_{logM_{HI}'} = \frac{\sqrt{(\frac{\sigma_{S_{21}'}}{S_{21}'})^2 + (\frac{2\sigma_D'}{D'})^2 + 0.1^2}}{ln(10)}.
\end{equation}

The WSRT HI mass $M_{HI}$ and its associated uncertainty were calculated using the WSRT data with the same equations.  In three of four cases, the masses yielded by both measurements were in agreement within uncertainties, but for AGC~229398, WSRT appears to have only recovered about 40\% of the mass measured by ALFALFA.

The mass of atomic gas in an AD was estimated by multiplying the HI mass by 1.33 in order to account for the presence of helium.  Molecular gas is not expected to be a significant contributor to mass in ADs; \cite{Wang2020} suggest that the main contributor to the low star formation rate of ADs is a low efficiency of conversion of atomic gas to molecular gas.  Measurements of atomic and molecular gas in TDGs in \cite{Lelli2015} indicate that molecular gas mass can range from a few percent up to about a third of the atomic gas mass, though these estimates are for young objects which presumably have a larger reservoir of $H_2$ gas to support their higher current star formation rate compared to the AD-TDG candidates.  Since we were not sure which (if any) of these objects are TDGs, we didn't want to overestimate the gas or baryonic content; therefore, the baryonic mass calculation did not consider molecular gas as a significant contributor, and so $M_{atomic}' = 1.33 \times M_{HI}' = M_{gas}'$ for data originating from ALFALFA and $M_{atomic} = 1.33 \times M_{HI} = M_{gas}$ for WSRT data.

\subsection{HI Kinematics}\label{sec:HI_kin}

\begin{figure}
    \centering
    \includegraphics[height=0.23\textheight]{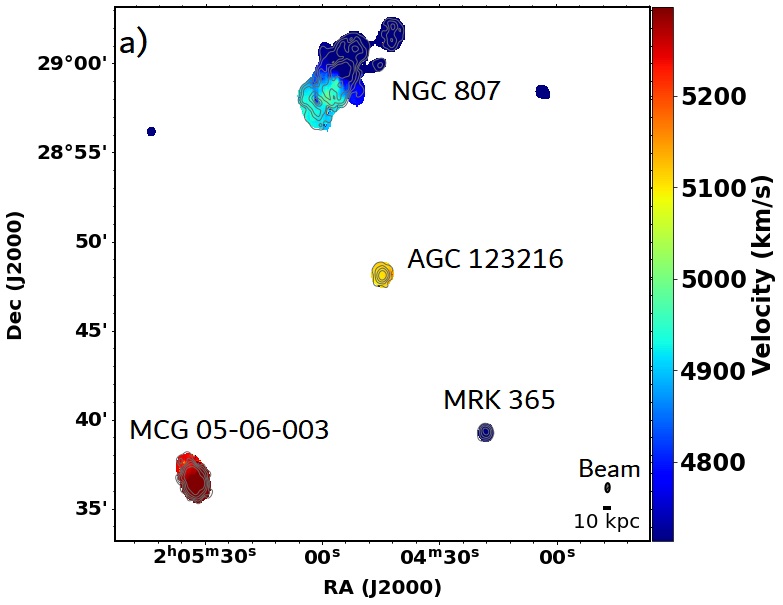}
    \includegraphics[height=0.23\textheight]{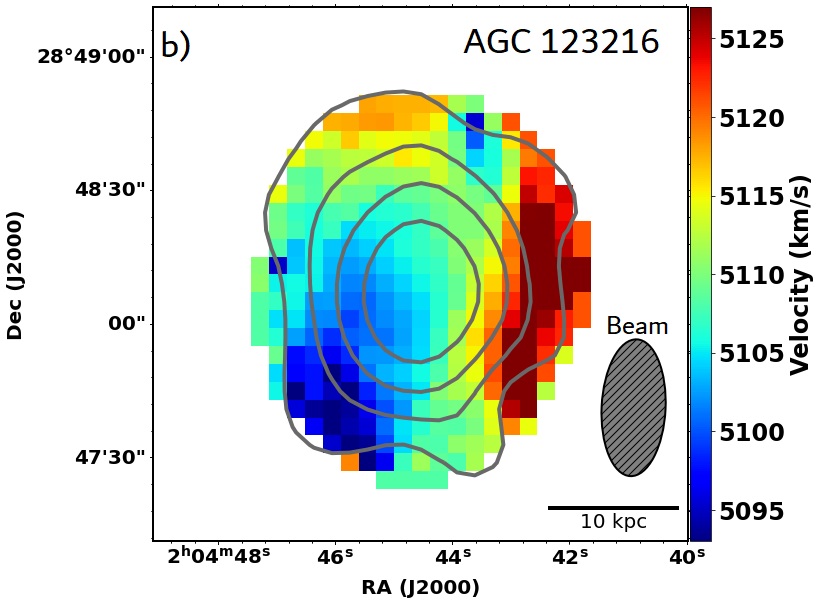}
    \includegraphics[height=0.23\textheight]{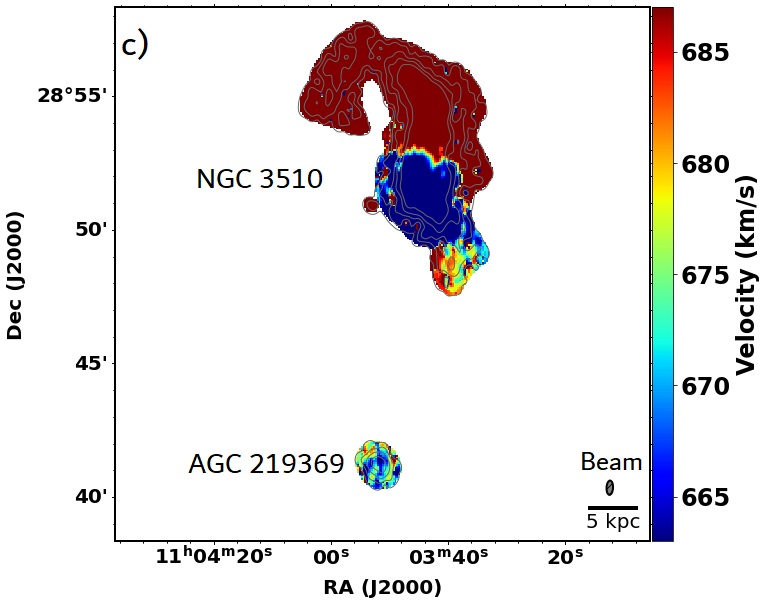}
    \includegraphics[height=0.23\textheight]{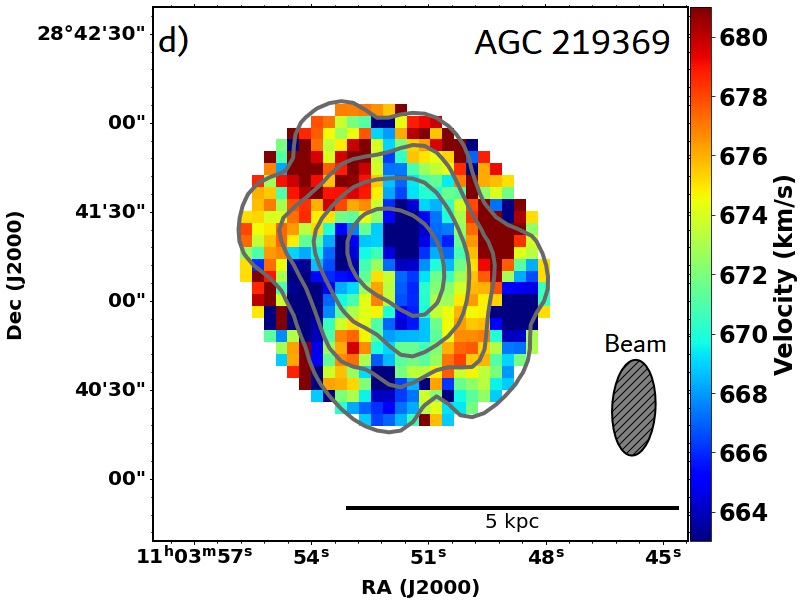}
    \includegraphics[height=0.23\textheight]{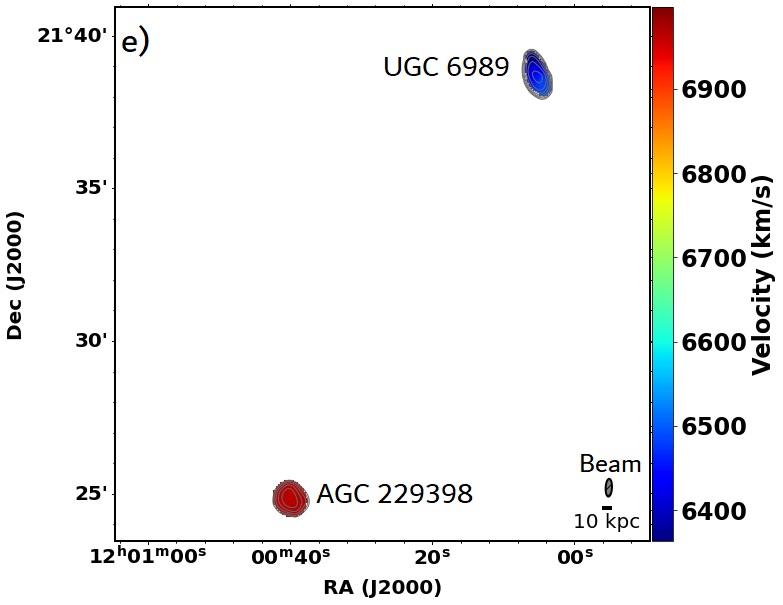}
    \includegraphics[height=0.23\textheight]{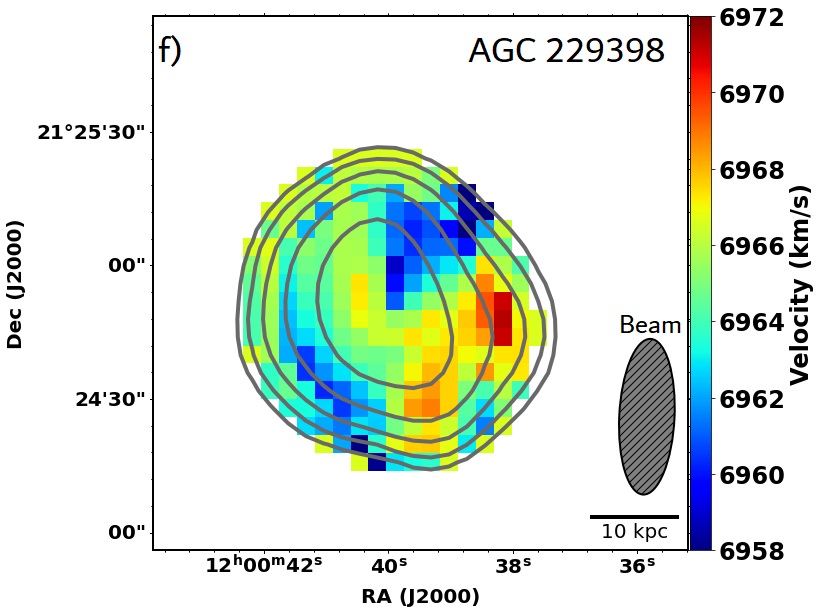}
    \includegraphics[height=0.23\textheight]{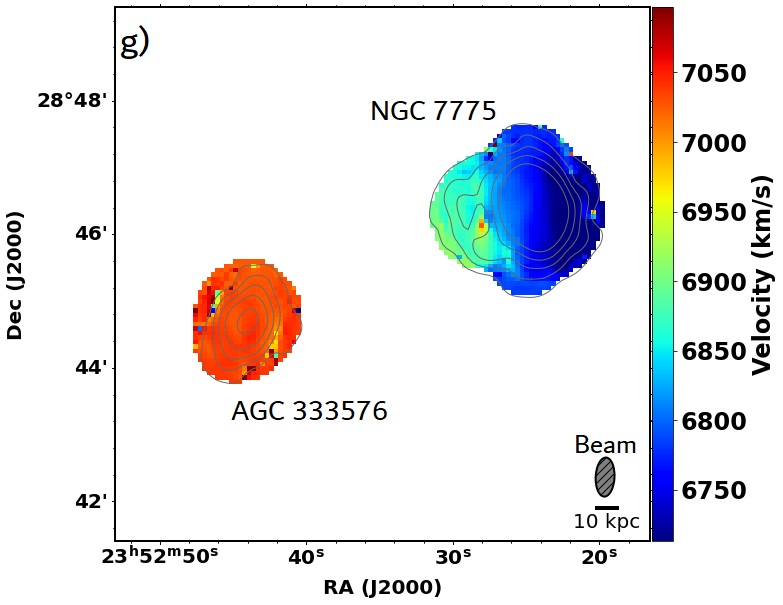}
    \includegraphics[height=0.23\textheight]{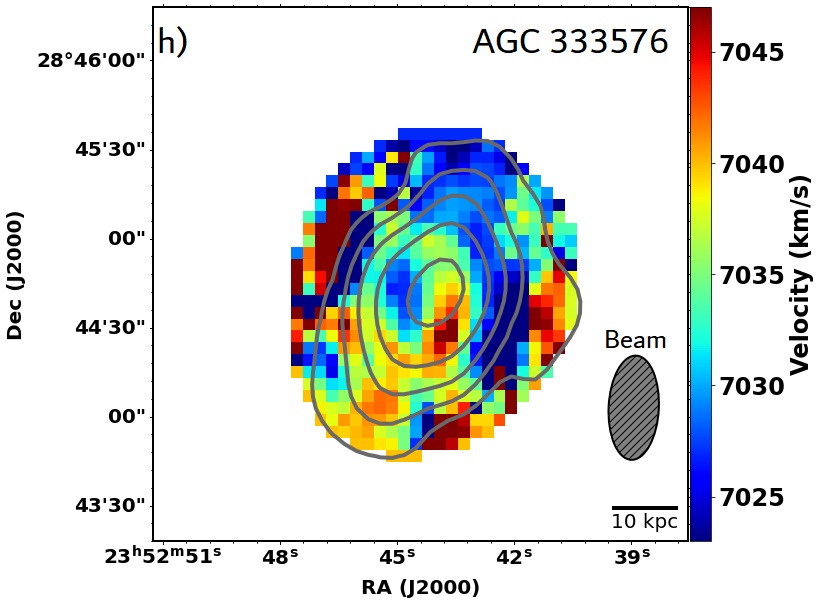}
    \caption{Moment one velocity maps from WSRT imaging, with WSRT HI column density contours superimposed in dark gray, for (from top to bottom) AGC~123216, AGC~219369, AGC~229398, and AGC~333576, and their nearby potential parent galaxies. The left panels show the full field around each AD, and the right panels show a close-up of the AD.  Beam size is shown as a shaded gray ellipse.  The figures for AGC~123216, AGC~219369, and AGC~333576 have column density contours at $N_{HI}$~=~(0.1, 0.5, 1.2, 2.4, 4.5) $\times$ $10^{20}$ $\mathrm{cm}^{-2}$, while the figures for AGC~229398 have contours at $N_{HI}$~=~(0.1, 0.2, 0.4, 0.8, 1.6) $\times$ $10^{20}$ $\mathrm{cm}^{-2}$, where the lowest contours are the first listed.  Note that KUG~1158+216 is not included in the full-field map for AGC~229398 because we did not have WSRT HI maps for it.}
    \label{fig:velmaps}
\end{figure}

The HI observations of our galaxies allow us to explore their kinematics and investigate whether the systems show signs of rotational support or if their ISM is more dominated by turbulence. TDGs are expected to be primarily supported by rotation (e.g. \citealt{Duc2007, Duc2012, Lelli2015}), but measuring rotation in TDGs is challenging, especially for lower mass objects \citep{Hibbard2001, LeeWaddell2012, Duc2014, LeeWaddell2014}.  Figure \ref{fig:velmaps} shows the WSRT moment one maps for each AD and its potential parent galaxies, as well as a close-up of the velocity field for each AD, with WSRT HI column density contours overlaid.  The distortions in the HI maps themselves will be discussed in the results sections for each object.  In the early stages of a TDG's evolution, we would expect a continuity in the velocity gradient between a parent and the TDG along the tidal tail connecting them.  However, this link will diminish over time as the tail fades, the parent system restabilizes, and the TDG moves away.  Because of this, while the presence of such a connection would support identification of a TDG, the absence of one does not disqualify it.  From the left panels of Fig. \ref{fig:velmaps}, there do not appear to be any HI tails connecting an AD to a potential parent galaxy.  AGC~219369 appears to be close to the velocity at the southern end of NGC~3510 and the velocity of AGC~333576 follows the direction of the gradient that is across NGC~7775, but otherwise there are no obvious relationships in velocity-space between the AD-TDG candidates and any of their potential parents. The velocity fields of the ADs themselves appear to be messy and disordered, with the exception of AGC~123216 which appears to show some ordered rotation.

Given the low spatial resolution of the WSRT observations, extracting kinematic information is not straightforward as the observations are heavily affected by beam smearing, which tends to erase velocity gradients and to increase the apparent gas velocity dispersion \citep{Swaters1999, Jozsa2007, DiTeodoro2015}. In order to account for this when extracting kinematic information from our data, we use the software \textsuperscript{3D}Barolo \citep{DiTeodoro2015}, which incorporates a forward modeling approach that considers the beam shape of our observations, and therefore is largely unaffected by beam smearing \citep{Jozsa2007, DiTeodoro2015, Iorio2017, ManceraPina2020}.

Prior to running \textsuperscript{3D}Barolo, we obtain independent constraints on the geometrical parameters (center, kinematic position angle, and inclination). The inclination, $i$, is a particularly crucial parameter given that $V_{rot} = V_{l.o.s}/sin(i)$. While \textsuperscript{3D}Barolo can constrain the inclination angle at high spatial resolution, this becomes challenging for low-resolution observations (and for slowly-rising rotation curves in general), and therefore we estimate the geometric parameters based only on the HI morphology of the galaxies and independently from the kinematics. The approach we follow has already been introduced in \cite{ManceraPina2020}, \cite{Fraternali2021} and \cite{ManceraPina2022}, and consists of generating a set of beam-convolved, azimuthally-averaged models of our galaxies with different geometrical parameters drawn from flat prior distributions using a MCMC Bayesian framework.\footnote{The software to do this is called Cannubi, and it is available at https://www.filippofraternali.com/cannubi.} Each model is compared against the real total intensity map, and residuals are minimized using a $\chi^2$ routine. Given that the inclination can have a degeneracy with the thickness of the disks, which we do not know precisely for our galaxies, we performed different tests assuming razor-thin and thick disks up to about 500 pc motivated by observations \citep{Bacchini2020b, ManceraPina2022b}, finding a very good agreement between both methods.  We use our MCMC method to estimate the center, position angle, and inclination of our galaxies based on their HI morphology, and we fix those parameters when modeling the data with \textsuperscript{3D}Barolo.

\begin{figure}
    \centering
    \includegraphics[width=0.98\columnwidth]{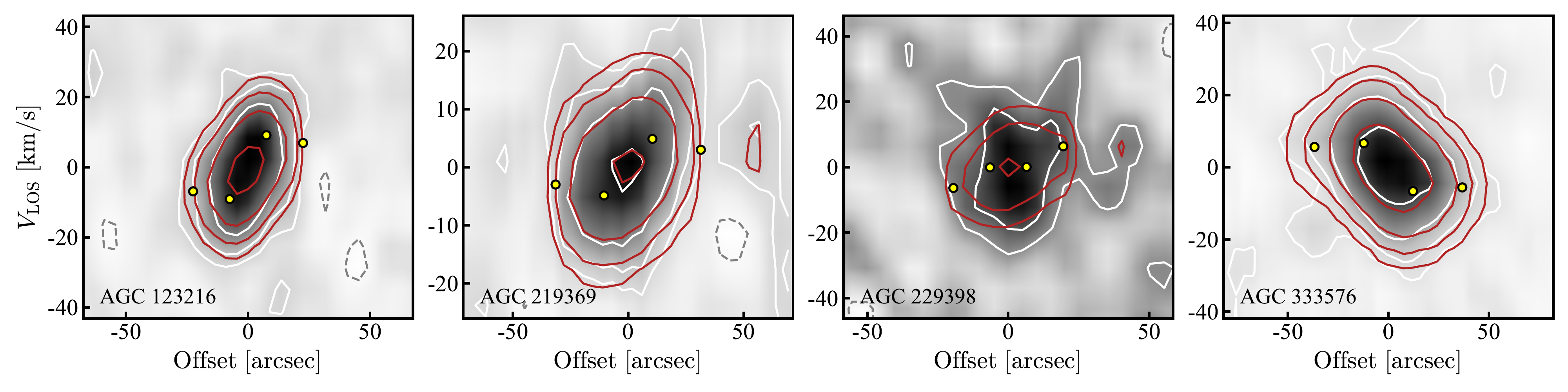}
    \caption{Position-velocity slices of 1-pixel width along the major-axis from the best-fitting model of each AD-TDG candidate, from WSRT data.  The data is shown in gray, with white contours at 2, 4, 8, and 16 times the signal to noise (S/N) of the data.  Dashed dark gray contours represent -2 $\times$ S/N.  The red contours represent the best-fit model, and the yellow points mark the line-of-sight rotation velocities from the model.}
    \label{fig:PV_diag}
\end{figure}

We show our kinematic models in Fig. \ref{fig:PV_diag}, where we display the major-axis PV slices of the data and best-fitting model. For these models, we have used a ring separation of length $\sqrt{b_{maj} \times b_{min}}$. AGC~123216, AGC~219369, and AGC~333576 show clear gradients in velocity, although the gas velocity dispersion is of comparable magnitude.  The results of the modeling indicate that the velocity gradients can be attributed to a disk with differential rotation for these three galaxies.  For AGC~229398, we did not manage to obtain a fully satisfactory model, but our preliminary results indicate that the system may be pressure supported.  More definitive models would require data with a much higher angular resolution.

For completeness, we include the emission-line channel maps of each of our galaxies and their best-fitting models in Appendix A. With each channel showing the line-of-sight velocities around the systemic velocity, the maps show evidence for rotation in AGC~123216, AGC~219369, and AGC~333576.

\subsection{Dynamical Mass} \label{sec:dynmass}

For a self-gravitating object, the dynamical mass of the HI region within the radius $r_{HI}$ can be estimated as
\begin{equation} \label{eq:dynmass}
    M_{dyn} [M_{\odot}] = 6.78 \times 10^4 r_{HI} D' (\frac{W_{20}'}{2sin(i)})^2,
\end{equation}
where $r_{HI}$ is the radius of the HI in arcminutes derived from the kinematic modeling, $D'$ and $W_{20}'$ are the distance to the source in Mpc and the width of the velocity profile at 20\% of the peak flux density in $\mathrm{km}\,\mathrm{s}^{-1}$, respectively \citep{Haynes2018}. The inclination $i$ is derived from the MCMC fitting of geometric parameters described in Section \ref{sec:HI_kin}.

TDGs are distinguished by their low dark matter content, which we evaluate using the dynamical-to-gas mass ratio, so we want to ensure that we are not underestimating the dynamical mass.  Due to the relatively narrow line widths of the HI for these objects, we used $W_{20}'$ rather than the more commonly used width at 50\% of the velocity profile peak $W_{50}'$, which will provide a more stringent upper limit on the dynamical mass. 
Additionally, using the emission line width as a proxy for the rotation velocity overestimates the dynamical mass due to artificial line broadening from beam-smearing effects \citep{Lelli2015}.  This effect is more pronounced in systems where the velocity dispersion is of comparable magnitude to the rotational velocity \citep{Lelli2015}, which is the case for these objects.  However, it is also important to consider that if the ADs are more pressure-supported than rotation-supported, these dynamical mass values may be comparatively underestimated.  These measurements do not account for turbulence and asymmetric drift, which may also increase the dynamical mass estimate.  We estimated a rough correction for turbulence and asymmetric drift using

\begin{equation}
    M_{dyn} [M_{\odot}] = 6.78 \times 10^4 r_{HI} D' ((\frac{\sqrt{W_{20}'^2 - \sigma^2}}{2sin(i)})^2 + 3\sigma^2),
\end{equation}

\noindent and a velocity dispersion ($\sigma$) of 11 $\mathrm{km}\,\mathrm{s}^{-1}$ to evaluate how it may affect the results.  We found that it increased the dynamical mass calculated by 23\% for AGC~123216, 40\% for AGC~219369, 45\% for AGC~229398, and 66\% for AGC~333576.  However, this did not change the conclusions about the nature of each object (see Section \ref{sec:origin}), and so the reported values do not reflect this correction as it was mostly exploratory.  As stated previously, a robust estimation of the correction through kinematic modeling would require data with a much higher spatial resolution \citep{Iorio2017}.  

This calculation assumes that the sources are in dynamical equilibrium, which is a reasonable assumption given the distances of the objects from any nearby perturbers.  The apparent rounded shapes of the HI distributions may be an additional indicator of dynamical equilibrium, but this may also be an effect of the elongated beam of WSRT, which could be artificially extending the emission along that axis.  Both the MCMC routine and \textsuperscript{3D}Barolo take this beam smearing into account for the geometric fitting and kinematic modeling.

\subsection{Photometry and Surface Brightness} \label{sec:phot}

We measured surface brightness profiles for the AD-TDG candidates using a procedure similar to those described in \cite{ManceraPina2019b} and \cite{Marasco2019}.  Briefly, a concentric set of ellipses were placed, centered on the optical component of the object.  The mean pixel brightness and standard deviation within each band (meaning the space between adjacent ellipses) was calculated and pixels with values more than a certain number of standard deviations above the mean were masked and excluded from further calculations.  For AGC~123216, AGC~229398, and AGC~333576, the threshold was five standard deviations, while for AGC~219369 it was four standard deviations.  Pixels within three pixels of the one above the threshold were also masked, in order to reduce contamination from an unassociated bright source that may have bled into the surrounding region. A new mean brightness and associated standard deviation were then calculated.  The final threshold and mask growth parameters were determined by evaluating the results of various combinations for a setting that was as conservative as possible while still removing objects that were clearly not part of the AD (usually objects that were much brighter, redder, and/or significantly more concentrated).  In a few cases, a region was manually masked in order to remain consistent between images (e.g. a very red object that did not have enough emission in $g$-band to exceed the threshold, but which should still not be included).  Due to the faint and irregular nature of the objects, fitting apertures to the objects was done manually. The visible portions of AGC~219369 and AGC~333576 are fairly round, so circular apertures were centered on the optical portion.  The optical counterparts of AGC~123216 and AGC~229398 appear to be quite elongated (in spite of their more circular HI gas distributions), so we measured the major and minor axis of the visible components to estimate a center and ellipticity, as well as a position angle.  For each of the four sources, ellipses were separated by two times the larger FWHM of the two filters for that source.  This separation allowed us to achieve a consistent and precise measurement of surface brightness across filters without the space between ellipses being smaller than the resolution scale of the image.  Ellipses were extended until the last ellipse before the signal to noise ratio fell below one in order to be sure we were capturing all of the light.  The apparent magnitude was calculated using the total amount of light contained within the outermost ellipse.

\begin{figure}
    \centering
    \includegraphics[width=0.59\linewidth]{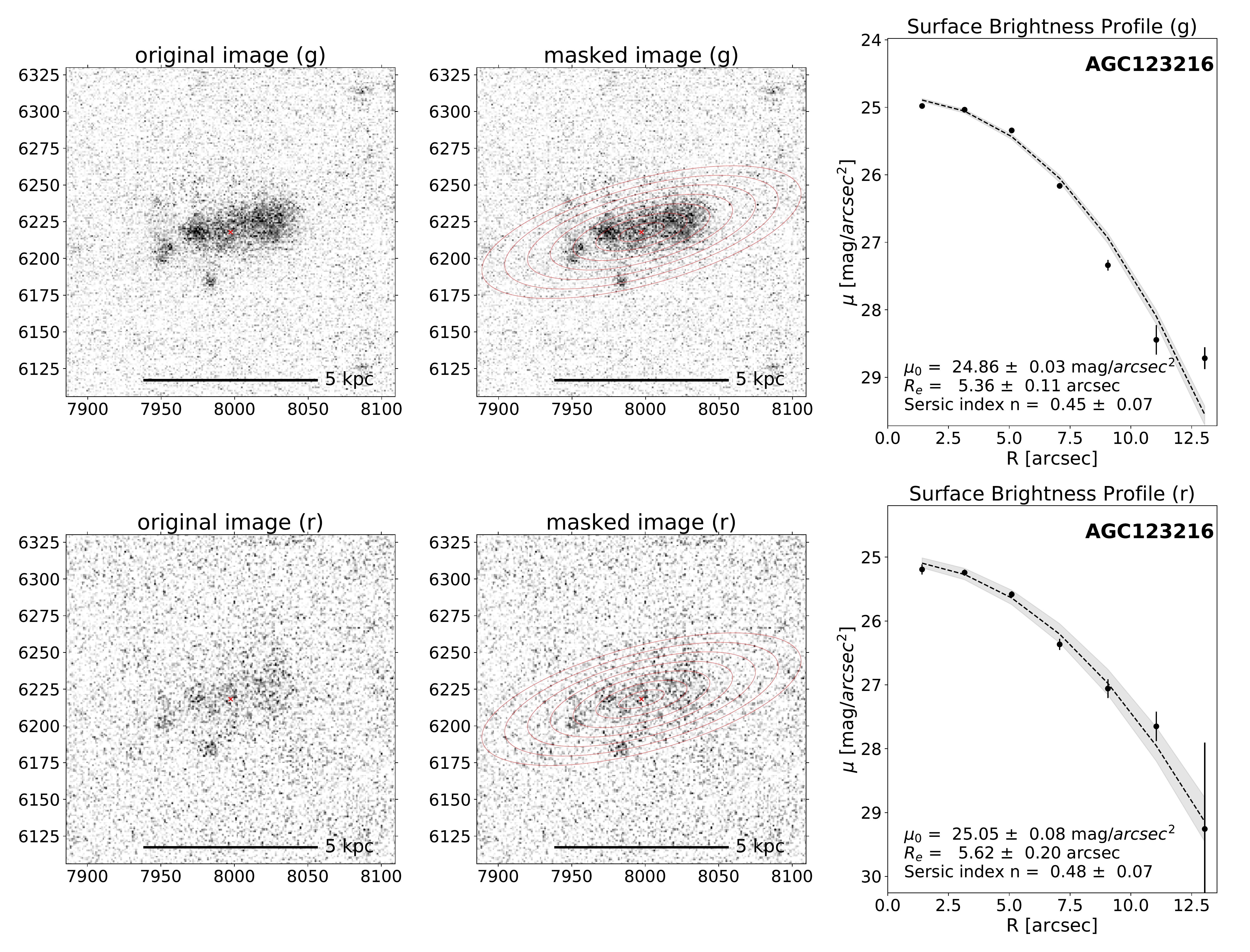} \hfill
    \caption{Masking, surface brightness profiles, and profile fit parameters for AGC~123216 in $g'$ (top row of each panel) and $r'$ (bottom row of each panel), with optical images from the WIYN 3.5m telescope.  For each object, the left images show the original image with a small red x marking the optical center; the center images show the ellipses that were placed in red and pixels that were masked in green, based on the masking procedure described in Section \ref{sec:phot}; and the right images show the best fitting surface brightness profile, along with the uncertainty shaded in light gray and the parameters for that fit listed below. For AGC~123216, there were no pixels flagged for masking in either filter.}
    \label{fig:AGC123216_sbprof}
\end{figure}

\begin{figure}
    \centering
    \includegraphics[width=0.59\linewidth]{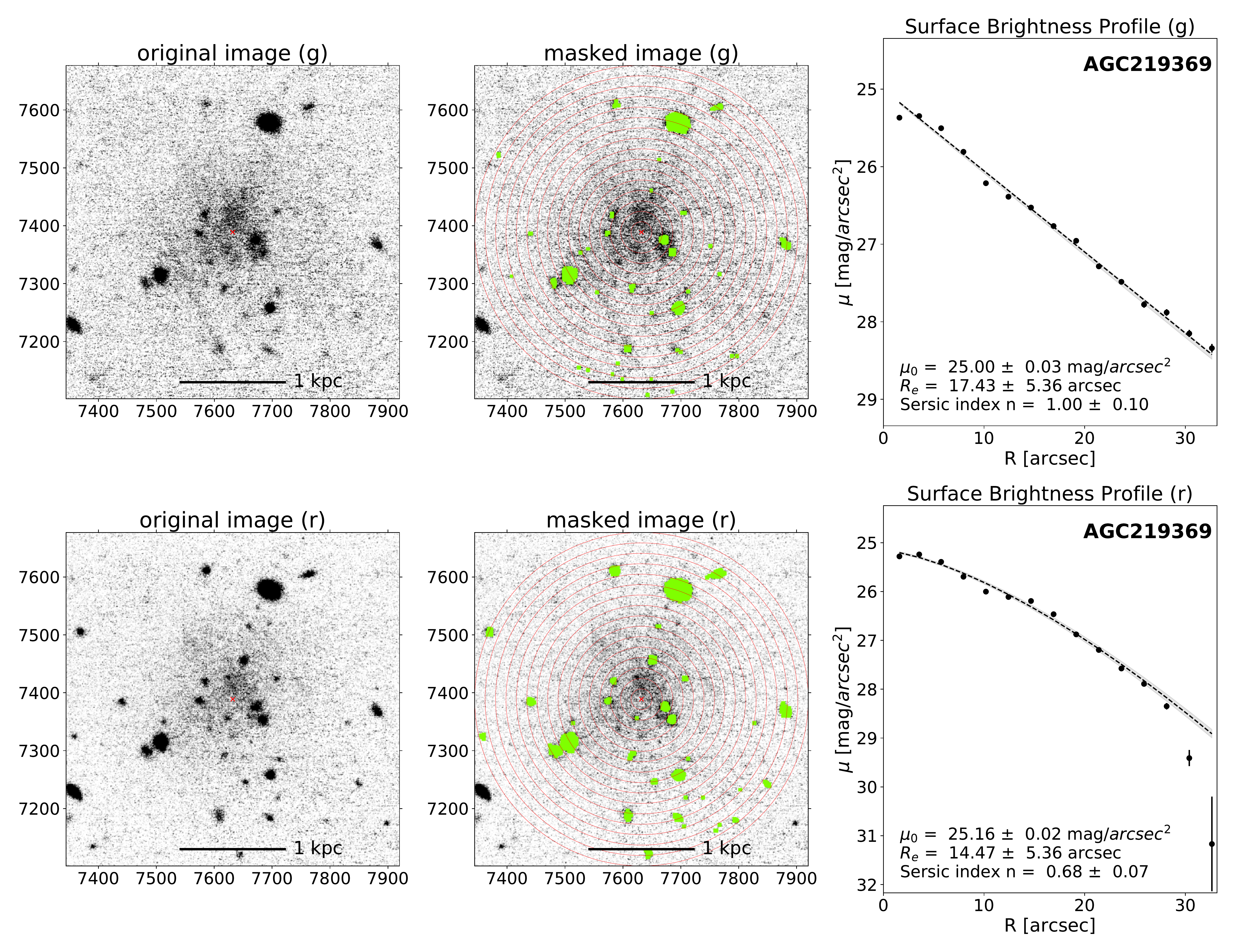}  
    \caption{Masking, surface brightness profiles, and profile fit parameters for AGC~219369 in $g'$ (top row of each panel) and $r'$ (bottom row of each panel), with optical images from the WIYN 3.5m telescope.  Figure description is the same as Fig. \ref{fig:AGC123216_sbprof}. The physical scale bar for AGC~219369 is calculated for a distance of 9.2 Mpc.}
    \label{fig:AGC219369_sbprof}
\end{figure}

\begin{figure}
    \centering
    \includegraphics[width=0.59\linewidth]{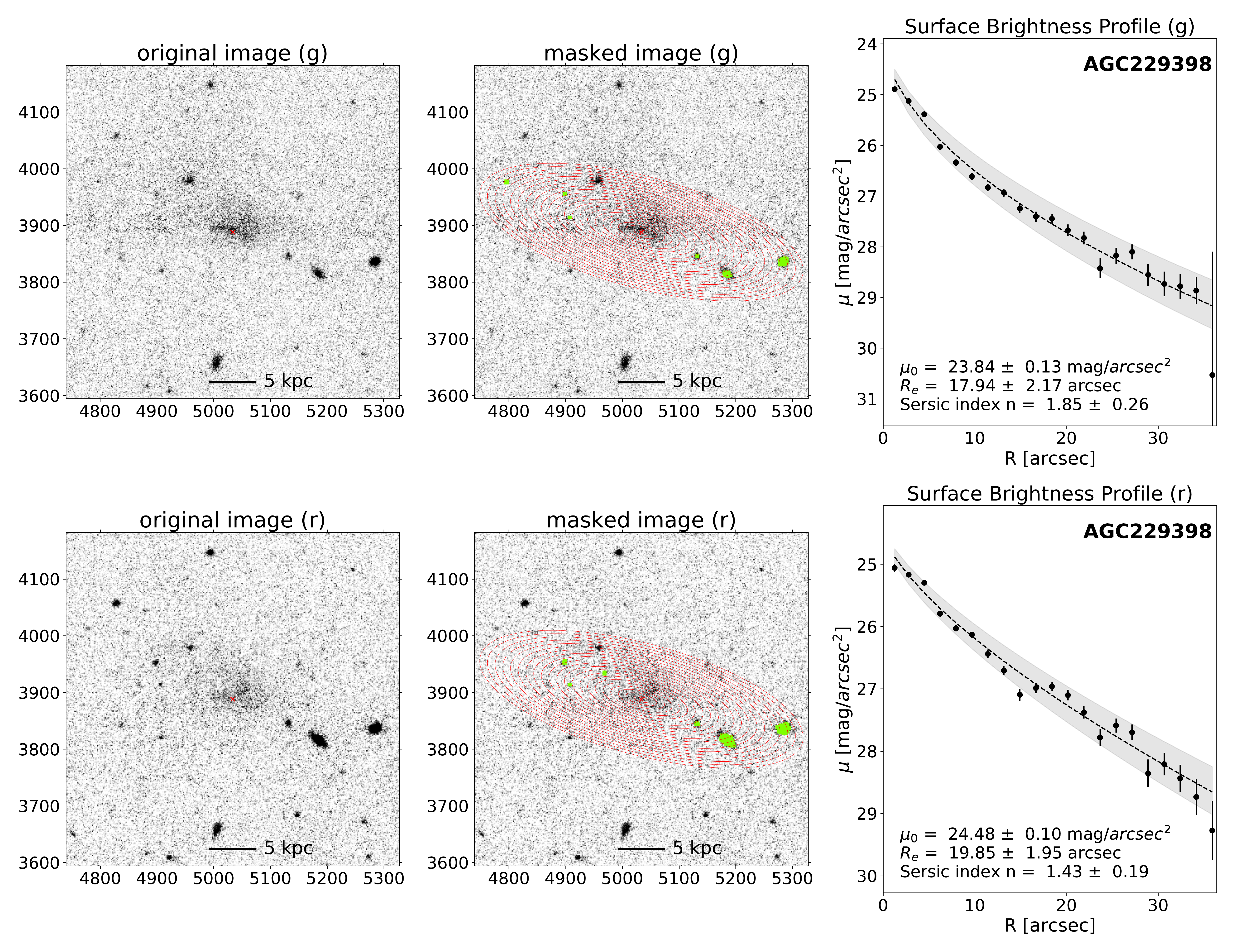}  \hfill
    \caption{Masking, surface brightness profiles, and profile fit parameters for AGC~229398 in $g'$ (top row of each panel) and $r'$ (bottom row of each panel), with optical images from the WIYN 3.5m telescope.  Figure description is the same as Fig. \ref{fig:AGC123216_sbprof}.}
    \label{fig:AGC229398_sbprof}
\end{figure}

\begin{figure}
    \centering
    \includegraphics[width=0.59\linewidth]{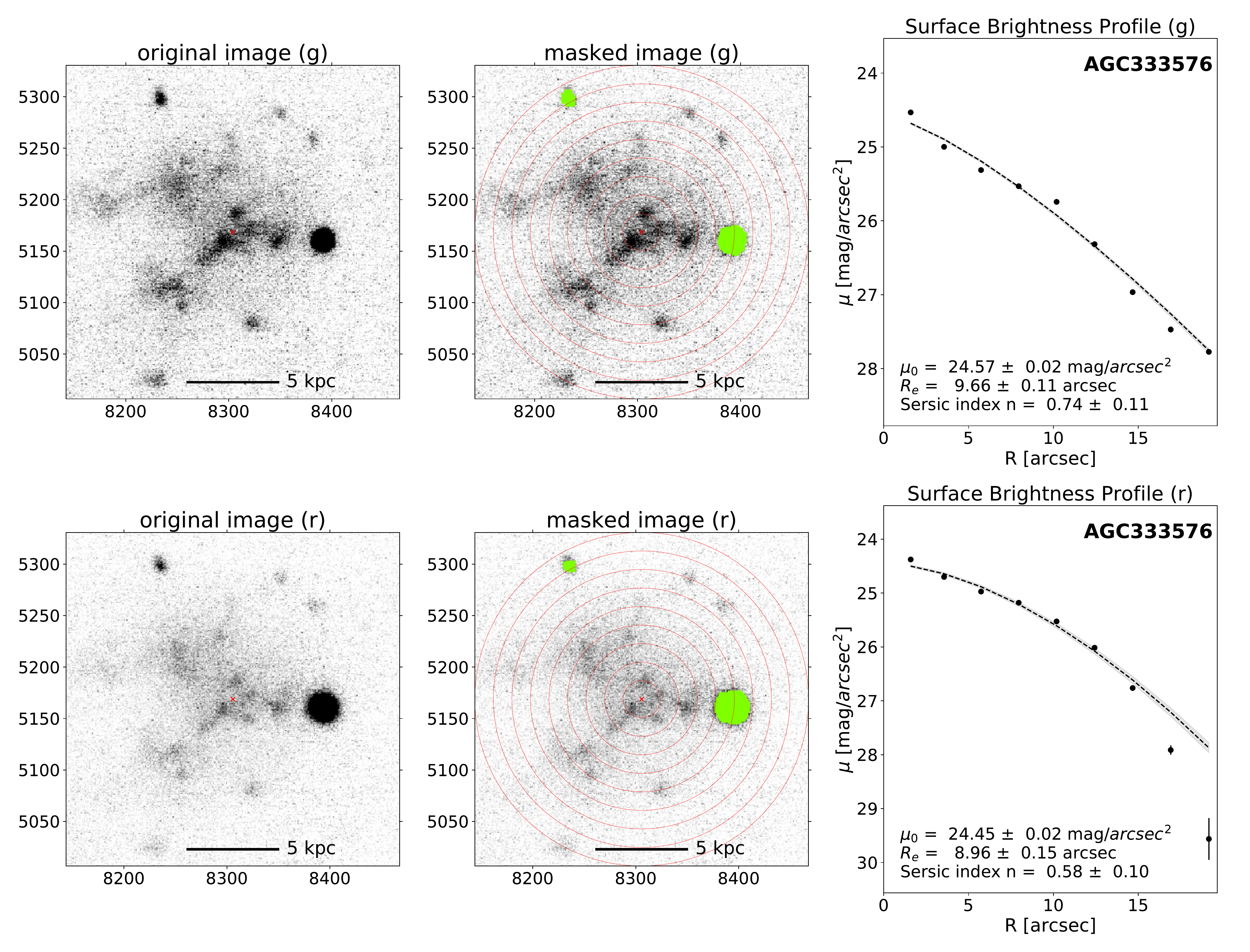}
    \caption{Masking, surface brightness profiles, and profile fit parameters for AGC~333576 in $g'$ (top row of each panel) and $r'$ (bottom row of each panel), with optical images from the WIYN 3.5m telescope.  Figure description is the same as Fig. \ref{fig:AGC123216_sbprof}.}
    \label{fig:AGC333576_sbprof}
\end{figure}

We obtained a central surface brightness and effective radius by fitting a S\'{e}rsic profile to the observed surface brightness profile using the equation 
 \begin{equation}
    \mu(r) = \mu_{0} + 1.0857*b_n*(r/r_e)^{1/n},
 \end{equation}
where $\mu(r)$ is the observed surface brightness at radius $r$, $\mu_{0}$ is the central surface brightness ($r$ = 0), $r_e$ is the effective radius, $n$ is the S\'{e}rsic index, and $b_n$ is approximated as $2n-\frac{1}{3}+\frac{4}{405n}+\frac{46}{25515n^2}+\frac{131}{1148175n^3}-\frac{2194697}{30690717750n^4}$ for $n$ $>$ 0.36 \citep{Ciotti1999}.  The scale radius $r_d$ was obtained with the relation $r_d = r_e/(b_n)^n$.  All of the fits were near-exponential, with $n$ ranging from 0.45 to 1.85.  The masking and resulting surface brightness profiles for each AD in $g'$ and $r'$ are shown in Figs. \ref{fig:AGC123216_sbprof} -- \ref{fig:AGC333576_sbprof}.

We measured the magnitudes and colors of SDSS stars in each field and used them to calculate zero-points and color terms for converting our measured instrumental magnitudes and colors to calibrated values.  We also corrected our photometry for extinction using the \cite{Schlegel1998} dust maps and the coefficients from \cite{Schlafly2011}.  Though we report results in both $g$ and $r$ bands, we generally focus on the $g$-band results, as these objects are quite blue and thus slightly better defined in $g$.  This is made clear by the difference between the leftmost panels of the surface brightness figures for each object, shown in Figs. \ref{fig:AGC123216_sbprof} -- \ref{fig:AGC333576_sbprof}.  Therefore, the effective radius and scale radius reported for each object are those calculated using the $g$-band image.

\subsection{Stellar Mass and Baryonic Mass}

We can photometrically estimate the stellar mass ratios and therefore stellar mass of a galaxy using color-stellar mass-to-light ratio relations (CMLRs) that relate the optical color and absolute magnitude to the stellar mass.  The CMLRs have been calibrated with model data and observational samples, but the extreme nature of the galaxies in this study means that it is unclear which fits are most appropriate. Using the measured $g-r$ color and total magnitudes from the largest ellipse, we explored the stellar mass-to-light ratio and resulting stellar mass estimates of two CMLRs: \cite{Du2020}, which is calibrated for low surface brightness galaxies, and \cite{Herrmann2016}, which is calibrated for dwarf irregular galaxies.  The \cite{Herrmann2016} relation consistently gave the highest stellar masses, and the \cite{Du2020} values are closer in agreement between the two optical bands than the \cite{Herrmann2016} ones.  In this paper, we report the average of the values derived in both optical bands between the two relations for the stellar masses of the optical counterparts because it is unclear which is a better fit for this unusual group of objects.  The stellar mass uncertainty is heavily dependent on the uncertainty in the color measurement, and was estimated by varying the color by adding or subtracting the associated error and calculating the resulting stellar mass.  

We also evaluated the stellar luminosity by using AB solar magnitudes from \cite{Willmer2018} to convert the absolute magnitudes in $g$ and $r$ to luminosity in $g$ and $r$ ($L_g$ and $L_r$).  This allowed us to calculate the HI-mass-to-stellar-luminosity ratios $M_{HI}'/L_g$ and $M_{HI}'/L_r$.

ALFALFA has a larger beam than WSRT, so it is more sensitive to diffuse, low surface brightness HI emission and therefore yields a more accurate measurement of the total HI content, so we estimate the baryonic mass as $M_{bary}'$ = $M_{gas}'$ $+$ $M_{*}$.

\section{Results} \label{sec:results}

\subsection{AGC~123216} \label{sec:A123216}

\begin{figure}
    \centering
    \includegraphics[width=.6\linewidth,height=.6\textheight]{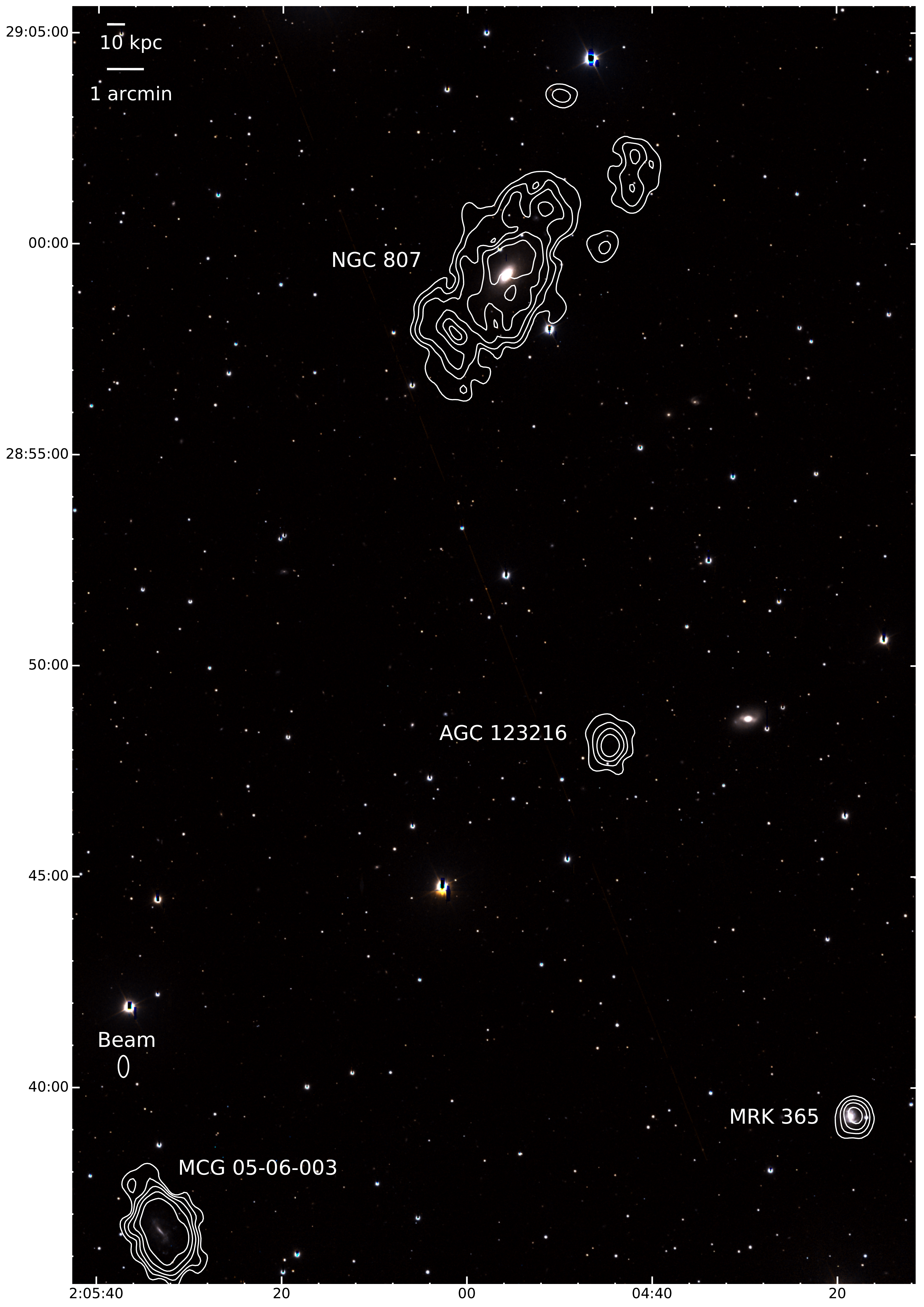}
    \caption{WIYN 3.5m $g'+r'$-band composite image showing the full-field view of AGC~123216 (center) with potential parents NGC~807 (top), MRK~365 (bottom right), and MCG~05-06-003 (bottom left).  There are WSRT HI column density contours at $N_{HI}$~=~(0.1, 0.5, 1.2, 2.4, 4.5) $\times$ $10^{20}$ $\mathrm{cm}^{-2}$ superimposed in white.}
    \label{fig:AGC123216full}
\end{figure}

There are three potential parent galaxies within 500 kpc in projected distance and with $v_{helio}$ within 500 $\mathrm{km}\,\mathrm{s}^{-1}$ of AGC~123216, all of which show distortions in their HI emission (Fig. \ref{fig:AGC123216full}).  We discuss them here in order of increasing angular separation from AGC~123216.

\begin{figure}
    \centering
    \includegraphics[width=0.5\textwidth]{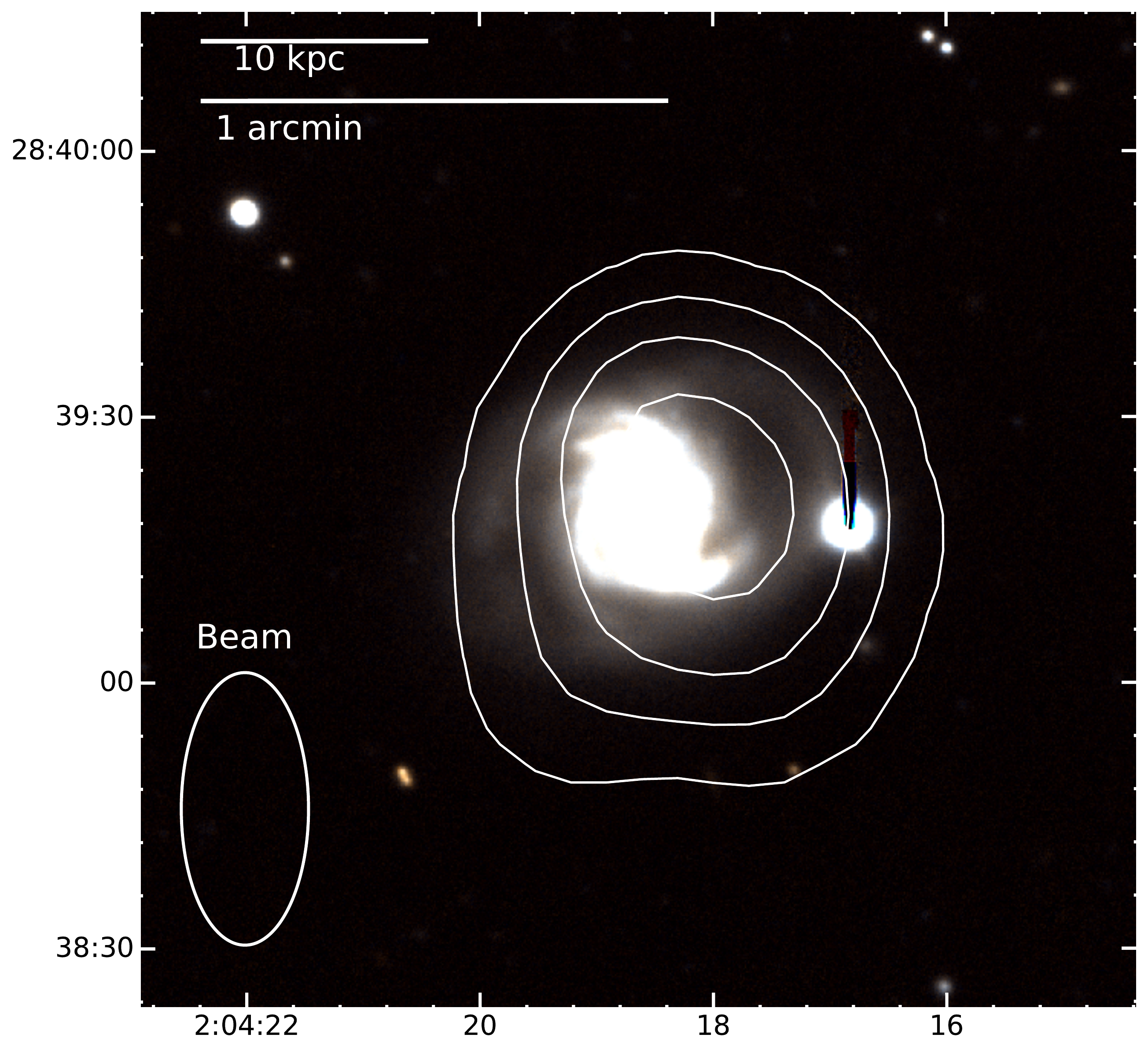}
    \caption{WIYN 3.5m $g'+r'$-band composite image of MRK~365 with WSRT HI column density contours at $N_{HI}$~=~(0.1, 0.5, 1.2, 2.4) $\times$ $10^{20}$ $\mathrm{cm}^{-2}$ in white.  The HI contours are slightly offset from the stellar disk, and the image is scaled to show a faint loop of stellar emission on the right side, between the first and second innermost contours.}
    \label{fig:MRK365}
\end{figure}

\begin{figure}
    \centering
    \includegraphics[width=0.49\textwidth]{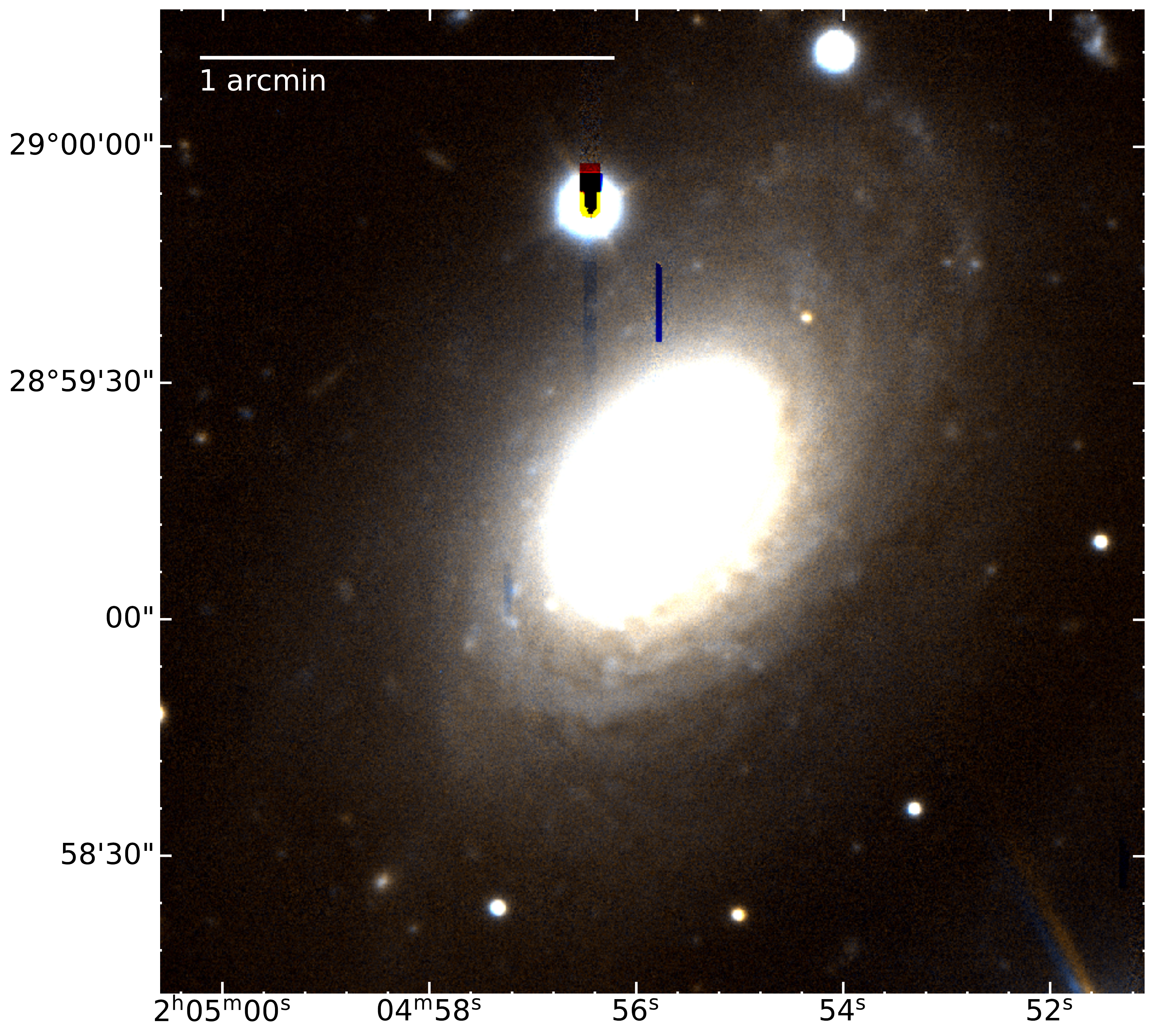}
    \includegraphics[width=0.49\textwidth]{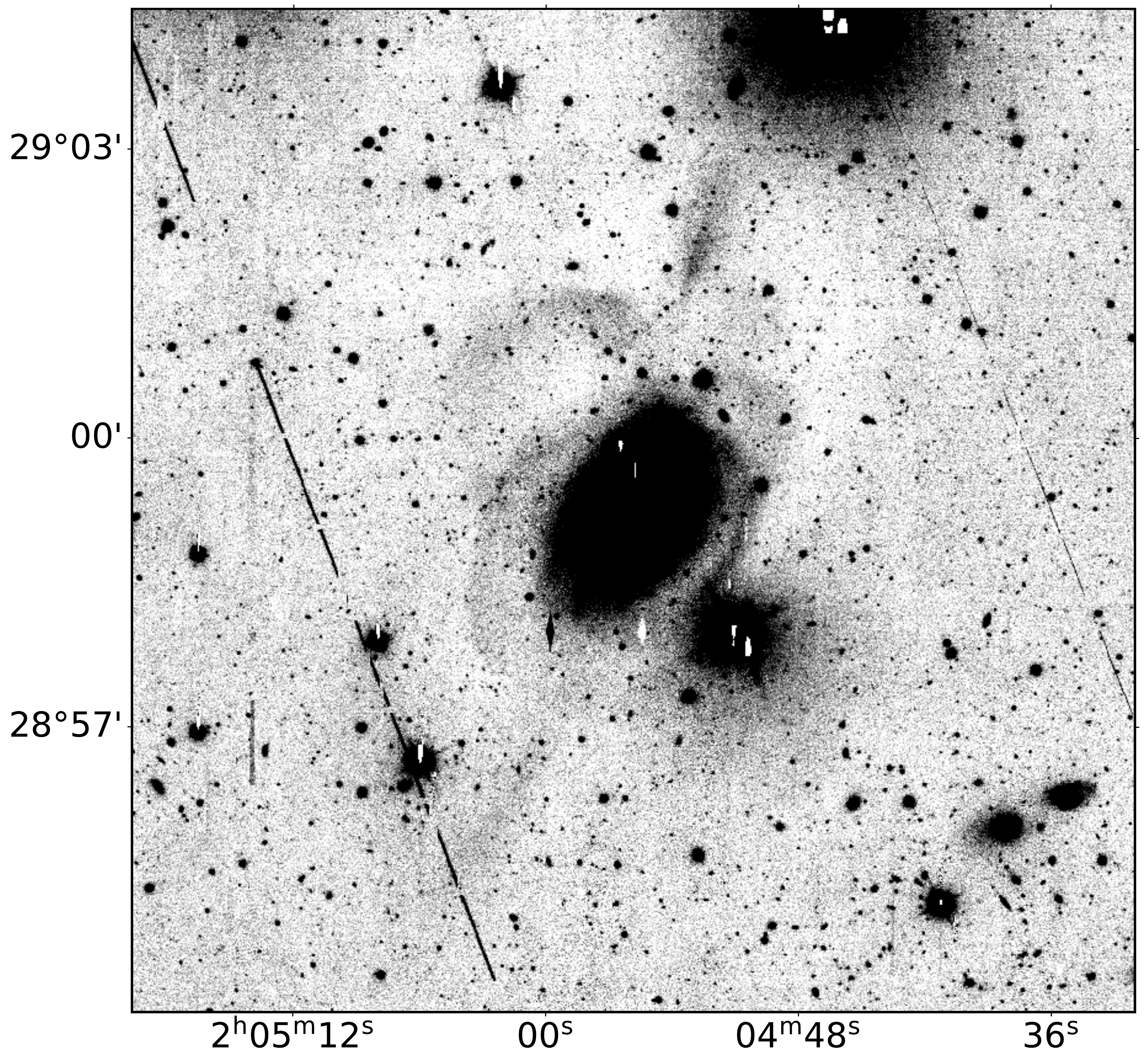}
    \caption{WIYN 3.5m $g'+r'$-band composite images of NGC~807.  Left: Color image revealing low surface brightness, dusty spiral arms surrounding a large and bright bulge.  The image has been scaled to show the full extent of the faint disk, which overemphasizes the bulge at the center.  Right: Composite $g'+r'$-band summed image in inverted grayscale showing arms of stellar emission extending in several directions, primarily out of the northern edge of the disk.  The image has been binned 3x3 to make the arms easier to see.  Note that this image is a different scale than the left panel, as the region of the disk in that image is the entire black area here.}
    \label{fig:NGC807}
\end{figure}

MRK~365 is a spiral galaxy to the southwest of AGC~123216.  MRK~365 is the potential parent that is closest to AGC~123216, with a minimum projected separation of 214 $\pm$ 7 kpc and a heliocentric velocity difference of 40 $\pm$ 15 $\mathrm{km}\,\mathrm{s}^{-1}$.  The HI gas in MRK~365 appears to be offset from the optical center (Fig. \ref{fig:MRK365}), and high-sensitivity images reveal a low surface brightness string of stellar emission wrapping around the outside of the galaxy.  It has been listed in the literature as having a double nucleus, which would suggest a previous or ongoing merger, though there is discussion about whether this is true \citep{Gimeno2004}.  \cite{Mezcua2014} constructed a luminosity profile and found that it was consistent with the presence of a single exponential disk; this is a sign that if there is a merger, it is in a late stage.  

NGC~807 is the next closest potential parent to AGC~123216, with a minimum projected separation of 231 $\pm$ 8 kpc and a heliocentric velocity difference of 361 $\pm$ 218 $\mathrm{km}\,\mathrm{s}^{-1}$.  It has been classified as an elliptical galaxy in multiple catalogs \citep{MCG1964, UGC1973, RC31991}.  Since the vast majority of TDGs originate from interactions between gas-rich galaxies, an elliptical galaxy would not normally be considered as a likely progenitor (especially with two other potential spiral parents nearby).  However, the extended amount of HI surrounding NGC~807 is unusual for an elliptical galaxy, and it has clearly been disturbed (Fig. \ref{fig:AGC123216full}).  Fig. \ref{fig:velmaps}(a) shows two small clumps of HI to the left and right of NGC~807 which appear to have similar heliocentric velocities to that galaxy.  The WSRT data is most sensitive towards the center of the image, so due to their location towards the edge of the image, these may not be statistically significant measurements of gas.  Additionally, they do not appear as sources in the ALFALFA catalog, which indicates that there is unlikely to be a deposit of low surface density gas at these locations that the smaller beam of WSRT is resolving out.  \cite{Dressel1987} found that the neutral hydrogen in NGC~807 has the double-horned line profile characteristic of a rotating disk of gas.  Further studies with deeper optical imaging have even revealed faint tidal arms \citep{Lucero2013}.  \cite{Young2002} mapped the CO emission of NGC~807 and found that it was strongly asymmetric, with 70\% of the CO mass being located in the southeast half of the galaxy; \cite{Lucero2013} also estimated that the southeastern tail comprised 70\% of the HI gas mass within the tails.  \cite{Young2002} mentions that the asymmetry of the CO gas would not be expected to persist for more than 1 Gyr due to shearing by differential rotation.  Inspection of Fig. \ref{fig:NGC807} reveals that the bright bulge is in the center of a low surface brightness spiral disk with dust lanes (left panel), and there are several low surface brightness features (tidal arms) extending outwards in several directions (right panel).  In this way, NGC~807 may have some similarities to UGC~1382, a galaxy believed to be a normal elliptical until deep UV and optical imaging revealed a set of low surface brightness spiral arms \citep{Hagen2016}. 

At a projected separation of 331 $\pm$ 11 kpc, MCG~05-06-003 is the farthest potential parent from AGC~123216.  Not much is known about MCG~05-06-003, which appears to be an edge-on spiral galaxy with a significant bar.  The HI map shows a slight extension of HI to the northeast, perpendicular to the direction of AGC~123216.  Given the closer proximity (see Table \ref{tab:ALFALFAdata}) along with extensive HI distortion or apparent recent interaction history of the other two potential parent galaxies, we consider MCG~05-06-003 to be the least likely of the three galaxies to be the progenitor of AGC~123216.

\begin{figure}
    \centering
    \includegraphics[width=0.6\linewidth]{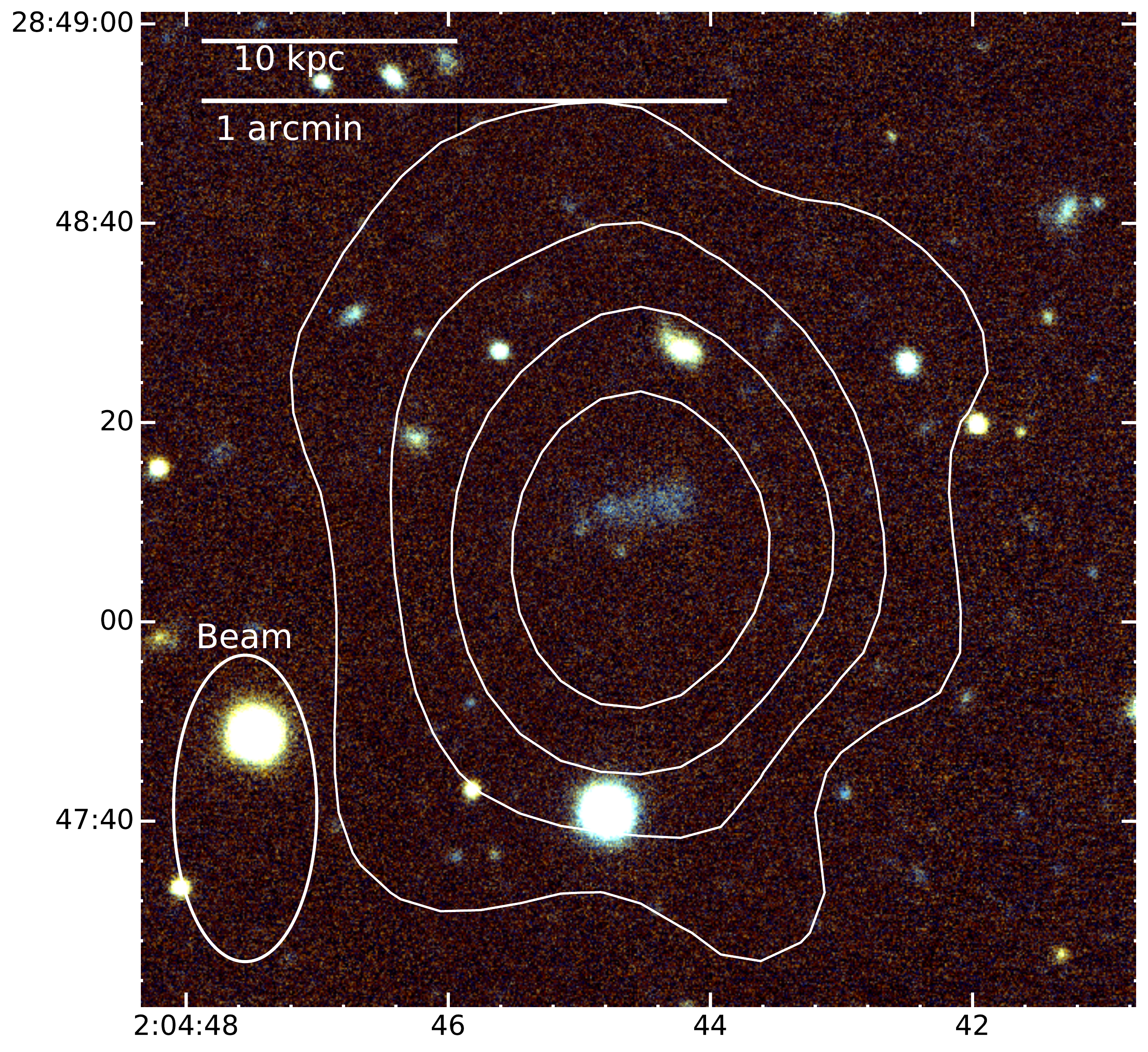}
    \caption{WIYN 3.5m $g'+r'$-band composite image of AGC~123216, with WSRT HI column density contours at $N_{HI}$~=~(0.1, 0.5, 1.2, 2.4) $\times$ $10^{20}$ $\mathrm{cm}^{-2}$.  The stellar component at the center is extremely blue and elongated in the east-west direction, while the HI distribution is rounder.}
    \label{fig:AGC123216closeup}
\end{figure}

AGC~123216 itself has an interesting appearance, with an optical component that is elongated in the east-west direction but surrounded by a relatively rounded concentration of HI gas, although this appearance is somewhat dependent on the elongated beam (Fig. \ref{fig:AGC123216closeup}).  The measured properties of AGC~123216 can be found in Table \ref{tab:AGC123216results}, and the surface brightness profiles are shown in Fig. \ref{fig:AGC123216_sbprof}.  It has a central surface brightness in $g$ of 24.86 $\pm$ 0.03 $\mathrm{mag}\,\mathrm{arcsec}^{-2}$ and an absolute total magnitude in $g$ of -13.44 $\pm$ 0.08 mag.  The $g-r$ color of -0.11 $\pm$ 0.12 makes it exceptionally blue.  The effective radius from the Sersic profile fit is 1.83 $\pm$ 0.04 kpc, which combined with the central surface brightness, places it in the regime of UDGs as defined by \cite{vanDokkum2015}.  The stellar mass of AGC~123216 is estimated at $(3.86 \pm 1.24) \times 10^6$ $M_{\odot}$, and its HI-to-stellar mass ratio is 115.28 $\pm$ 42.20.  While it is extremely gas-rich, it has apparently formed relatively few stars for its gas mass.  It has a dynamical mass of $(2.33 \pm 0.82) \times 10^9$ $M_{\odot}$, which yields a dynamical-to-gas mass ratio of 4.40 $\pm$ 1.70.

\begin{deluxetable}{lr@{\hspace{1in}}lr}[ht]
    \centering
    \caption{Measured and derived properties of AGC~123216}
    \tablehead{ \multicolumn{1}{c}{Property [Units]} & \multicolumn{1}{c@{\hspace{1in}}}{Value} & \multicolumn{1}{c}{Property [Units]} & \multicolumn{1}{c}{Value} }
    \startdata
        RA [h m s, J2000] & 02 04 44.6 & Dec [$^{\circ}$ ' ", J2000] & +28 48 11.28  \\ 
        $V_h'$ [$\mathrm{km}\,\mathrm{s}^{-1}$] & 5111 $\pm$ 14 & $N_{HI,peak}$ [$10^{20}$ $\mathrm{cm}^{-2}$] & 5.02 \\ 
        $S_{21}'$ [Jy $\mathrm{km}\,\mathrm{s}^{-1}$] & 0.38 $\pm$ 0.05 & $S_{21}$ [Jy $\mathrm{km}\,\mathrm{s}^{-1}$] & 0.34 $\pm$ 0.03 \\ 
        $M_{HI}'$ [$10^8$ $M_{\odot}$] & 4.45 $\pm$ 0.79 & $M_{HI}$ [$10^8$ $M_{\odot}$] & 3.98 $\pm$ 0.62 \\ 
        $M_{gas}'$ [$10^8$ $M_{\odot}$] & 5.91 $\pm$ 1.05 & $M_{gas}$ [$10^8$ $M_{\odot}$] & 5.30 $\pm$ 0.83 \\ 
        $R_{HI}$ [kpc] & 7.69 $\pm$ 1.06 & Inclination [$^{\circ}$] & 45 $\pm$ 5 \\
        $R_e$ [kpc] & 1.83 $\pm$ 0.04 & $R_d$ [kpc] & 2.32 $\pm$ 0.05 \\ 
        $\mu_{g,0}$ [$\mathrm{mag}\,\mathrm{arcsec}^{-2}$] & 24.86 $\pm$ 0.03 & $\mu_{r,0}$ [$\mathrm{mag}\,\mathrm{arcsec}^{-2}$] & 25.05 $\pm$ 0.08 \\ 
        $m_g$ [mag] & 20.80 $\pm$ 0.04 & $m_r$ [mag] & 20.91 $\pm$ 0.11 \\
        $M_g$ [mag] & -13.44 $\pm$ 0.08 & $M_r$ [mag] & -13.33 $\pm$ 0.13 \\
        (g-r) [mag] & -0.11 $\pm$ 0.12 & $M_{*}$ [$10^6$ $M_{\odot}$] & 3.86 $\pm$ 1.24 \\ 
        $M_{bary}'$ [$10^8$ $M_{\odot}$] & 5.96 $\pm$ 1.05 & $M_{dyn}$ [$10^9$ $M_{\odot}$] & 2.33 $\pm$ 0.82 \\
        $M_{HI}'/L_g$ & 17.04 $\pm$ 3.39 & $M_{HI}'/L_r$ & 28.80 $\pm$ 6.32 \\ 
        $M_{HI}'/M_{*}$ & 115.28 $\pm$ 42.20 & $M_{dyn}/M_{gas}$ & 4.40 $\pm$ 1.70 \\ 
    \enddata
    \tablecomments{RA and Dec are listed for the optical component of AGC~123216.  Properties marked with a $'$ are derived from ALFALFA measurements.  Inclination is measured from the HI gas distribution, not the optical component. $M_{dyn}$ is calculated using $W_{20}'$ and $D'$ from ALFALFA data and $R_{HI}$ and inclination from WSRT data, as described in Sections \ref{sec:HI_kin} and \ref{sec:dynmass}.}
    \label{tab:AGC123216results}
\end{deluxetable}

\subsection{AGC~219369} \label{sec:A219369}

\begin{figure}
    \centering
    \begin{minipage}{0.48\textwidth}
        \centering
        \includegraphics[width=\textwidth]{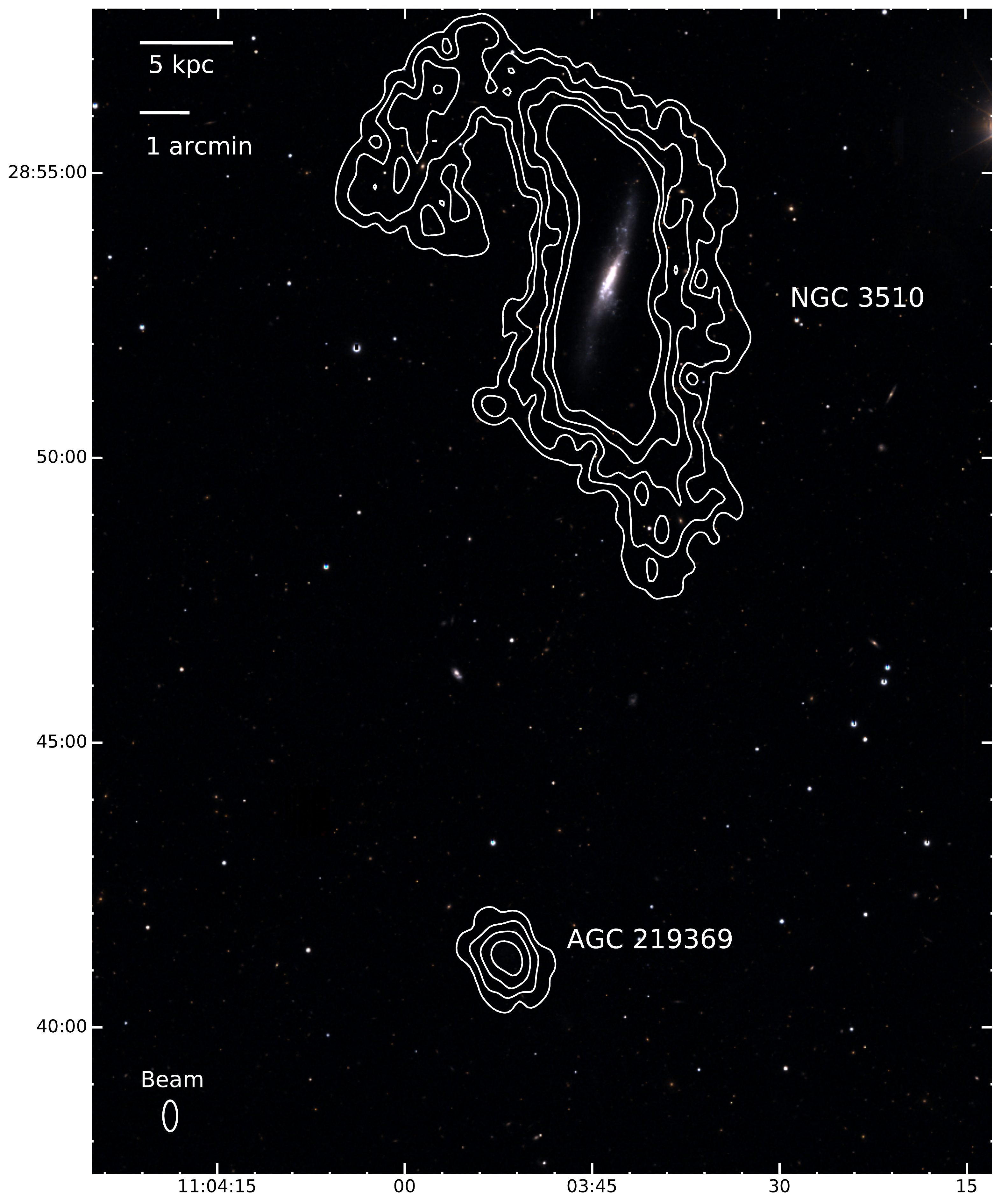}
        \caption{WIYN 3.5m $g'+r'$-band composite image showing the full-field view of AGC~219369 (bottom), including potential parent NGC~3510 (top).  There are WSRT HI column density contours at $N_{HI}$~=~(0.1, 0.5, 1.2, 2.4, 4.5) $\times$ $10^{20}$ $\mathrm{cm}^{-2}$ superimposed in white.  The physical scale bar is calculated for a distance of 9.2 Mpc.}
        \label{fig:AGC219369full}
    \end{minipage}\hfill
    \begin{minipage}{0.5\textwidth}
        \centering
        \includegraphics[width=\textwidth]{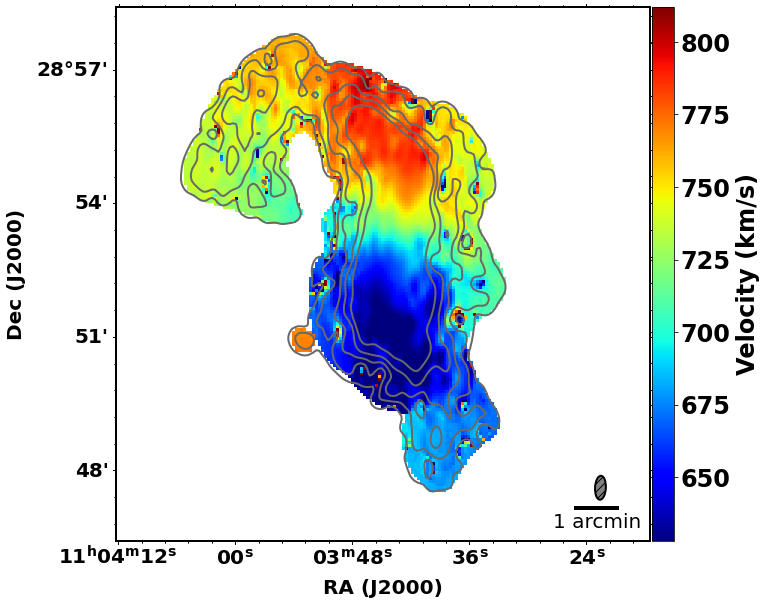}
        \caption{NGC~3510 moment one map from WSRT imaging, with WSRT HI column density contours superimposed in dark gray.  Contour levels are the same as in Fig. \ref{fig:AGC219369full}.}
        \label{fig:NGC3510_velmap}
    \end{minipage}
\end{figure}

\begin{figure}
    \centering
    \includegraphics[width=0.5\textwidth]{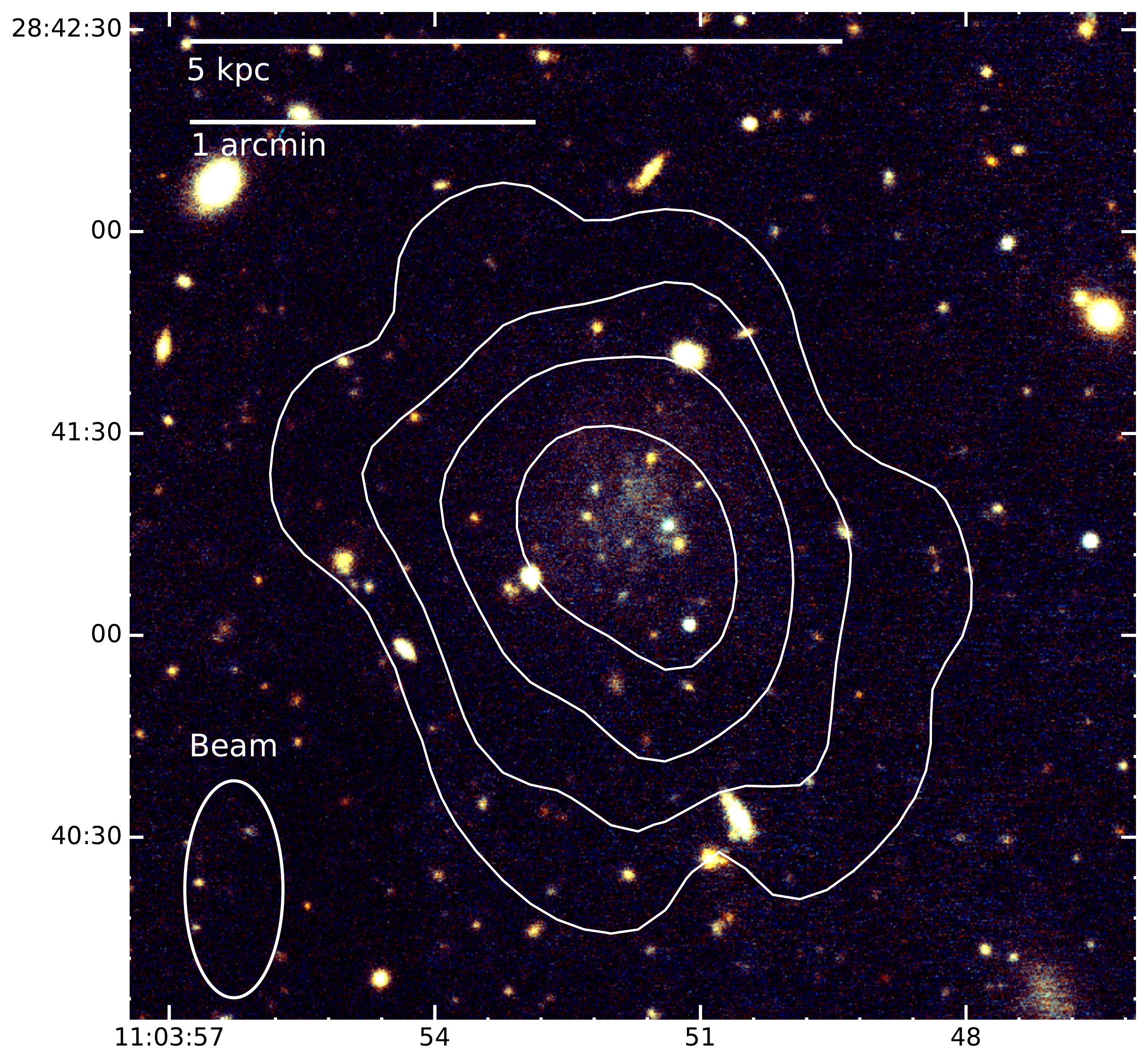}
    \caption{WIYN 3.5m $g'+r'$-band composite image of AGC~219369, with WSRT HI column density contours at $N_{HI}$~=~(0.1, 0.5, 1.2, 2.4) $\times$ $10^{20}$ $\mathrm{cm}^{-2}$ superimposed in white. The physical scale bar is calculated for a distance of 9.2 Mpc.}
    \label{fig:AGC219369closeup}
\end{figure}

AGC~219369 has a heliocentric velocity of 667 $\pm$ 11 $\mathrm{km}\,\mathrm{s}^{-1}$, which implies a distance of 9.2 $\pm$ 2.2 Mpc according to the \cite{Masters2005} flow model that was used to construct the ALFALFA catalog.  However, AGC~219369 is located in the “triple-value region" of the sky, where proximity to the Virgo Cluster affects the radial velocities of nearby objects such that objects at three different distances can have the same radial velocity \citep{Marinoni1998,Shaya2017}.  This makes measuring the distance to AGC~219369 with only the heliocentric velocity difficult.  Redshift-independent Tully-Fisher distance measurements for the nearby spiral galaxy NGC~3510 (which has a similar heliocentric velocity to AGC~219369) place it at 16.8 $\pm$ 2.5 Mpc \citep{Cosmicflows3}, and this is in agreement with the \cite{Haynes2018} distance of 16.7 $\pm$ 3.3 Mpc from the \cite{Masters2005} flow model.  NGC~3510 has clearly undergone an interaction that stretched its stellar arms and distorted its HI (Fig. \ref{fig:AGC219369full}), and AGC~219369 is the nearest object to it (angular separation of 12.2 arcminutes, which corresponds to only 59 $\pm$ 12 kpc at a distance of 16.7 Mpc), so given the ambiguity on the distance it is reasonable to consider that AGC~219369 may have a distance of 16.7 Mpc as well.  

The HI tail is significantly more substantial on the north side of NGC~3510, but tidal interactions usually produce two tidal tails which may or may not be similar sizes based on the parameters of the interaction \citep{Duc2008a}.  We can consider whether the HI that formed the southern tail may have concentrated into a more massive object at its end and separated from it, producing AGC~219369, or whether AGC~219369 originally came from the larger tail and the southern tail was always short.  In spite of the motion that the curvature of the HI tails suggests, NGC~3510 is very edge on, and Fig. \ref{fig:NGC3510_velmap} confirms that the rotation axis is roughly perpendicular to the line of sight.  This makes it particularly difficult to evaluate the state of the stellar disk, which otherwise may have provided clues about the interaction.  Although they are not pictured in Fig. \ref{fig:AGC219369full}, there are three other galaxies with similar heliocentric velocities nearby: UGC~6102 ($V_h$ = 697 $\pm$ 35 $\mathrm{km}\,\mathrm{s}^{-1}$ from \citealt{Haynes2018}), UGCA~225 ($V_h$ = 647 $\pm$ 22 $\mathrm{km}\,\mathrm{s}^{-1}$ from \citealt{Haynes2018}), and NGC~3486 ($V_h$ = 678 $\pm$ 1 $\mathrm{km}\,\mathrm{s}^{-1}$ from \citealt{Springob2005}).  UGC~6102 is a dwarf irregular galaxy with an angular separation of about 27 arcmin directly west from AGC~219369.  UGCA~225 is a Blue Compact Dwarf located about 31 arcmin to the northeast of AGC~219369, and is actively forming stars \citep{Cairos2001}.  NGC~3486 is a spiral galaxy located 49 arcmin to the northwest of AGC~219369, and is of comparable HI mass to NGC~3510 \citep{Haynes2018}.

\begin{deluxetable}{lr@{\hspace{1in}}lr}[ht]
    \centering
    \caption{\protect\label{tab:AGC219369_dist_indep} Distance-Independent Measured Properties of AGC~219369}
    \tablehead{ \multicolumn{1}{c}{Property [Units]} & \multicolumn{1}{c@{\hspace{1in}}}{Value} & \multicolumn{1}{c}{Property [Units]} & \multicolumn{1}{c}{Value} }
    \startdata
        RA [h m s, J2000] & 11 03 51.8 & Dec [$^{\circ}$ ' ", J2000] & +28 41 17.8 \\ 
        $V_h'$ [$\mathrm{km}\,\mathrm{s}^{-1}$] & 667 $\pm$ 4 & $N_{HI,peak}$ [$10^{20}$ $\mathrm{cm}^{-2}$] & 4.40 \\ 
        $S_{21}'$ [Jy $\mathrm{km}\,\mathrm{s}^{-1}$] & 0.90 $\pm$ 0.04 & $S_{21}$ [Jy $\mathrm{km}\,\mathrm{s}^{-1}$] & 0.60 $\pm$ 0.03 \\ 
        $R_{HI}$ [arcsec] & 32 $\pm$ 3 & Inclination [$^{\circ}$] & 53 $\pm$ 2 \\
        $R_e$ [arcsec] & 17.43 $\pm$ 5.36 & $R_d$ [arcsec] & 10.35 $\pm$ 3.18 \\ 
        $\mu_{g,0}$ [$\mathrm{mag}\,\mathrm{arcsec}^{-2}$] & 25.00 $\pm$ 0.03 & $\mu_{r,0}$ [$\mathrm{mag}\,\mathrm{arcsec}^{-2}$] & 25.16 $\pm$ 0.02 \\
        $m_g$ [mag] & 18.12 $\pm$ 0.02 & $m_r$ [mag] & 18.04 $\pm$ 0.02 \\ 
        (g-r) [mag] & 0.08 $\pm$ 0.03 & $M_{HI}'/M_{*}$ & 12.42 $\pm$ 5.23 \\ 
        $M_{HI}'/L_{g}$ & 3.38 $\pm$ 1.93 & $M_{HI}'/L_{r}$ & 4.78 $\pm$ 2.73 \\ 
    \enddata
    \tablecomments{RA and Dec are listed for the optical component of AGC~219369. Properties marked with a $'$ are derived from ALFALFA measurements.  Inclination is measured from the HI gas distribution, not the optical component.}
\end{deluxetable}

\begin{deluxetable}{ccc}[ht]
    \centering
    \caption{\protect\label{tab:AGC219369_dist_dep} Distance-Dependent Measured Properties of AGC~219369}
    \tablehead{ \colhead{Distance [Mpc]} & \colhead{9.2 $\pm$ 2.2} & \colhead{16.7 $\pm$ 3.3} }
    \startdata
        $M_{HI}'$ [$10^7$ $M_{\odot}$] & 1.79 $\pm$ 0.88 & 5.91 $\pm$ 2.43 \\
        $M_{gas}'$ [$10^7$ $M_{\odot}$] & 2.39 $\pm$ 1.17 & 7.87 $\pm$ 3.23 \\ 
        $M_{HI}$ [$10^7$ $M_{\odot}$] & 1.20 $\pm$ 0.59 & 3.94 $\pm$ 1.62 \\ 
        $M_{gas}$ [$10^7$ $M_{\odot}$] & 1.59 $\pm$ 0.78 & 5.24 $\pm$ 2.15 \\ 
        $R_{HI}$ [kpc] & 1.40 $\pm$ 0.36 & 2.55 $\pm$ 0.56 \\ 
        $R_e$ [kpc] & 0.78 $\pm$ 0.24 & 1.41 $\pm$ 0.52 \\ 
        $R_d$ [kpc] & 0.46 $\pm$ 0.14 & 0.84 $\pm$ 0.31 \\
        $M_g$ [mag] & -11.70 $\pm$ 0.52 & -12.99 $\pm$ 0.43 \\ 
        $M_r$ [mag] & -11.78 $\pm$ 0.52 & -13.08 $\pm$ 0.43 \\ 
        $M_{*}$ [$10^6$ $M_{\odot}$] & 1.44 $\pm$ 0.14 & 4.76 $\pm$ 0.44 \\ 
        $M_{bary}'$ [$10^7$ $M_{\odot}$] & 2.53 $\pm$ 1.17 & 8.34 $\pm$ 3.23 \\ 
        $M_{dyn}$ [$10^8$ $M_{\odot}$] & 2.60 $\pm$ 1.06 & 4.72 $\pm$ 1.81 \\ 
        $M_{dyn}/M_{gas}$ & 16.34 $\pm$ 10.42 & 9.00 $\pm$ 5.06 \\ 
    \enddata
    \tablecomments{Properties marked with a $'$ are derived from ALFALFA measurements.  $M_{dyn}$ is calculated using $W_{20}'$ ALFALFA data and $R_{HI}$ and inclination from WSRT data, as described in Sections \ref{sec:HI_kin} and \ref{sec:dynmass}.}
\end{deluxetable}

In Fig. \ref{fig:AGC219369closeup}, we can see that the HI distribution of AGC~219369 appears to be round, as is the optical component, which would suggest dynamical stability.  The surface brightness profiles of AGC~219369 are plotted in Fig. \ref{fig:AGC219369_sbprof}, with scale bars for physical units corresponding to the reported ALFALFA distance of 9.2 Mpc \citep{Haynes2018}.  The surface brightness profile in $g$ follows an exponential law very closely, and results in a central surface brightness in $g$ of 25.00 $\pm$ 0.03 $\mathrm{mag}\,\mathrm{arcsec}^{-2}$.  The $(g-r)$ color of 0.08 $\pm$ 0.03 indicates it is quite blue.  Due to the uncertainty in the distance to AGC~219369, properties that do not depend on distance are listed in Table \ref{tab:AGC219369_dist_indep}, and properties which are distance-dependent have been calculated for both 9.2 Mpc and 16.7 Mpc (the distance of NGC~3510) and are listed in Table \ref{tab:AGC219369_dist_dep}.  While both sets of values are listed for completeness, AGC~219369 is assumed to be at the distance of NGC~3510 for the rest of this work. Under this assumption, the effective radius of AGC~219369, 17.43 $\pm$ 5.36$\arcsec$, would correspond to 1.41 $\pm$ 0.52 kpc.  AGC~219369 is the smallest AD-TDG candidate in this study, with an ALFALFA HI mass of only (5.91 $\pm$ 2.43) $\times$ $10^7$ $M_{\odot}$ and an estimated stellar mass of (4.76 $\pm$ 0.44) $\times$ $10^6$ $M_{\odot}$.  This gives an HI-to-stellar mass ratio of 12.42 $\pm$ 5.28.  A dynamical mass of (4.72 $\pm$ 1.81) $\times$ $10^8$ $M_{\odot}$ leads to a dynamical-to-gas mass ratio of 9.00 $\pm$ 5.06.

\subsection{AGC~229398} \label{sec:A229398}

\begin{figure}
    \centering
    \includegraphics[width=.7\linewidth,height=.7\textheight]{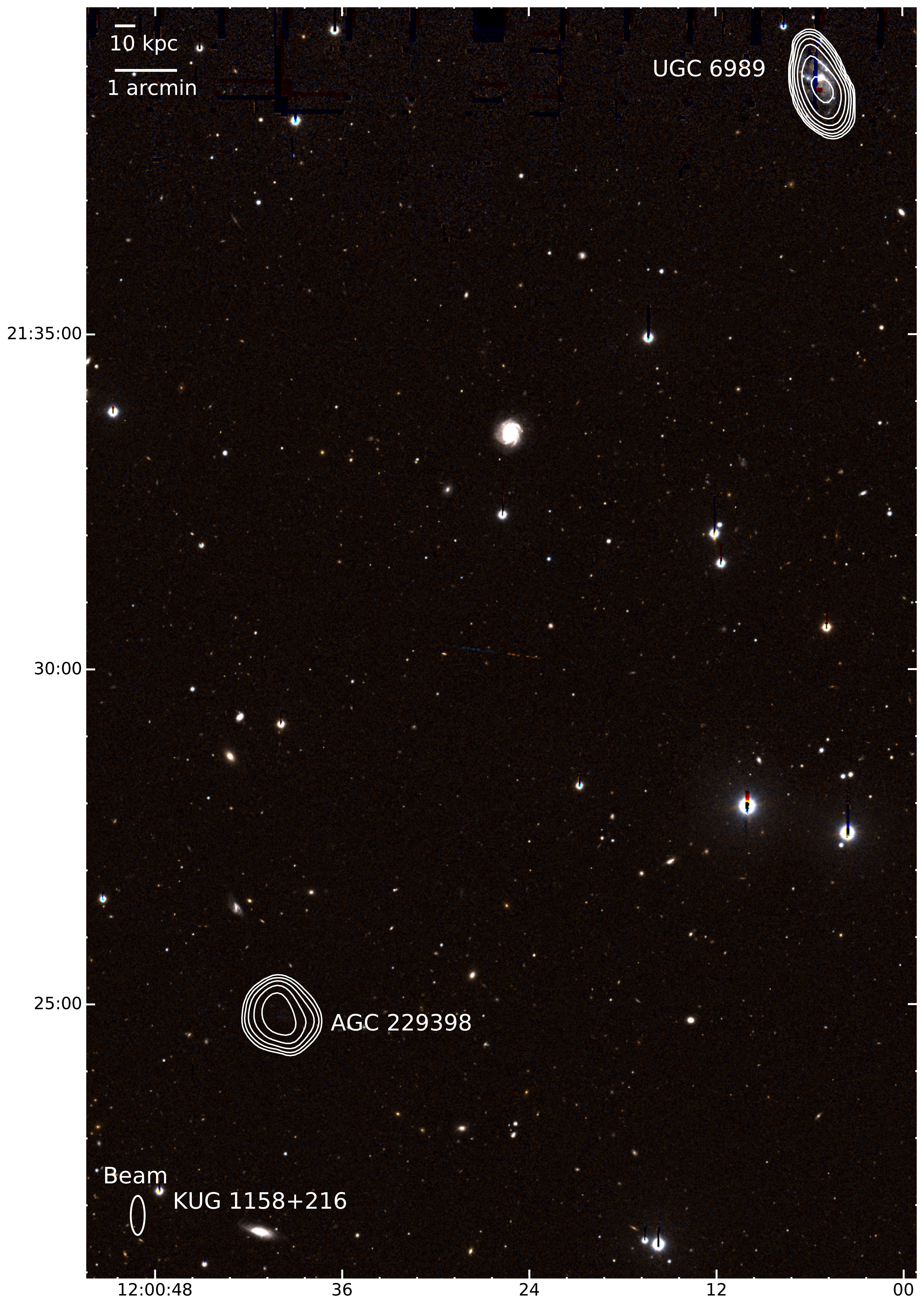}
    \caption{WIYN 3.5m $g'+r'$-band composite image showing the full-field view of AGC~229398 (lower left) with potential parents UGC~6989 (upper right) and KUG~1158+216 (below label). There are WSRT HI column density contours at $N_{HI}$~=~(0.1, 0.2, 0.4, 0.8, 1.6, 3.2, 6.4) $\times$ $10^{20}$ $\mathrm{cm}^{-2}$ superimposed in white.  Note that the prominent spiral galaxy near the center of the image is CGCG~127-132, which has a heliocentric velocity of 14,457 $\pm$ 4 $\mathrm{km}\,\mathrm{s}^{-1}$ and is not associated with the objects of interest in this paper \citep{SDSSDR13}.}
    \label{fig:AGC229398full}
\end{figure}

\begin{figure}
    \centering
    \includegraphics[width=0.6\textwidth]{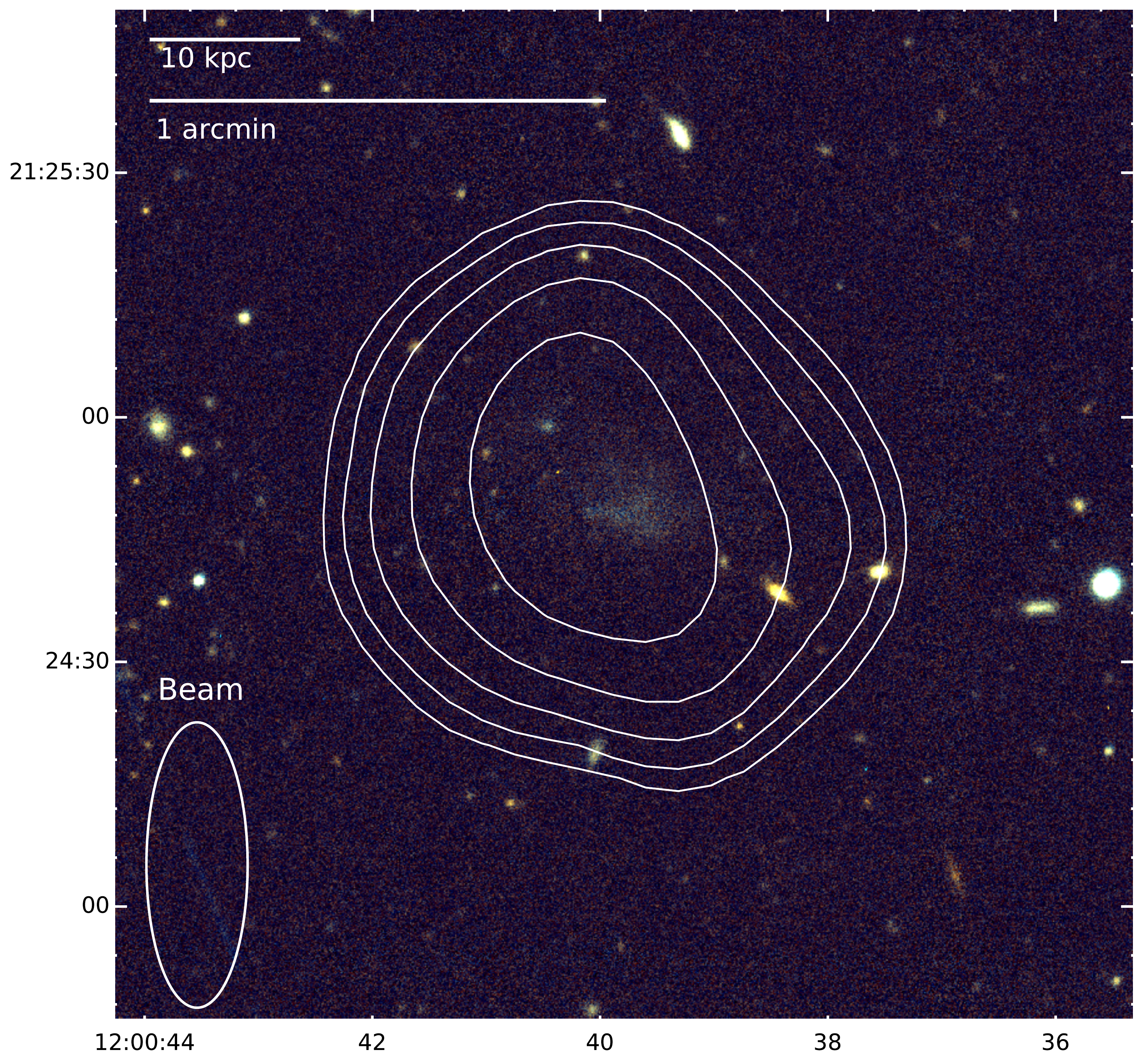}
    \caption{WIYN 3.5m $g'+r'$-band composite image of AGC~229398, with WSRT HI column density contours at $N_{HI}$~=~(0.1, 0.2, 0.4, 0.8, 1.6) $\times$ $10^{20}$ $\mathrm{cm}^{-2}$ in white.}
    \label{fig:AGC229398close}
\end{figure}

\begin{figure}
    \centering
    \includegraphics[width=0.49\textwidth]{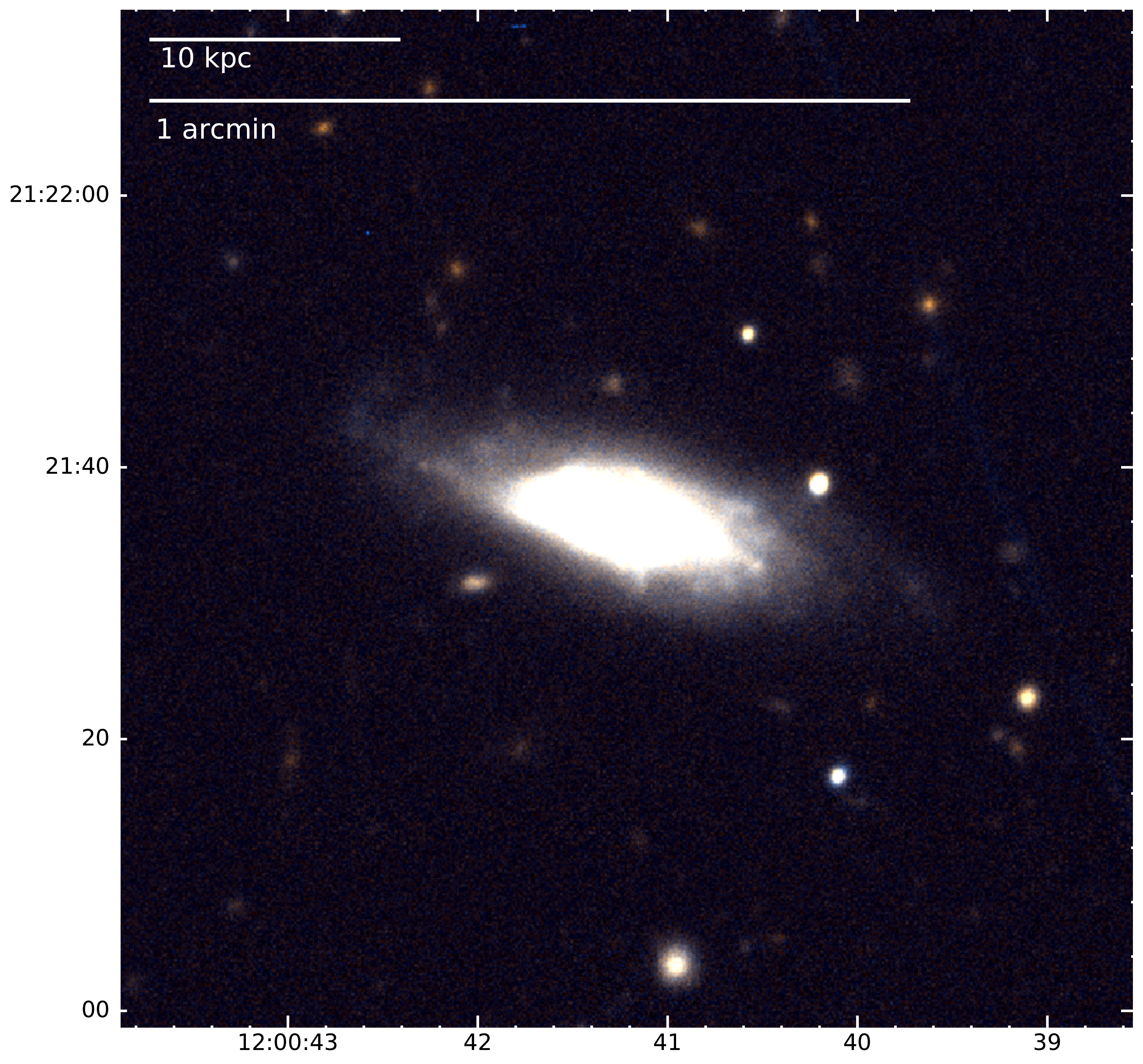}
    \includegraphics[width=0.49\textwidth]{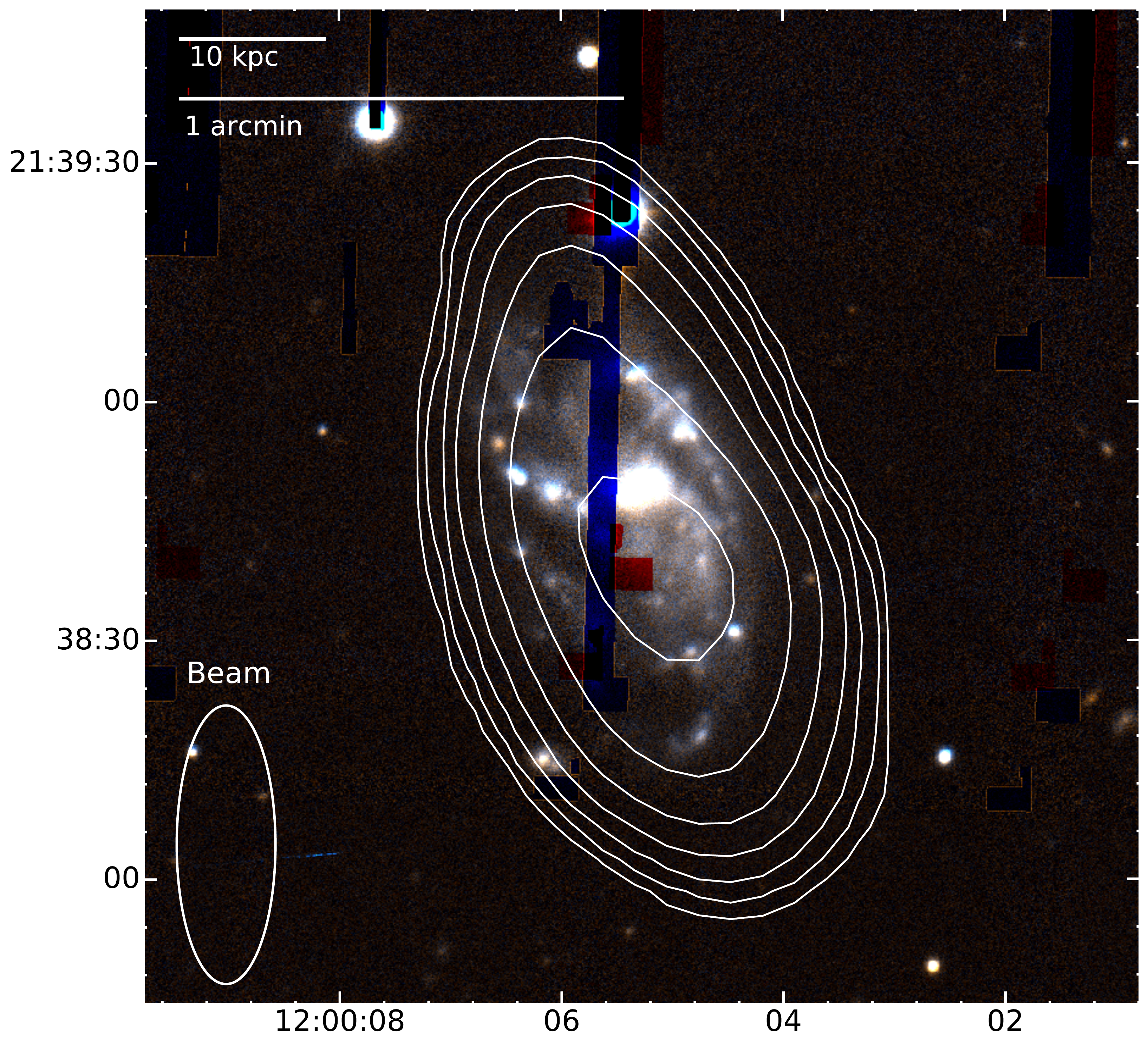}
    \caption{WIYN 3.5m $g'+r'$-band composite imaging for the potential parents of AGC~229398.  Left: KUG~1158+216, scaled to show the faint extension of stellar emission on the left. Right: UGC~6989, with WSRT HI column density contours at $N_{HI}$~=~(0.1, 0.2, 0.4, 0.8, 1.6, 3.2, 6.4) $\times$ $10^{20}$ $\mathrm{cm}^{-2}$ in white.  Due to the large angular separation from AGC~229398 and the smaller field of view of pODI, UGC~6989 was located at the edge of the CCD, which resulted in several detector artifacts in the image.}
    \label{fig:AGC229398parents}
\end{figure}

AGC~229398 is the farthest and most massive AD of the set, with a distance of 104.0 $\pm$ 2.3 Mpc and an HI mass of (1.91 $\pm$ 0.22) $\times$ $10^9$ $M_{\odot}$ \citep{Haynes2018}.  It has two potential parent galaxies, of which UGC~6989 is the most distant potential parent in the sample (Fig. \ref{fig:AGC229398full}).  While the HI distribution is round, the optical component of AGC~229398 itself is elongated in the east-west direction (Fig. \ref{fig:AGC229398close}).  There also appears to be a small knot of stellar emission to the north-east.  

The spiral galaxy KUG~1158+216, which is located to the south of AGC~229398, is the closest potential parent galaxy in distance, heliocentric velocity, and angular separation.  The distance given by the ALFALFA survey is 110.1 $\pm$ 2.3 Mpc, the heliocentric velocity difference is 423 $\pm$ 126 $\mathrm{km}\,\mathrm{s}^{-1}$, and the projected separation on the sky is 102 $\pm$ 2 kpc \citep{Haynes2018}.  There is a faint stellar tail extending from the eastern side that may be the remains of a tidal interaction (left panel of Fig. \ref{fig:AGC229398parents}).  We do not have maps of the HI emission for KUG~1158+216, so we are unable to examine the neutral gas distribution for signs of tidal distortion in the direction of AGC~229398.

UGC~6989 is a spiral galaxy some distance away from AGC~229398, with a projected separation of 482 $\pm$ 11 kpc.  The ALFALFA survey lists the distance of UGC~6989 as 96.0 $\pm$ 2.4 Mpc, but there are some discrepancies between other measurement methods.  The redshift-independent Tully-Fisher measurement from the Cosmicflows-4 survey, which uses the relationship between HI linewidths (i.e. rotation rates) and galaxy luminosities to measure the distances to almost 10,000 spiral galaxies, is 79.8 $\pm$ 8.9 Mpc \citep{Kourkchi2020}, which would remove it from consideration as a possible parent.  The HI distribution and spiral disk of UGC~6989 do not show much sign of disturbance (right panel of Fig. \ref{fig:AGC229398parents}), though AGC~229398 is far enough away that if it had come from UGC~6989, enough time may have passed for it to restabilize.

\begin{deluxetable}{lr@{\hspace*{1in}}lr}[ht]
    \centering
    \caption{\protect\label{tab:AGC229398results} Measured and Derived Properties of AGC~229398}
    \tablehead{ \multicolumn{1}{c}{Property [Units]} & \multicolumn{1}{c@{\hspace{1in}}}{Value} & \multicolumn{1}{c}{Property [Units]} & \multicolumn{1}{c}{Value} }
    \startdata
        RA [h m s, J2000] & 12 00 39.8 & Dec [$^{\circ}$ ' ", J2000] & +21 24 47.6 \\ 
        $v_{h}'$ [$\mathrm{km}\,\mathrm{s}^{-1}$] & 6965 $\pm$ 14 & $N_{HI,peak}$ [$10^{20}$ $\mathrm{cm}^{-2}$] & 3.95 \\
        $S_{21}'$ [Jy $\mathrm{km}\,\mathrm{s}^{-1}$] & 0.75 $\pm$ 0.04 & $S_{21}$ [Jy $\mathrm{km}\,\mathrm{s}^{-1}$] & 0.32 $\pm$ 0.03 \\ 
        $M_{HI}'$ [$10^9$ $M_{\odot}$] & 1.91 $\pm$ 0.23 & $M_{HI}$ [$10^8$ $M_{\odot}$] & 8.15 $\pm$ 1.11 \\
        $M_{gas}'$ [$10^9$ $M_{\odot}$] & 2.54 $\pm$ 0.31 & $M_{gas}$ [$10^9$ $M_{\odot}$] & 1.09 $\pm$ 0.15 \\ 
        $R_{HI}$ [kpc] & 9.83 $\pm$ 1.53 & Inclination [$^{\circ}$] & 66 $\pm$ 8 \\
        $R_e$ [kpc] & 9.05 $\pm$ 1.10 & $R_d$ [kpc] & 0.95 $\pm$ 0.12 \\
        $\mu_{g,0}$ [$\mathrm{mag}\,\mathrm{arcsec}^{-2}$] & 23.84 $\pm$ 0.13 & $\mu_{r,0}$ [$\mathrm{mag}\,\mathrm{arcsec}^{-2}$] & 24.48 $\pm$ 0.10 \\ 
        $m_g$ [mag] & 19.74 $\pm$ 0.11 & $m_r$ [mag] & 19.39 $\pm$ 0.10 \\ 
        $M_g$ [mag] & -15.35 $\pm$ 0.12 & $M_r$ [mag] & -15.69 $\pm$ 0.11 \\ 
        (g-r) [mag] & 0.35 $\pm$ 0.15 & $M_{*}$ [$10^7$ $M_{\odot}$] & 9.78 $\pm$ 4.82 \\
        $M_{bary}'$ [$10^9$ $M_{\odot}$] & 2.64 $\pm$ 0.31 & $M_{dyn}$ [$10^9$ $M_{\odot}$] & 1.64 $\pm$ 0.39 \\ 
        $M_{HI}'/L_g$ & 12.51 $\pm$ 1.99 & $M_{HI}'/L_r$ & 13.88 $\pm$ 2.10 \\
        $M_{HI}'/M_{*}$ & 19.54 $\pm$ 9.92 & $M_{dyn}/M_{gas}$ & 1.52 $\pm$ 0.41 \\ 
    \enddata
    \tablecomments{RA and Dec are listed for the optical component of AGC~229398.  Properties marked with a $'$ are derived from ALFALFA measurements.  Inclination is measured from the HI gas distribution, not the optical component. $M_{dyn}$ is calculated using $W_{20}'$ and $D'$ from ALFALFA data and $R_{HI}$ and inclination from WSRT data, as described in Sections \ref{sec:HI_kin} and \ref{sec:dynmass}.}
\end{deluxetable}

The measured properties of AGC~229398 can be found in Table \ref{tab:AGC229398results}, and Fig. \ref{fig:AGC229398_sbprof} shows the surface brightness profiles.  It has the largest effective radius, with $r_{e}$ = 9.05 $\pm$ 1.10 kpc, and a central surface brightness in $g$ of 23.84 $\pm$ 0.13 $\mathrm{mag}\,\mathrm{arcsec}^{-2}$.  The $(g-r)$ color of 0.35 $\pm$ 0.15 mag is the reddest of the AD-TDG candidates.  The HI mass measured by WSRT is significantly smaller than the one measured by ALFALFA, which suggests that ALFALFA may be detecting diffuse gas that is distributed over a larger area.  AGC~229398 also has the second highest stellar mass of (9.78 $\pm$ 4.82) $\times$ $10^7$ $M_{\odot}$, though this is still small compared to the gas content, giving a fairly high HI-to-stellar mass ratio of 19.54 $\pm$ 9.92.  The estimated dynamical mass is (1.64 $\pm$ 0.39) $\times$ $10^9$ $M_{\odot}$, and it has a low dynamical-to-gas mass ratio of 1.52 $\pm$ 0.41.

\subsection{AGC~333576} \label{sec:A333576}

\begin{figure}
    \centering
    \includegraphics[width=0.7\textwidth]{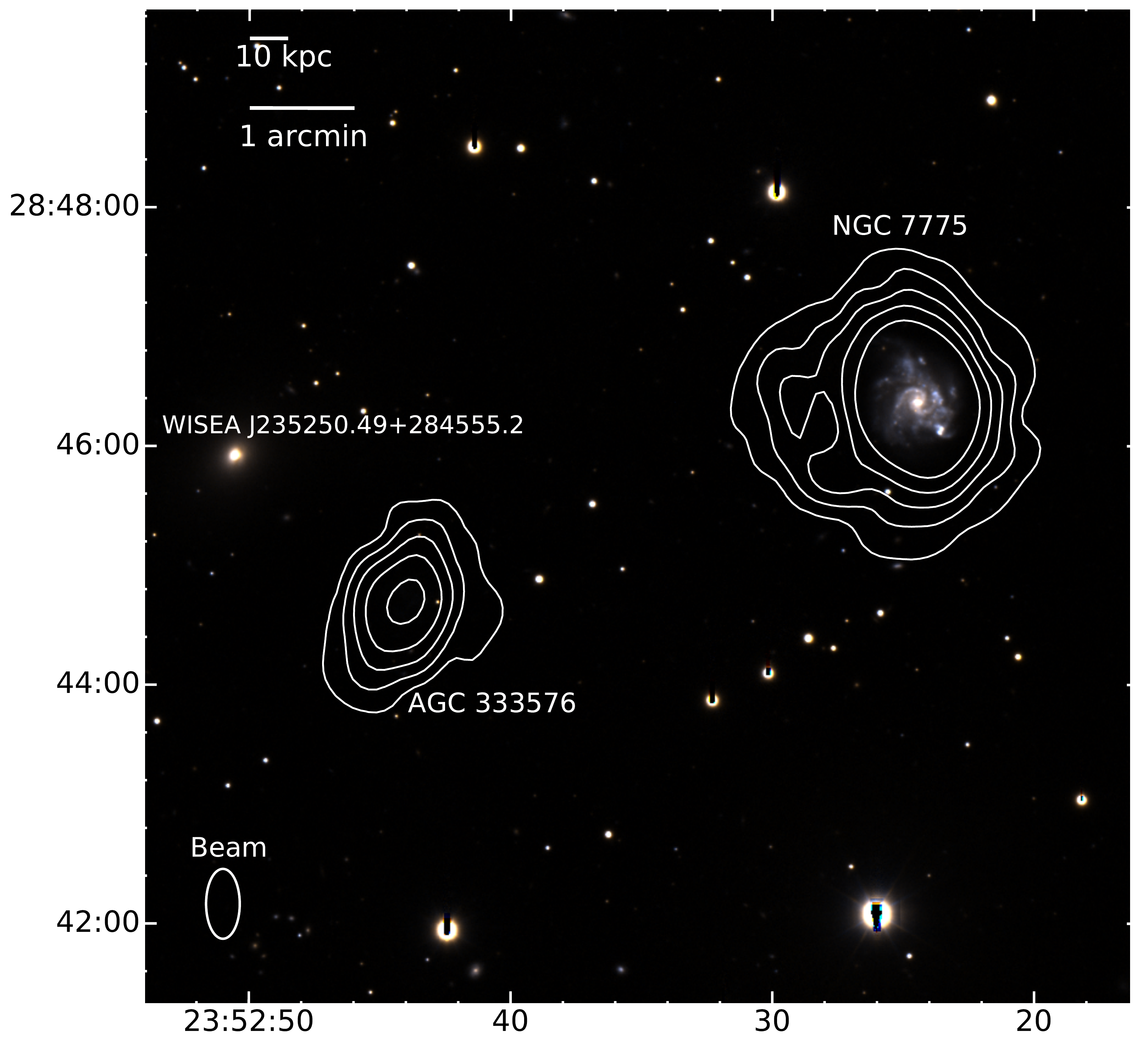}
    \caption{WIYN 3.5m $g'+r'$-band composite image showing the full-field view of AGC~333576 and potential parent NGC~7775.  There are WSRT HI column density contours at $N_{HI}$~=~(0.1, 0.5, 1.2, 2.4, 4.5) $\times$ $10^{20}$ $\mathrm{cm}^{-2}$ superimposed in white. The nearby elliptical galaxy WISEA~J235250.49+284555.2 is also labeled.}
    \label{fig:AGC333576_full}
\end{figure}

\begin{figure}[ht]
    \centering
    \includegraphics[width=0.47\textwidth,height=.35\textheight]{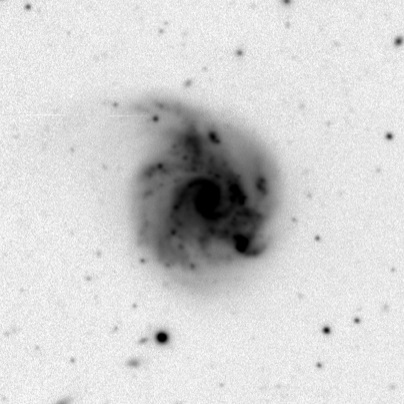} 
    \includegraphics[width=0.49\textwidth]{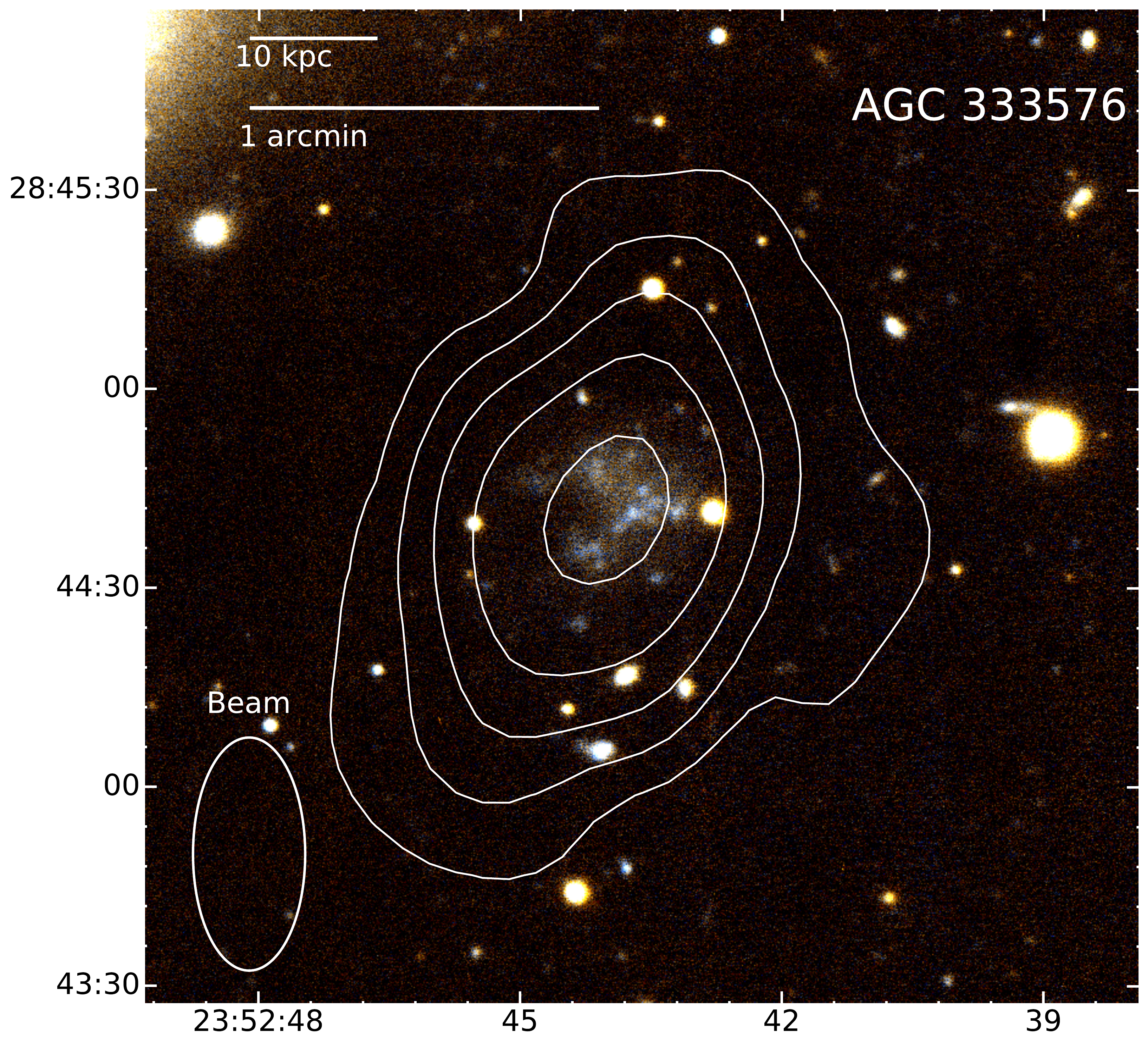}
    \caption{Left: WIYN 3.5m composite $g'+r'$-band summed image of NGC~7775 in grayscale, inverted to highlight the tidal tail at the top.  Right: WIYN 3.5m $g'+r'$-band composite image of AGC~333576, with same HI column density contours as Fig. \ref{fig:AGC333576_full} in white.}
    \label{fig:AGC333576_closeup}
\end{figure}

AGC~333576 has one obvious potential parent galaxy, NGC~7775.  NGC~7775 is a spiral galaxy with chaotic stellar arms and a bright area in the lower right corner of the galaxy which may be a massive star formation region (Fig. \ref{fig:AGC333576_full}).  NGC~7775's HI distribution reveals a clear distortion in the direction of AGC~333576 (Fig. \ref{fig:AGC333576_full}) and the optical image shows a tidal tail extending from the north side and curving toward AGC~333576 (Fig. \ref{fig:AGC333576_closeup}, left).  According to the HI heliocentric velocities from \cite{Haynes2018}, both objects are 93.9 $\pm$ 4.3 Mpc away, and this is in agreement with other redshift-independent measurements for NGC~7775 (96.50 $\pm$ 19 Mpc from \citealt{Theureau2007}).  The optical component of AGC~333576 is relatively rounded, though there appear to be two distinct branches on the left side and there are clear knots of brighter stars (right panel of Fig. \ref{fig:AGC333576_closeup}).  The surrounding HI is slightly elongated along the northwest/southeast axis.  In the course of investigating this system, we have also found the elliptical galaxy to the upper left of AGC~333576, WISEA~J235250.49+284555.2.  Not much is known about this galaxy, and it should be noted that this galaxy has been mistakenly cross-identified with AGC~333576 in the NASA/IPAC Extragalactic Database, so some of the information listed there is actually data associated with the AD.  It has a measured heliocentric velocity of 7157 $\pm$ 24 $\mathrm{km}\,\mathrm{s}^{-1}$ \citep{Huchra2012}, which places it at a similar distance to the other two galaxies in this system, and it is 105~$\pm$~0.125$\arcsec$ away from AGC~333576 on the sky, or roughly 48 $\pm$ 2 kpc.  Because WISEA~J235250.49+284555.2 is an elliptical, it likely did not have enough HI gas to be included in the ALFALFA catalog and thus be identified as a potential parent galaxy; however, it may still be involved in the interactions of this system.

\begin{deluxetable}{lr@{\hspace*{1in}}lr}[ht]
    \centering
    \caption{\protect\label{tab:AGC333576results} Measured and Derived Properties of AGC~333576}
    \tablehead{ \multicolumn{1}{c}{Property [Units]} & \multicolumn{1}{c@{\hspace{1in}}}{Value} & \multicolumn{1}{c}{Property [Units]} & \multicolumn{1}{c}{Value} }
    \startdata
        RA [h m s, J2000] & 23 52 43.6 & Dec [$^{\circ}$ ' ", J2000] & +28 44 42.5 \\ 
        $V_{h}'$ [$\mathrm{km}\,\mathrm{s}^{-1}$] & 7031 $\pm$ 13 & $N_{HI,peak}$ [$10^{20}$ $\mathrm{cm}^{-2}$] & 5.89 \\ 
        $S_{21}'$ [Jy $\mathrm{km}\,\mathrm{s}^{-1}$] & 0.60 $\pm$ 0.04 & $S_{21}$ [Jy $\mathrm{km}\,\mathrm{s}^{-1}$] & 0.64 $\pm$ 0.05 \\ 
        $M_{HI}'$ [$10^9$ $M_{\odot}$] & 1.25 $\pm$ 0.19 & $M_{HI}$ [$10^9$ $M_{\odot}$] & 1.33 $\pm$ 0.21 \\ 
        $M_{gas}'$ [$10^9$ $M_{\odot}$] & 1.66 $\pm$ 0.25 & $M_{gas}$ [$10^9$ $M_{\odot}$] & 1.77 $\pm$ 0.28 \\ 
        $R_{HI}$ [kpc] & 16.73 $\pm$ 1.57 & Inclination [$^{\circ}$] & 57 $\pm$ 2 \\ 
        $R_e$ [kpc] & 4.40 $\pm$ 0.05 & $R_d$ [kpc] & 3.94 $\pm$ 0.04 \\ 
        $\mu_{g,0}$ [$\mathrm{mag}\,\mathrm{arcsec}^{-2}$] & 24.57 $\pm$ 0.02 & $\mu_{r,0}$ [$\mathrm{mag}\,\mathrm{arcsec}^{-2}$] & 24.45 $\pm$ 0.02 \\ 
        $m_g$ [mag] & 18.48 $\pm$ 0.02 & $m_r$ [mag] & 18.27 $\pm$ 0.02 \\ 
        $M_g$ [mag] & -16.38 $\pm$ 0.10 & $M_r$ [mag] & -16.60 $\pm$ 0.10 \\
        (g-r) [mag] & 0.21 $\pm$ 0.03 & $M_{*}$ [$10^8$ $M_{\odot}$] & 1.64 $\pm$ 0.12 \\
        $M_{bary}'$ [$10^9$ $M_{\odot}$] & 1.82 $\pm$ 0.25 & $M_{dyn}$ [$10^9$ $M_{\odot}$] & 1.90 $\pm$ 0.37 \\ 
        $M_{HI}'/L_g$  & 3.18 $\pm$ 0.59 & $M_{HI}'/L_r$ & 4.00 $\pm$ 0.75 \\ 
        $M_{HI}'/M_{*}$ & 7.60 $\pm$ 1.28 & $M_{dyn}/M_{gas}$ & 1.07 $\pm$ 0.27 \\
    \enddata
    \tablecomments{RA and Dec are listed for the optical component of AGC~333576.  Properties marked with a $'$ are derived from ALFALFA measurements.  Inclination is measured from the HI gas distribution, not the optical component. $M_{dyn}$ is calculated using $W_{20}'$ and $D'$ from ALFALFA data and $R_{HI}$ and inclination from WSRT data, as described in Sections \ref{sec:HI_kin} and \ref{sec:dynmass}.}
\end{deluxetable}

The measured properties of AGC~333576 are listed in Table \ref{tab:AGC333576results}, and Fig. \ref{fig:AGC333576_sbprof} displays the surface brightness profiles.  It has a central surface brightness in $g$ of 24.57 $\pm$ 0.02 $\mathrm{mag}\,\mathrm{arcsec}^{-2}$, an effective radius of 4.40 $\pm$ 0.05 kpc, and a $g-r$ color of 0.21 $\pm$ 0.03 mag.  AGC~333576 is estimated as having a stellar mass of (1.64 $\pm$ 0.12) $\times$ $10^8$ $M_{\odot}$, the largest in the set, which gives it a HI-to-stellar mass ratio of 7.60 $\pm$ 1.28.  The dynamical mass of (1.90 $\pm$ 0.37) $\times$ $10^9$ $M_{\odot}$ gives a very low dynamical-to-gas mass ratio of 1.07 $\pm$ 0.27.

\section{Discussion} \label{sec:discussion}

\subsection{Basic TDG Properties}

\cite{Kaviraj2012} report several general characteristics of TDGs based on their statistical study of candidate TDGs.  This study selects objects that appear in SDSS images of galaxy mergers and have been identified as separate photometric objects by the SDSS processing software and have a tidal stellar tail that connects the object to the parent.  These criteria mean that younger objects are favored.  Some of these objects likely will not become independent self-gravitating objects, so they are only TDG candidates.  It is important to consider which characteristics may change as a TDG ages.  The clearest example of this is that \cite{Kaviraj2012} report that 95\% of TDG candidates were located within a projected distance of $\sim$20 kpc from their progenitors.  It is likely that they have simply not had enough time to move much farther, as tidal tails are only present for a few hundred million years after the first pericentric encounter of the interacting galaxies \citep{Bournaud2006}.

TDGs tend to be bluer than their parents (with a median offset of 0.3 mag) due to the star formation triggered by the turbulence of the tidal interaction \citep{Kaviraj2012}.  It may be the case that as TDGs age, this color differential becomes less pronounced, though they may also remain quite blue.  All of the ADs in this sample are markedly blue, and bluer than their potential parents, which may indicate that the stellar populations present were recently formed.  However, low-surface-brightness galaxies as a population tend to be blue for reasons independent of hosting a primarily young stellar population \citep{McGaugh1992, deBlok1995}.

The systems in the \cite{Kaviraj2012} sample primarily involve galaxies that are clearly in the process of interacting | there are at least two galaxies fairly close to each other on the sky, showing some sort of stellar distortion (see merger classification details in \citealt{Darg2010a}).  \cite{Bournaud2006} conducted a set of numerical simulations of galaxy interactions with a variety of interaction parameters, in order to examine the merger scenarios in which TDGs are most likely to form and to survive long-term (which they defined as $>$1 Gyr).  Based on the \cite{Bournaud2006} simulations, the interactions most likely to produce long-lived TDGs are those between spiral galaxies with prograde orbits, where the mass of the companion galaxy is large enough to drive matter out of the main progenitor galaxy (at least one-fourth the mass of the progenitor), but not so massive that the material falls back onto the companion (less than eight times more massive than the progenitor).  In cases where a clearly merging system is present, we might be able to use these expected properties to evaluate whether an interaction is likely to have produced a long-lived TDG.  However, most of the proposed parents in this study appear to be single galaxies, well-separated from the other potential parents and with generally undisturbed stellar distributions.  In some cases the interactions could have progressed enough that the parent systems have merged completely; learning about the interaction would then require further analysis of stellar populations, star formation histories, and gas dynamics.  

Simulations indicate that the masses of TDGs are usually a few percent of the baryonic mass of their parents, and rarely exceed 10\% of that mass \citep{Bournaud2006}. We can estimate the baryonic mass of the parent galaxies as follows:  $M_{atomic}$ is estimated to account for the presence of helium and other elements, so $M_{atomic}'$ = 1.33 $\times$ $M_{HI}'$, where $M_{HI}'$ is from the ALFALFA survey \citep{Haynes2018} as calculated by Eq. \ref{eq:HImass}.  The stellar masses were estimated by \cite{Durbala2020}.  They calculated stellar masses using CMLRs between photometry and mass for SDSS optical and unWISE infrared photometry, and checked the agreement with stellar masses derived from SED fitting in the GALEX-SDSS-WISE Legacy Catalog 2 (GSWLC-2; \citealt{Salim2016, Salim2018}).  We used their estimates from the Taylor relationship, noting that they are generally more conservative than those from other calibrations, which means that the mass ratios calculated are on the higher end.  As most of the potential parent galaxies are spiral galaxies, molecular gas mass must be taken into account when estimating the baryonic mass.  \cite{Young1991} report median $H_2$-to-HI mass ratios for different types of spiral galaxies, which we used to estimate $M_{molecular}'$ for each potential parent based on the type listed on NED (\citeyear{NED}). For galaxies that did not have a specific type listed, a mass ratio of one was used.  \cite{Young1991} note that early-type galaxies have higher $H_2$-to-HI mass ratios than late-types, so we used a ratio of 2.5 for NGC~807.  Then, $M_{bary}'$ = $M_{atomic}'$ + $M_{*} $ + $M_{molecular}'$. This calculation was not done for MRK~365, as we do not have mass estimates for it.  Most of the AD-TDG candidates do not exceed 5\% of the baryonic mass of any individual potential parent.  However, AGC~229398 has high mass ratios with respect to its potential parents (15.3\% with UGC~6989 and 23.9\% with KUG~1158+216), which seems to make it less likely that either galaxy is the parent; such an enormous removal of mass would cause major disruption to the parent, but these spirals appear to be relatively intact. 

\subsection{Context of Other TDGs}

The sample of candidate TDGs to compare these candidates to is small, and the sample of confidently identified TDGs that have completely separated from their parent systems is even smaller.  We have selected a comparison sample of likely TDGs identified in other studies that have similar measurements to those we have conducted for the AD-TDG candidates.  These are AGC~749170, AGC~208457, NGC~5557-E1, NGC~5557-E2, NGC~5291N, NGC~5291S, NGC~5291SW, NGC~7252E, NGC~7252NW, VCC~2062, and HCG~16-LSB1.

\begin{itemize}
    \item In Lee-Waddell et al. (\citeyear{LeeWaddell2012}, \citeyear{LeeWaddell2014}, \citeyear{LeeWaddell2016}), two ADs have been identified as TDGs, and so are included in the discussion of AD-TDGs as a group.  The analyses in those papers were carried out with measurements from the Giant Metrewave Radio Telescope (GMRT), which has a higher spatial resolution than the Arecibo telescope.  A higher resolution beam does a better job of mapping the fine gas structure, but does not capture the more diffuse surrounding gas.  \cite{LeeWaddell2012} note that the GMRT maps capture less than 40\% of the flux density of the ALFALFA measurements and therefore the ALFALFA gas mass.  In the interest of comparing these sources to the work done here, we use the ALFALFA survey data from \cite{Haynes2018} when discussing the total gas mass content of the galaxy and the GMRT values reported by Lee-Waddell et al. (\citeyear{LeeWaddell2012}, \citeyear{LeeWaddell2014}, \citeyear{LeeWaddell2016}) for the dynamical-to-gas mass ratio.  The estimates reported for these objects do not include an asymmetric drift correction.
    \begin{itemize}
        \item AGC~749170 has an extremely high gas-mass-to-light ratio ($>$1000; \citealt{LeeWaddell2014}).  \cite{LeeWaddell2014} used stellar population models to estimate its age at 6.3 - 12 Myr, which is very young for a galaxy's global population and suggests that there are very few pre-existing stars in AGC~749170.  This stellar population age estimate depends on an estimated $g-i$ range due to a non-detection in the $i$-band.  Additionally, in order to reach its present separation of at least 90 kpc (from the angular distance projection on the sky, though the true distance is likely greater) in only 12 Myr, AGC~749170 would need to travel at a speed of at least 7300 $\mathrm{km}\,\mathrm{s}^{-1}$, which would be unreasonably fast. The lack of a gaseous tail stretching from AGC~749170 also seems to imply an older age, because forming such a large mass of HI would require a significant tail which would take at least a few hundred million years to fade below the detection limits.  On the other hand, the tail may be present but very diffuse, and therefore not captured in the higher resolution GMRT images.
        \item AGC~208457 appears to have the last remnants of an HI tail connecting it to its potential parent, the spiral galaxy NGC~3169, based on HI maps from \cite{LeeWaddell2012}.  This, in combination with the elongation of the HI in AGC~208457, suggests that the TDG may not have reached full dynamical equilibrium yet, so the estimated dynamical mass (and thus the dark matter content) may not be accurate.  AGC~208457 has an estimated age range of 6 - 2600 Myr, based on the age of the stellar component using a Chabrier IMF \citep{LeeWaddell2016}.  Follow-up spectroscopy of AGC~208457 confirmed that it has the enriched metallicity indicative of a tidal origin \citep{LeeWaddell2018b}.
    \end{itemize}
    \item \cite{Duc2014} identified two objects likely to be TDGs around the elliptical galaxy NGC~5557: NGC~5557-E1 and NGC~5557-E2.  Using spectrophotometric methods, they analyzed NGC~5557-E1 and came to the conclusion that its near-solar oxygen abundance was typical of galaxies 10 times more massive, and therefore the material was pre-enriched.  Deep optical imaging from MegaCam on the Canada-France-Hawaii Telescope (CFHT) revealed that NGC~5557-E1 was located in a large and diffuse stellar stream.  They considered that NGC~5557-E1 could be the remnant of a pre-existing satellite galaxy that had been tidally disturbed, but the remains of such an interaction would be redder, and the surrounding tidal tails would likely have an “S-shape" \citep{Duc2014}.  The blue color of NGC~5557-E1 and the straightness of the tail precluded this as a likely origin scenario, and so \cite{Duc2014} concluded that it was a TDG formed during a tidal interaction.  NGC~5557-E2 is located farther along this tail and likely originated in the same interaction.  They estimated the age of these objects to be between 2 and 4 Gyr, which would make them much older than any TDGs previously identified.  Both “older" TDGs have larger effective radii than other dwarf galaxies of comparable stellar mass, in accordance with other, younger TDGs that have been identified by their tidal tails \citep{Duc2014}.  \cite{Duc2014} suggested that the larger radii may be used as an additional tool to aid in distinguishing TDGs from dwarf satellite galaxies without spectroscopy.
    \item \cite{Lelli2015} performed new observations and analysis for a sample of six previously identified candidate TDGs: NGC~5291N, NGC~5291S, NGC~5291SW, NGC~7252E, NGC~7252NW, and VCC~2062.  They also cited additional photometric properties of the latter three compiled in \cite{Duc2014}.  \cite{Lelli2015} found that NGC~7252E and NGC~7252NW are comprised of pre-enriched material and lack a significant dark matter mass, and are therefore genuine TDGs, and reference similar results for the three objects around NGC~5291 \citep{Bournaud2007} and for VCC~2062 \citep{Duc2007}.
    \item HCG~16-LSB1 is located at the end of an HI tail extending out of the galaxy HCG~16c, which is part of the complex galaxy interactions occurring in HCG~16.  \cite{Roman2021} concluded that HCG~16-LSB1 has a high metallicity that is compatible with it being a newly-formed TDG.  They estimated a dynamical mass range of (1.2 - 4.1) $\times$ $10^9$ $M_{\odot}$, with the further caveat that HCG~16-LSB1 may be too young to be gravitationally bound \citep{Roman2021}.  Even the low end of this range would indicate a higher than expected amount of dark matter for a TDG, yet the object has clearly formed from enriched material at the end of a tidal tail.  The mass and location of HCG~16-LSB1 at the end of a tidal tail make it very likely to have longevity based on the results of the simulations of \cite{Bournaud2006}, so it seems feasible that it will become gravitationally independent as a proper TDG in the future.
\end{itemize}

\subsection{Dark Matter Content}
\begin{figure}
    \centering
    \includegraphics[height=.35\textheight]{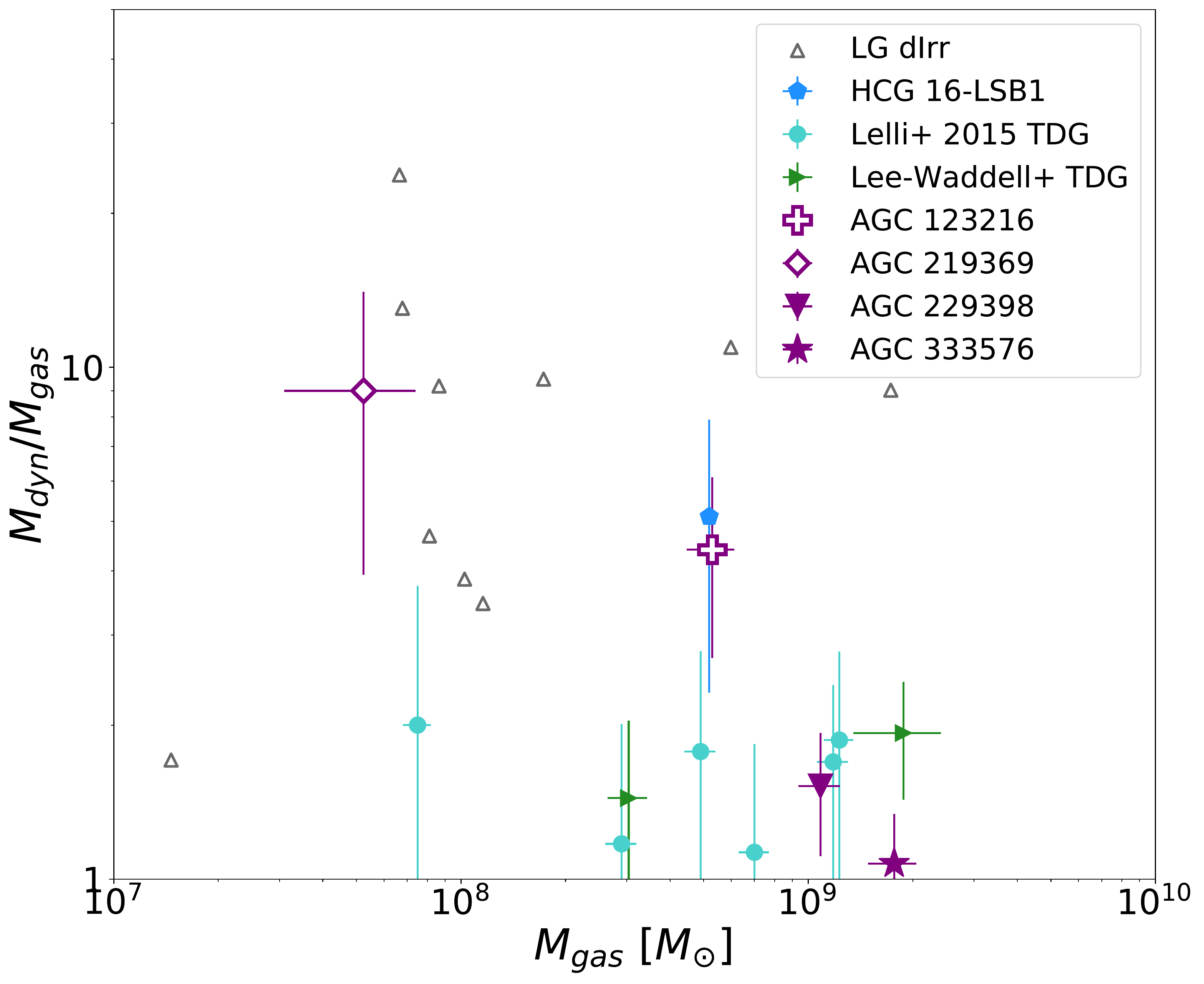}
    \caption{Comparison of dynamical-to-gas mass ratios for objects from this work (purple, various symbols) to the young TDG candidate HCG~16-LSB1 (\citealt{Roman2021}; blue pentagon), AD-TDG candidates from \citeauthor{LeeWaddell2016} (\citeyear{LeeWaddell2016}; green right-pointing triangles), and TDG candidates from \citeauthor{Lelli2015} (\citeyear{Lelli2015}; turquoise circles). For the comparison of TDG properties to those of normal (non-TDG) dwarf galaxies, we have plotted a sample of Local Group dIrrs from \citeauthor{McConnachie2012} (\citeyear{McConnachie2012}; dark gray triangles).  $M_{gas}$ for the LG dIrrs is calculated using 1.33 $\times$ the HI mass reported in \cite{McConnachie2012}, and $M_{dyn}$ uses the dynamical masses from the same paper supplemented by the total masses reported in \cite{Mateo1998} when necessary.  $M_{gas}$ for the TDGs from \cite{Lelli2015} is the sum of $M_{atomic}$ and $M_{molecular}$ reported in that work.  All objects that have been identified as TDGs have filled symbols, and those that are not TDGs have unfilled symbols.}
    \label{fig:LW2016_14}
\end{figure}

The ratio of the dynamical (total) mass to the gas mass provides a measure of the dark matter content in a galaxy.  To calculate the numerator in this ratio, we use the dynamical mass described in Sec. \ref{sec:dynmass}, which is calculated with Eq. \ref{eq:dynmass} and uses the ALFALFA velocity width and distance ($W_{20}'$ and $D'$ and the inclination and radius ($R_{HI}$) derived from the WSRT data. Because the spatial values used to calculate dynamical mass (inclination and radius) are derived from spatial measurements from the WSRT data, we use the WSRT gas mass as the denominator in the dynamical-to-gas mass ratio.  For our ADs, the stellar contribution to the mass is minimal, so we consider only the gas mass for the current discussion. TDGs are primarily composed of gas from the disk material of their progenitors, and so should contain very little dark matter and therefore are expected to have $M_{dyn}/M_{gas}$ near 1.  Fig. \ref{fig:LW2016_14} shows the dynamical-to-gas mass ratios of the four candidate AD-TDGs in this sample in comparison to the values for the TDGs in \cite{Roman2021}, \cite{Lelli2015}, and \cite{LeeWaddell2016}.  To provide a comparison sample of non-TDG dwarf galaxies we have also plotted values for the dwarf irregular galaxies in the Local Group compiled by \cite{McConnachie2012}.  

AGC~229398 and AGC~333576 have ratios that clearly place them in the regime of previously identified TDGs in the lower right corner of the plot, while AGC~219369 appears to be dark matter dominated (Fig. \ref{fig:LW2016_14}). This suggests that AGC~219369 is unlikely to be a TDG.  Instead, it may simply be a small dwarf galaxy that happens to be near a tidally disturbed spiral in the sky, or may have even interacted with it previously.  AGC~123216's dynamical mass to gas mass ratio indicates a higher dark matter content than expected for an object of tidal origin, though it still has a lower amount of dark matter than most of the Local Group dwarf irregular galaxies.  It is worth noting that while HCG~16-LSB1, which is a clearly tidal object, is located in the dark matter dominated portion of the plot, it is such a young object that it may not be dynamically stable yet.  This means the currently estimated dynamical mass may be larger than the actual total dynamical mass.

\subsection{Stellar Content}

\begin{figure}
    \centering
    \includegraphics[height=.35\textheight]{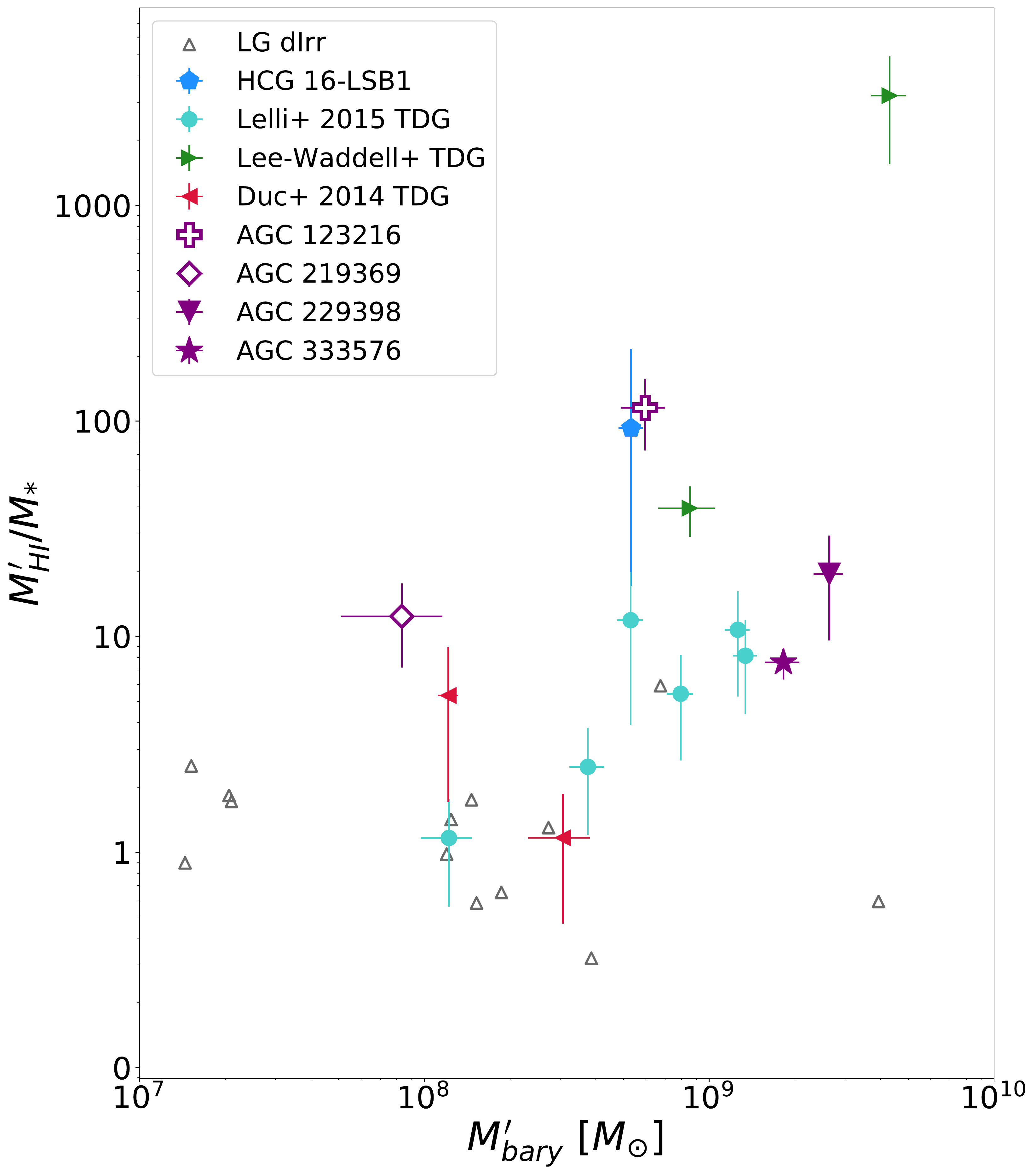}
    \caption{Comparison of HI-to-stellar mass vs. baryonic mass for objects from this work (purple, various symbols) to the young TDG candidate HCG~16-LSB1 (\citealt{Roman2021}; blue pentagon), AD-TDG candidates from \citeauthor{LeeWaddell2016} (\citeyear{LeeWaddell2016}; green right-pointing triangles), TDG candidates from \citeauthor{Lelli2015} (\citeyear{Lelli2015}; turquoise circles), and likely older TDGs identified in \citeauthor{Duc2014} (\citeyear{Duc2014}; red left-pointing triangles).  $M_{HI}'$ and $M_{bary}'$ for the objects from this work and from \cite{LeeWaddell2016} are calculated with the ALFALFA measurements \citep{Haynes2018}.  The HI masses for the \cite{Duc2014} TDGs used in the baryonic mass calculation are from \cite{Duc2011}.  For the comparison of TDG properties to those of normal (non-TDG) dwarf galaxies, we have plotted Local Group dIrrs from \citeauthor{McConnachie2012} (\citeyear{McConnachie2012}; dark gray triangles).  As with Fig. \ref{fig:LW2016_14}, filled symbols correspond to objects that have been identified as TDGs.}
    \label{fig:HIvsStar}
\end{figure}

The four AD-TDG candidates studied in this work were selected from the larger ALFALFA AD sample in part because of their large HI-to-optical-light ratios.  As previously mentioned, it is likely that ADs may not efficiently convert neutral HI to cold molecular gas \citep{Wang2020}, impacting their star formation rate.  According to simulations, the critical column density threshold of HI for star formation is thought to range from (3-10) $\times$ $10^{20}$ $\mathrm{cm}^{-2}$ \citep{Schaye2004}, depending on the local conditions.  The critical surface density corresponds to a critical minimum pressure to trigger a transition to a cold phase, where the temperature of the gas drops and the $H_2$ formation rate increases in order to maintain the pressure, thereby increasing the fraction of molecular gas \citep{Schaye2004}.  The phase transition also causes gravitational instability in the gas cloud, which results in fragmentation and, eventually, star formation \citep{Schaye2004}.  If there is not a high enough column density (and therefore pressure) of neutral gas to cause the phase transition and increase the molecular gas fraction, the gas cloud does not become unstable and collapse to form stars.  The peak column densities for the four objects in this work are on the low-to-mid end of the range (3.95 $\times$ $10^{20}$ $\mathrm{cm}^{-2}$ for AGC~229398 at the least, 5.89 $\times$ $10^{20}$ $\mathrm{cm}^{-2}$ for AGC~333576 at the most), so one could suggest that they did not have a high enough gas density for efficient star formation as an explanation for their high HI-to-optical-light ratios. 

On the other hand, TDGs are expected to have a burst of star formation during their formation, due to the instability of the gas from turbulence, and then fade over time, which would also result in a higher HI-to-optical-light ratio as they age.  For example, \cite{Roman2021} predicted the evolution of the surface brightness for HCG~16-LSB1 based on its metallicity and the aging of its stellar population. They expect that the dimming of HCG~16-LSB1 will be rather extreme; in 400 Myr, the effective surface brightness will decrease by more than 1 $\mathrm{mag}\,\mathrm{arcsec}^{-2}$, making it fall below the optical threshold of the survey used in their work (DECaLS).  Stellar populations with higher metallicities will experience more dimming, which means that TDGs will be particularly affected, though they are assumed to continue forming stars at a lower rate \citep{Duc2014}.  This means that an evolved TDG could have a stellar population mainly formed in an initial burst followed by a lower rate of formation, but a high HI-to-optical-light ratio due to dimming. 

While this idea does not easily distinguish TDGs from dIrr galaxies that have produced stars normally, objects that are faint due to a lack of stars proportional to their HI contents may be considered less likely to be TDGs because they likely didn't have this initial burst of star formation. Fig. \ref{fig:HIvsStar}, which plots the HI-to-stellar mass ratio against the baryonic mass of each object, demonstrates how this appears to be true for AGC~123216, which has an extreme $M_{HI}'/M_*$ of 115.28 $\pm$ 42.20.  This could mean that AGC~123216 lacked the star formation burst at its formation that would be expected for a TDG.  HCG~16-LSB1 and AGC~208457 also fall higher than the distribution of the rest of the TDGs, even though they are confirmed tidal objects.  Considering the young age of HCG~16-LSB1 ($57.9^{+21.7}_{-9.4}$ Myr according to \citealt{Roman2021}) and the possible still-extant tidal tail and estimated age range of 6-2600 Myr for AGC~208457 \citep{LeeWaddell2016}, they might have a high HI-to-stellar mass ratio because they are so young that they have not had sufficient time to convert gas into stars.  The two older TDGs from \cite{Duc2014} have relatively average HI-to-stellar mass ratios compared to the TDGs from \cite{Lelli2015}.  Fig. \ref{fig:HIvsStar} also shows that the other three AD-TDG candidates appear to have $M_{HI}'/M_*$ that is comparable to these TDGs and about an order of magnitude higher than dwarf irregulars.  This suggests that their high HI-to-optical-light ratios may be the result of dimming rather than a deficit of stars in proportion to their HI content.

\begin{figure}
    \centering
    \includegraphics[height=.4\textheight]{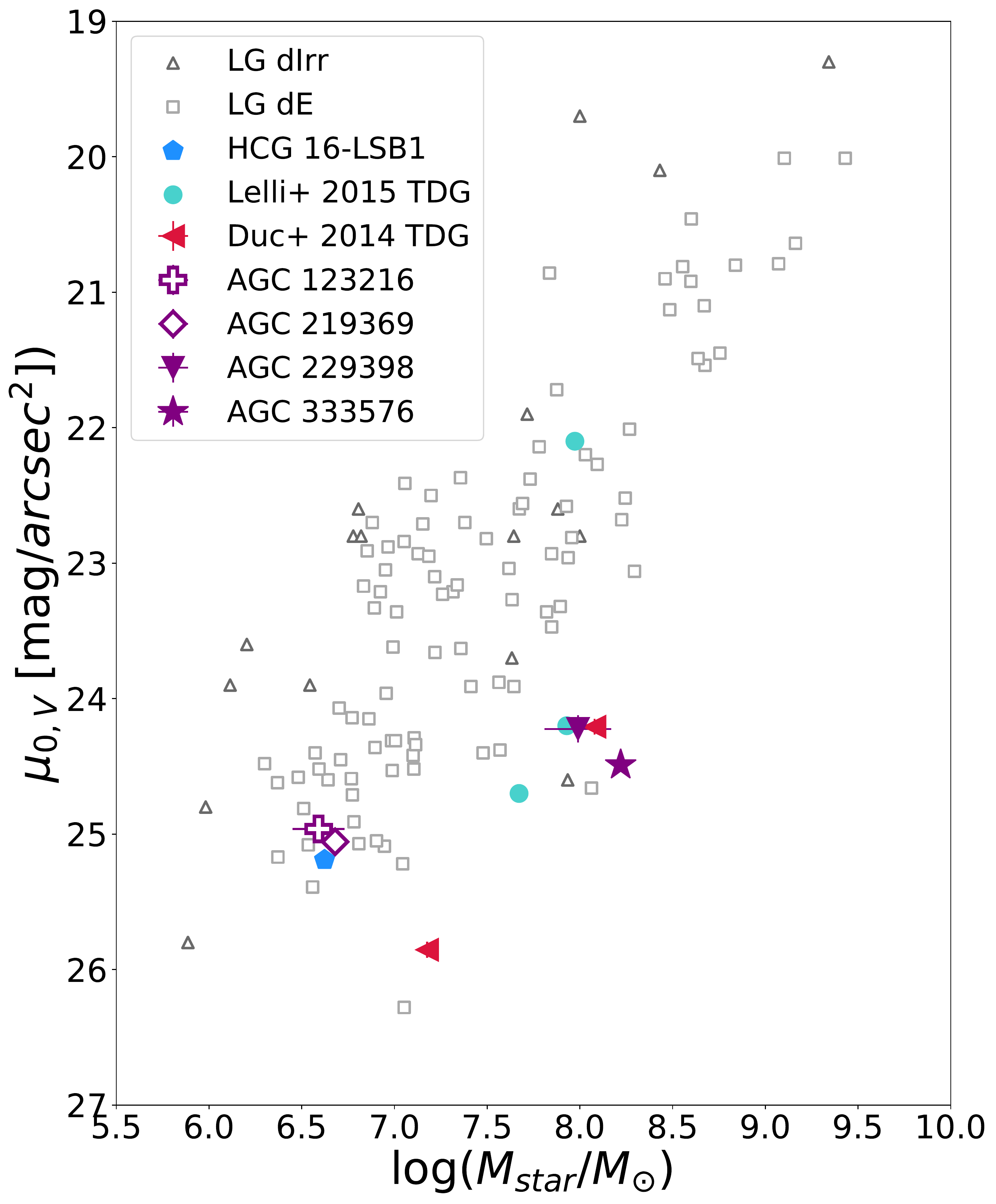}
    \caption{Comparison of central surface brightness vs. stellar mass for objects from this work (purple, various symbols) to the young TDG candidate HCG~16-LSB1 (\citealt{Roman2021}; blue pentagon), TDG candidates discussed in \citeauthor{Lelli2015} (\citeyear{Lelli2015}; turquoise circles), and likely older TDGs identified in \citeauthor{Duc2014} (\citeyear{Duc2014}; red left-pointing triangles).  The non-TDG dwarf galaxy comparison samples plotted are Local Group dwarf irregulars from \citeauthor{McConnachie2012} (\citeyear{McConnachie2012}; dark gray triangles) and Hydra I cluster dwarf ellipticals whose properties were compiled by \cite{Misgeld2008} and \citeauthor{Misgeld2011} (\citeyear{Misgeld2011}; light gray squares).  Conversions from surface brightness in $g$ to V were made with the \cite{Jester2005} transformation equations and the central color of the object where possible.  If the central color was not known, the total $g-r$ color of the object was used, and if no color information was given, an average estimate of 0.4 $\pm$ 0.1 for dwarf galaxies based on the colors of \cite{Honey2018} was used.  As with Fig. \ref{fig:LW2016_14}, filled symbols correspond to objects that have been identified as TDGs.}
    \label{fig:Duc2014_7}
\end{figure}

Due to this dimming, we can also expect TDGs to increasingly be found on the lower end of the central surface brightness versus stellar mass relationship as they age. Fig. \ref{fig:Duc2014_7} shows that this is the case for the older TDGs from \cite{Duc2014}, while younger TDGs (HCG~16-LSB1, the sample from \citealt{Lelli2015}), though still low, are more scattered along the relation shown by the comparative non-TDG sample, which is comprised of dwarf irregulars in the Local Group and dwarf ellipticals from the Hydra I cluster \citep{Misgeld2008, Misgeld2011, McConnachie2012}.  AGC~229398 and AGC~333576 also have low central surface brightnesses for their stellar masses, appearing quite similar to NGC~5557-E1.  AGC~123216 and AGC~219369 are located within the distribution of dwarf ellipticals, so their central surface brightnesses are low compared to dwarf irregular galaxies of similar stellar mass.  HCG~16-LSB1 is similarly placed, though it is unknown whether this is because it is so young that it simply may not have had enough time to form a larger stellar population yet.  Over time, it is expected to eventually move farther right and lower in the plot as it forms stars and dims.  It is important to note that though UDGs are not shown here, because they are defined as having $\mu_{g,0} >$ 24 $\mathrm{mag}\,\mathrm{arcsec}^{-2}$ \citep{vanDokkum2015} they will also occupy the lower portion of the plot and may overlap with the evolved TDGs.

\begin{figure}
    \centering
    \includegraphics[height=.4\textheight]{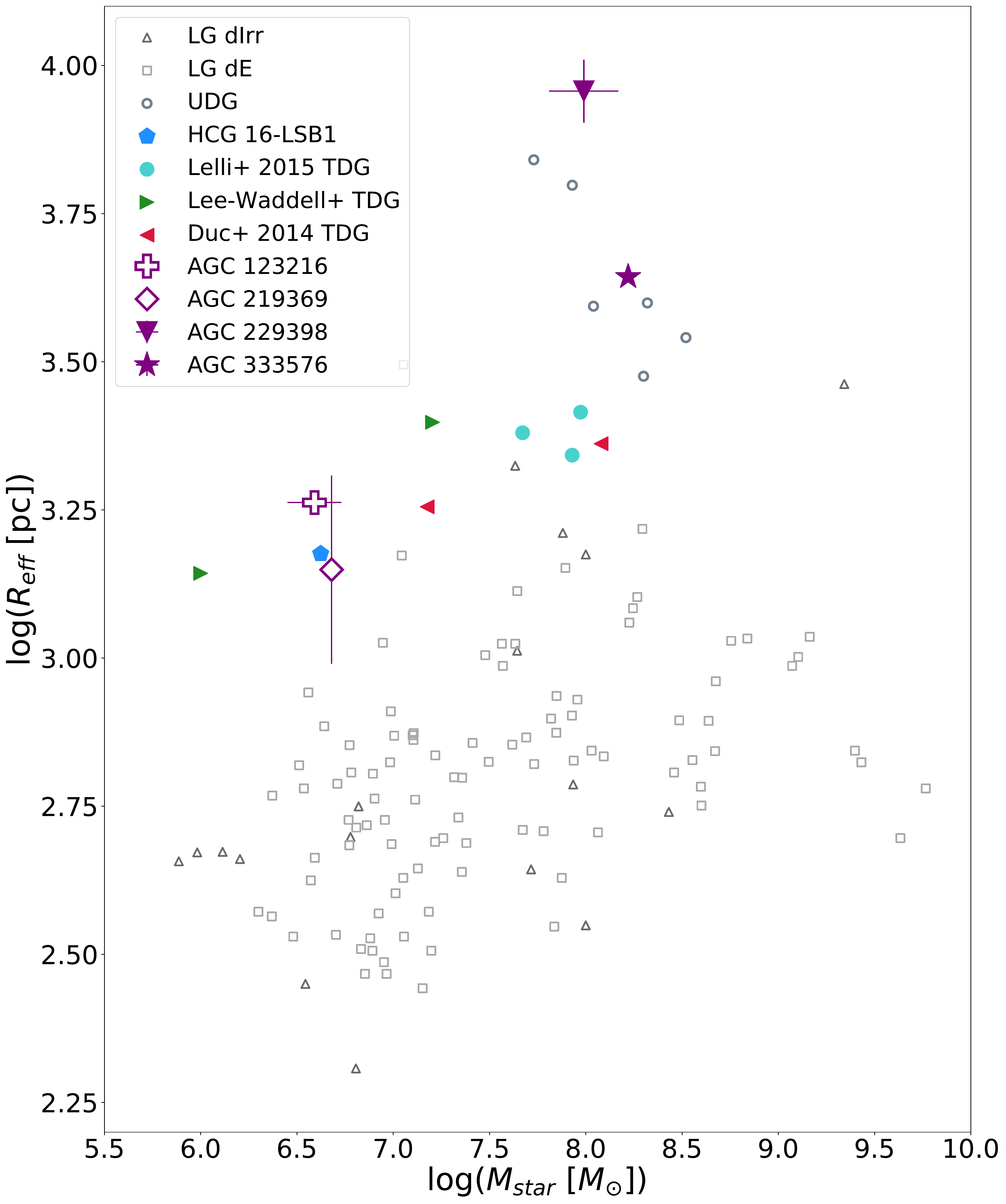}
    \caption{Comparison of effective radius vs. stellar mass for objects from this work (purple, various symbols) to the young TDG candidate HCG~16-LSB1 (\citealt{Roman2021}; blue pentagon), AD-TDG candidates from \citeauthor{LeeWaddell2016} (\citeyear{LeeWaddell2016}; green right-pointing triangles), TDG candidates from \citeauthor{Lelli2015} (\citeyear{Lelli2015}; turquoise circles), and likely older TDGs identified in \citeauthor{Duc2014} (\citeyear{Duc2014}; red left-pointing triangles).  We have also plotted the sample of UDGs from \citeauthor{ManceraPina2020} (\citeyear{ManceraPina2020}; dark gray circles).  For the comparison of TDG properties to those of normal (non-TDG) dwarf galaxies, we have plotted Local Group dIrrs from \citeauthor{McConnachie2012} (\citeyear{McConnachie2012}; dark gray triangles), and Hydra I cluster dwarf ellipticals compiled by \citeauthor{Misgeld2011} (\citeyear{Misgeld2011}; light gray squares).  As with Fig. \ref{fig:LW2016_14}, filled symbols correspond to objects that have been identified as TDGs.}
    \label{fig:Duc2014_6}
\end{figure}

\cite{Duc2014} considered a comparison of the effective radius to the stellar mass as a tool for the optical identification of TDG candidates.  TDGs tend to have effective radii larger than those of dwarf irregulars or dwarf ellipticals of a comparable stellar mass.  In this sense, they resemble UDGs because their stellar mass is more spread out than usual.  In the comparison of effective radius to stellar mass in Fig. \ref{fig:Duc2014_6}, it is clear that other identified TDGs occupy the top part of the distribution.  AGC~229398 and AGC~333576 are located well above the distribution, along with the TDGs from \cite{LeeWaddell2016}.  AGC~123216 and AGC~219369 appear to be on the upper end of the distribution as well (though not as dramatically as the other two objects).

\subsection{Estimation of Travel Time} \label{sec:age}

It is common to use the combination of stellar population models and photometry in multiple bands to estimate the age and metallicity of a galaxy \citep{Bruzual2003}.  Our photometry in $g$ and $r$ bands for our TDG candidates is not sufficient to break the age-metallicity degeneracy associated with some broadband filter combinations, and a third filter (e.g. $i$-band) would be required.  Therefore, we must turn to other methods to evaluate the evolutionary histories of the sample galaxies.  

The simulations from \cite{Bournaud2006} give us a few markers to estimate the evolutionary progress of the TDG systems.  They show that tidal tails fade within 300-500 Myr, so based on the absence of tidal tails, our TDG candidates would be at least a few hundred million years old if they are genuine TDGs.  This alone would make them older than the majority of previously studied TDGs.  TDGs face many obstacles to their survival: a significant number simply fall back onto their parent galaxy, or are torn apart by tidal disruption from the progenitor galaxy or by feedback from star formation.  \cite{Bournaud2006} found that only 25\% of tidal knots more massive than $10^8$ $M_{\odot}$ survive longer than 800 Myr, and that 45\% of knots surviving at least 1 Gyr have masses exceeding $10^9$ $M_{\odot}$. Note that “surviving" in this simulation meant remaining above the detection threshold of $10^8$ $M_{\odot}$ within a diameter of 3 kpc or $3\times10^8$ $M_{\odot}$ within a diameter of 6 kpc, due to computational constraints.  The tidal knots may undergo mass loss that puts them below this threshold while still appearing as distinct, intact knots that would continue to be classified as candidate TDGs in a higher resolution simulation than that from \cite{Bournaud2006}. 

If the ADs in this study are indeed TDGs, they are located much farther from their parents than the majority of previously proposed candidate TDGs, so it is necessary to show that it is possible for them to travel the observed distance in a reasonable time scale. The logic is similar to the logic we used to evaluate whether AGC~749170 could have reasonably reached its present location in 12 Myr, though now we consider rotational velocities and orientations of the potential parents.  For each TDG-parent pair, we estimate a travel time as follows: TDGs are expected to have velocities with respect to their parent galaxies comparable to the rotational velocities of those parent galaxies \citep{Bournaud2006}.  We approximate the rotational velocity using
 
\begin{equation}\label{eq:vrot}
    V_{rot}' = \frac{W_{20}'}{2sin(i)}
\end{equation}
\noindent with the $W_{20}'$ value measured by the ALFALFA survey. Inclination $i$ here is estimated from reported optical semimajor and semiminor axis measurements of the galaxy, using 
\begin{equation} \label{eq:inc}
    cos^2(i) = \frac{(b/a)^2-q_0^2}{1-q_0^2},
\end{equation}

\noindent with $q_0$ = 0.13 as is appropriate for spirals (all of the potential parent galaxies are classified as spirals with the exception of NGC~807, which nevertheless shows evidence of a disk).  The resulting rotational velocities (and thus TDG candidate travel velocities) are on the order of 100-300 $\mathrm{km}\,\mathrm{s}^{-1}$, which is typical for spiral galaxies and consistent with the velocities of the TDGs simulated by \cite{Bournaud2006}, so this is a reasonable estimate.  TDGs in simulations tend to have low orbital inclination ($<$40$^{\circ}$) with respect to the rotational plane of their progenitor \citep{Bournaud2006}, so for simplicity we assume that the TDG candidate remained in the rotation plane of the parent.  Therefore, we can use the inclination axis of the parent to deproject the angular separation on the sky and estimate a total travel distance for the TDG candidate.  For this calculation, we consider the TDG candidates to be at the distance of the parent galaxy, give or take a negligible distance from any component of the velocity along the line-of-sight, which is consistent with the observations within uncertainties for all sources (given the previous assumption that AGC~219369 is at the distance of NGC~3510).  With these approximations for velocity and travel distance, we can estimate the time it would take for a newly-formed TDG to travel to its current position from each potential parent.  Ranges for the travel velocities, travel distance, and travel time estimates for each TDG-parent pair are included in Table \ref{tab:traveltimes}.  In most cases, it is reasonable for the candidate TDGs to travel the observed distance in several hundred megayears to a few gigayears.  This timescale is also consistent with the absence of a tidal tail.  The older TDGs identified in \cite{Duc2014} are estimated to be 2-4 Gyr, so our rough travel time estimates suggest that it is reasonable to think that our AD galaxies may indeed be older TDGs.  Another way to evaluate whether these systems are in dynamical equilibrium would be to obtain resolved kinematics and compare the dynamical timescales with these travel times | if a dynamical time is much less than the estimated travel time, it becomes more likely that that system is dynamically stable.

\begin{deluxetable}{ccccc}[ht]
    \centering
    \caption{\protect\label{tab:traveltimes} Estimated Travel Time for TDG-Parent Candidate Pairs}
    \tablehead{ \colhead{Source} & \colhead{Parent} & \colhead{Estimated Travel Velocity [$\mathrm{km}\,\mathrm{s}^{-1}$]} & \colhead{Estimated Travel Distance [kpc]}
    & \colhead{Estimated Travel Time [Gyr]}}
    \startdata
        AGC~123216 & MRK~365 & $\sim$210 & 260-310 & 1.2-1.4 \\
         & NGC~807 & 310-360 & 300-350 & 0.8-1.1 \\ 
         & MCG~05-06-003 & 100-110 & 900-1050 & 8.1-9.6 \\ \hline
        AGC~219369 & NGC~3510 & 100-110 & 170-790 & 1.5-7.7 \\ \hline
        AGC~229398 & KUG~1158+216 & 150-160 & 210-270 & 1.3-1.8 \\
         & UGC~6989 & 120-170 & 570-1020 & 3.2-8.2 \\ \hline
        AGC~333576 & NGC~7775 & 140-220 & 150-170 & 0.7-1.2 \\ \hline
    \enddata
\end{deluxetable}

\subsection{The Origins of the Almost-Dark Galaxies in This Work} \label{sec:origin}
Much of the previous work on TDGs has been carried out with objects formed in merging galaxy systems, since ongoing interactions are easier to spot than past ones and the TDGs in these cases are readily identifiable as tidal debris.  According to the simulations of \cite{Bournaud2006}, the interactions most likely to form long-lived TDGs are those that involve two gas-rich spiral galaxies with mass ratios of at least 1:8.  Following the first pericenter passage of a gas-rich merging system, two opposing tidal tails of gas stretch out, and denser knots form along the tails relatively quickly.  Knots located within the tails suffer significant mass loss and are more likely to fall back into the parent galaxies or disperse in under 1 Gyr, while those formed at the end (which also tend to be the most massive) have a better chance at long-term survival, becoming fully-fledged TDGs \citep{Bournaud2006}.

Another proposed formation mechanism for TDGs is a high velocity encounter.  \cite{Duc2008a} suggest that high velocity encounters can produce independent HI clouds with minimal stellar content, resulting in tidal debris that is gas-rich but extremely optically faint.  This scenario would explain the lack of a disrupting companion and the apparent minimal perturbation of the stellar disk for many of the proposed parents. \cite{Duc2008a} simulated the results of various encounters at $\sim$300-1000 $\mathrm{km}\,\mathrm{s}^{-1}$ and showed that fly-bys of 1000 $\mathrm{km}\,\mathrm{s}^{-1}$ or more are capable of producing primary tidal tails that are just as long as those in merger scenarios.  Rather than the two large tails produced in a standard major merger, the tidal tails formed in the high velocity fly-by consist of a long tail and a fainter, shorter countertail that quickly falls back onto the parent's disk.  The tails have a lower stellar content than tails produced in a lower velocity interaction (i.e. a galaxy merger) because most stars remain in the disk.  While the total mass of gas removed from the parent in a high velocity fly-by is much lower than in a more traditional merger scenario, this is due to the shortness of the countertail; the gas mass of the primary tidal tail is comparable to the mass of an individual tail in a slower encounter.  According to \cite{Duc2008a}, the velocity of the tail is determined by the mass of the progenitor galaxy, not the speed of the interloper, and is on the same order of magnitude as the rotational velocity of the parent, so the reasoning behind the travel times as calculated in Section \ref{sec:age} should still be valid for TDGs formed in this way.  With this understanding of how TDGs can form, we now move on to examining the potential origins of each system presented here.

AGC~123216 is the most ambiguous and complex case in the sample.  AGC~123216 has three potential parents, two of which show signs of a previous interaction (the elongation of the HI around NGC~807, and the slight offset of HI from the optical center of MRK~365).  The disturbance around NGC~807 is much more pronounced, and so it would be considered the primary candidate for providing most of AGC~123216's material.  The estimated travel time for AGC~123216 from NGC~807 (assuming its relative travel speed is similar to NGC~807's rotation velocity) is around 1 Gyr.  This is also approximately the timescale that \cite{Young2002} estimates for the asymmetry of the CO gas to settle, though they note that the molecular gas may be farther from the center than it appears, which would extend this timescale.  \cite{Young2002} also mentions that the gas in this galaxy may have come from a merger between two gas-rich spiral galaxies, an event which potentially could have produced a TDG.  If AGC~123216 is a TDG and was instead created by a later fly-by interaction, MRK~365 is near the direction of travel that we would expect for an interloper, and would only need a velocity of $\sim$600 $\mathrm{km}\,\mathrm{s}^{-1}$ relative to NGC~807 to reach its current position in 1 Gyr. AGC~123216 has a large effective radius for its stellar mass, which would be expected for a TDG.  However, the dark matter content of AGC~123216 appears to be rather high for a TDG candidate.  The object is well separated from the surrounding HI sources, so it is valid to assume that it is dynamically stable.  \cite{Wang2020} studied other ADs with properties similar to AGC~123216 and found that they had low molecular-to-atomic gas mass ratios, so it seems unlikely that there is a large deposit of unseen molecular gas.  Therefore, the majority of the unseen mass is likely to be non-baryonic dark matter.  It is important to remember that the results here represent an upper limit for the dynamical mass ratio, because the kinematic modeling indicated that the velocity dispersion was on the same order of magnitude as the rotation velocity and so the emission line width would be artificially broadened \citep{Lelli2015}.  This would mean the dynamical mass ratio (which is already somewhat low compared to most of the LG dwarf irregular galaxies) is potentially lower than we have estimated here (Fig. \ref{fig:LW2016_14}).  This makes it challenging to definitively say that AGC~123215 is not a TDG based solely on the estimate of the dark matter content.  AGC~123216 also has an unusually high HI-to-stellar mass ratio compared to other TDGs, which could indicate it never experienced the increased rate of star formation expected to occur for TDGs when they form.  AGC~123216 has many properties in common with the ALFALFA AD sources in the HI1232+20 system investigated by \cite{Janowiecki2015}.  That work concluded that AGC~229383, AGC~229384, and AGC~229385 were too isolated to be likely tidal features, and all three objects have some of the highest HI-to-stellar mass ratios measured to date for the AD sample \citep{Janowiecki2015}.  If the objects in \cite{Janowiecki2015} are indeed non-tidal, they could potentially be genuine “dark" galaxies- otherwise ordinary dwarf galaxies with unusually low stellar masses for the gaseous matter present.  \cite{Janowiecki2015} also found that these objects appear to rotate slower than their measured baryonic mass would imply based on the baryonic Tully-Fisher relation (BTFR).  The width of the velocity profile of AGC~123216 is rather small for a rotating disk of similar baryonic mass, and the velocity gradient map appears to show some minor ordered rotation (Fig. \ref{fig:velmaps}b), which is supported by the kinematic model. AGC~123216's BTFR location, effective radius, and low surface brightness are also similar to those of UDGs \citep{vanDokkum2015, ManceraPina2019b}, though the distinction between a UDG and an “almost-dark" galaxy is not well-defined.

AGC~219369 is possibly located very near to NGC~3510, which clearly has significant disruption to its HI distribution, so at first glance AGC~219369 is a good candidate to be a TDG. The edge-on orientation of NGC~3510 (Fig. \ref{fig:AGC219369full} and Fig. \ref{fig:NGC3510_velmap}) makes it difficult to evaluate the state of the stellar disk, but the stretched stellar limbs suggest that it has been severely disturbed.  The orientation of NGC~3510 also makes it difficult to estimate the travel time for AGC~219369 (see the wide range in Table \ref{tab:traveltimes}; additionally, if the angular separation is used as a minimum travel distance, it could be as low as 0.5 Gyr). The very blue color of NGC~3510 suggests that there has been fairly recent star formation activity, which may help constrain the length of time since the interaction.  A more face-on orientation of NGC~3510 may have revealed a second interacting galaxy.  NGC~3510 appears to be in a group with three other nearby galaxies (UGCA~225, UGC~6102, and NGC~3486), so a fly-by scenario for the disruption of NGC~3510 (potentially resulting in the formation of AGC~219369) is also possible.  However, the high dynamical mass ratio indicates AGC~219369 is more likely to be a dark matter dominated dwarf irregular galaxy.  It is possible that AGC~219369 was a pre-existing dwarf irregular galaxy which interacted with NGC~3510 in the past and caused the disruption of its HI, and not a post-interaction product such as a TDG.  The large effective radius may be explained by the fact that pre-existing dwarf galaxies may experience a temporary increase in size following tidal threshing, as shown in \cite{Paudel2013}.  On the other hand, it seems unlikely that an object as small as AGC~219369 could have caused such a large disturbance to the much more massive NGC~3510 and remained intact.  We must also consider that AGC~219369 may be closer than 16.7 Mpc and the disturbed state of NGC~3510 may be entirely unrelated.  If AGC~219369 is at a distance of 9.2 Mpc, the dynamical mass ratio would be even higher and the effective radius and stellar mass would be smaller.  This means the central surface brightness and effective radius would be closer to those of other dwarf irregular galaxies of similar stellar mass.  For these reasons, we consider AGC~219369 unlikely to be a TDG.  While the central surface brightness is fainter than 24 $\mathrm{mag}\,\mathrm{arcsec}^{-2}$, the effective radius is at most 1.41 $\pm$ 0.52 kpc, which is not large enough to clearly classify it as a UDG \citep{vanDokkum2015}.  Therefore, we will simply identify AGC~219369 as a low surface brightness dwarf irregular galaxy.

AGC~229398 and AGC~333576 appear to have low dark matter contents (Fig. \ref{fig:LW2016_14}), and very large effective radii for their stellar masses (Fig. \ref{fig:Duc2014_6}), both of which are indicators that they may be TDGs.  They also have fainter central surface brightnesses compared to dwarf irregular and dwarf elliptical galaxies of similar stellar mass (Fig. \ref{fig:Duc2014_7}), which suggests their stellar populations may have dimmed over time, as is expected for TDGs.  AGC~229398 has two potential parents identified by the velocity and angular separation selection criteria: KUG~1158+216 and UGC~6989.  Based on mass ratio predictions for TDGs and their parents, AGC~229398 seems too massive to have come out of KUG~1158+216, so we turn our attention to UGC~6989.  The mass ratio is still higher than that which is predicted by simulations, but more viable than for KUG~1158+216.  However, UGC~6989 is 482 kpc away from AGC~229398 on the sky, much farther than any other TDG-parent system studied.  The TDGs simulated in \cite{Bournaud2006} could have travel velocities ranging from 50-400 $\mathrm{km}\,\mathrm{s}^{-1}$ relative to a parent with a virial velocity of 220 $\mathrm{km}\,\mathrm{s}^{-1}$, so we could consider the case of a TDG travel speed of 400 $\mathrm{km}\,\mathrm{s}^{-1}$ and the minimum travel distance of 445 kpc (using the \citealt{Haynes2018} distance of UGC~6989) to estimate a minimum travel time of 1.09 Gyr between UGC~6989 and the current location of AGC~229398.  This would be an extreme case, so it is likely that AGC~229398 would be older than this; the travel times estimated in Table \ref{tab:traveltimes} are at least 3.2 Gyr.  The $g-r$ color of AGC~229398 is 0.35, making it the reddest TDG candidate in our sample; this is consistent with it also being the oldest.  UGC~6989's undisturbed spiral structure suggests that it is unlikely to have undergone a recent major merger, but a high velocity fly-by could have pulled off gas without disrupting the interior disk.  There is no obvious interloper candidate in the images, but UGC~6989 was already at the limits of the selection criteria; whatever speed AGC~229398 is traveling at relative to UGC~6989, the interloper would only need to go a little more than twice that speed to end up outside of the angular separation selection criteria and be overlooked.  \cite{Duc2008a} notes that in 2 Gyr, an interloper traveling at 1000 $\mathrm{km}\,\mathrm{s}^{-1}$ relative to the parent galaxy could be a projected distance of 2 Mpc away | at the distance of AGC~229398, this is over 1 degree of separation, outside the field of view of the pODI images.

AGC~333576 seems to have a more obvious progenitor: the nearby spiral galaxy NGC~7775.  The HI gas surrounding NGC~7775 is distorted in the direction of AGC~333576, and the distorted stellar arms and bright region of star formation in the southwest imply a recent interaction of some kind.  A major merger would likely have disrupted the structure of the arms even further, but a fly-by interaction might not have, so we searched the field surrounding AGC~333576 and found the elliptical galaxy WISEA~J235250.49+284555.2.  As an elliptical galaxy, WISEA~J235250.49+284555.2 would not have possessed very much gas and so would not have been the main contributor to a tidal dwarf, but it might have flown by NGC~7775 and pulled out the material from that disk, producing AGC~333576.  The proximity of these galaxies implies that the interaction would not have happened a very long time ago, with travel time estimates for AGC~333576 from NGC~7775 between 0.7 and 1.2 Gyr; this is enough time for the majority of the gaseous tidal tail to have disappeared, but the end of a stellar tail connected to NGC~7775 is still slightly visible on the north side (Fig. \ref{fig:AGC333576_closeup}, left).  Depending on the parameters of the interaction, the countertail may have been very short, which could explain why there is no strong evidence of one; the star formation region is located roughly opposite the stellar tail and might have been triggered by the same tidal forces.

\section{Summary}

We have investigated the HI and optical properties and possible origins of four Almost-Dark galaxies selected from the ALFALFA survey.  After exploring HI kinematics and analyzing deep optical images, we concluded that AGC~229398 and AGC~333576 are the likeliest candidates to be TDGs based on their apparent low dark matter content and large effective radii compared to other dwarf irregular and elliptical galaxies of similar stellar masses.  AGC~123216 is ambiguous | it is near other galaxies with large HI reservoirs that have clearly had an interaction of some kind, and it has a large effective radius. Its central surface brightness is also low compared to dwarf irregular galaxies of similar stellar mass.  However, the upper limit on the dark matter content is too high to positively confirm it as a TDG, and it has an unusually high HI-to-stellar mass ratio for a TDG that should have formed stars when it was created.  AGC~123216 also meets the criteria outlined in \cite{vanDokkum2015} for a UDG.  AGC~219369 is most likely to be a low surface brightness dwarf irregular galaxy.  While AGC~219369 appears in the sky near the spiral galaxy NGC~3510 (which has undergone some kind of interaction that disrupted its HI and stellar distributions), it appears to be dominated by dark matter and therefore is unlikely to have formed via the same tidal processes that create TDGs.  All four of the objects would benefit from higher resolution HI observations to better constrain their kinematics and dynamical masses.

In this study, we distinguished TDG candidates from irregular dwarf galaxies based on their low dark matter content and larger than expected effective radius for their stellar content.  These candidate TDGs are located farther from any potential parents than the vast majority of previously studied TDGs, and the tidal tail from which they would have formed is nowhere to be found.  While nearby surrounding galaxies often have disrupted HI distributions, none show clear signs of a recent merger that would be the expected source for a TDG.  This suggests that either enough time has passed for the parent galaxy to mostly restabilize, or that the interaction was one that did not cause obvious disruption to the stellar disk.  The overall picture provided by the HI and optical measurements of these objects suggests that they are older and more evolved than most of the TDGs that have previously been identified.  Blind HI surveys with wide-field coverage like ALFALFA provide an opportunity to detect TDGs that have moved away from their parents and become optically fainter over their evolution, and which may be missed in more traditional optical surveys that prioritize stellar emission.  Such surveys provide us with a more complete picture of galaxy formation and evolution at the low-mass end of the HI mass function.

\begin{acknowledgments}
The authors acknowledge the ALFALFA collaboration's work in observing, processing, and extracting sources.  The authors would also like to thank the anonymous referee for their thorough comments, which greatly improved the quality of the manuscript.

LMG acknowledges support from the Indiana Space Grant Consortium and thanks Laura Hunter for her explanations about the nuances of radio astronomy.  
This work was supported in part by NSF grant AST-1615483 to KLR.
L.L. acknowledges support from NSF grant AST-2045369.
MPH acknowledges support from NSF/AST-1714828 and grants from the Brinson Foundation.
EAKA is supported by the WISE research programme, which is financed by the Dutch Research Council (NWO).
JMC and JF are supported by NSF grant AST-2009894.
We thank the Indiana University (IU) College of Arts and Sciences for funding IU's share of the WIYN telescope.

We thank the staff of the WIYN Observatory and Kitt Peak National Observatory for their help and support during our WIYN pODI observing runs. We are grateful to the staff members at WIYN, NOIRLab, and Indiana University Pervasive Technology Institute for designing and implementing the ODI-PPA and assisting us with the pODI data reduction.  We owe special thanks to Wilson Liu for helping us to troubleshoot and optimize the ODI image stacking process.

Funding for SDSS-III and SDSS-IV has been provided by the Alfred P. Sloan Foundation, the Participating Institutions, the National Science Foundation, and the U.S. Department of Energy Office of Science.  SDSS acknowledges support and resources from the Center for High-Performance Computing at the University of Utah.  The SDSS web site is www.sdss.org.
\end{acknowledgments}

%


\facilities{Arecibo, WIYN(pODI and ODI), WSRT}





\appendix
\restartappendixnumbering
\section{HI Channel Maps}

\begin{figure}
    \centering
    \includegraphics[width=0.52\textwidth]{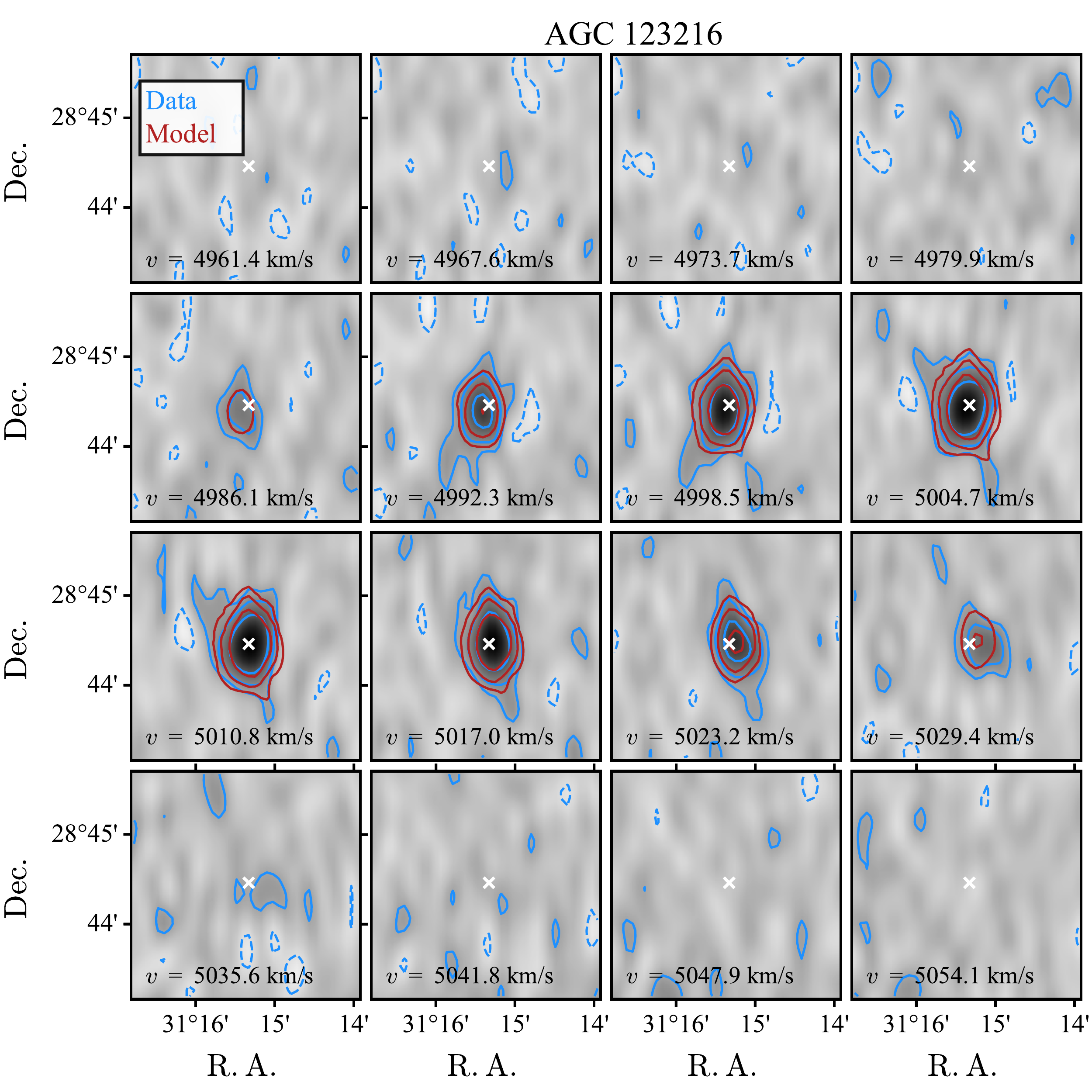}
    \caption{HI emission-line channel maps for AGC~123216 from WSRT data.  The white cross marks the center of the galaxy and the velocity for each channel map is listed in the lower left corner.  The emission of the galaxy is displayed by gray background and blue contour lines.  The best-fitting model is shown by the red contour lines.  Solid contours are at 2, 4, and 8 times the S/N of the data, and dashed contours represent -2 $\times$ S/N.}
    \label{fig:AGC123216_chan}
\end{figure}

\begin{figure}
    \centering
    \includegraphics[width=0.52\textwidth]{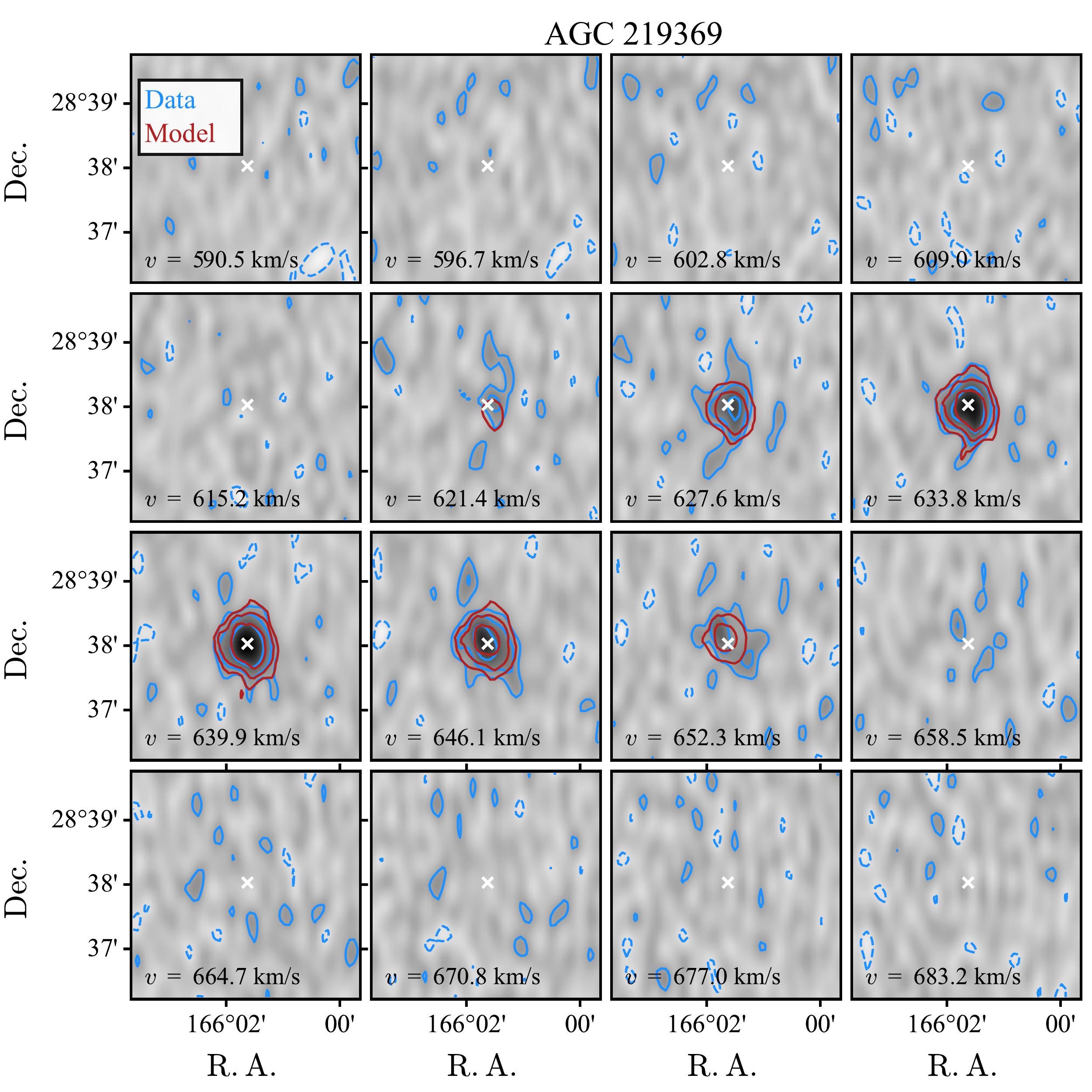}
    \caption{HI emission-line channel maps for AGC~219369 from WSRT data.  The white cross marks the center of the galaxy and the velocity for each channel map is listed in the lower left corner.  The emission of the galaxy is displayed by gray background and blue contour lines.  The best-fitting model is shown by the red contour lines.  Solid contours are at 2, 4, and 8 times the S/N of the data, and dashed contours represent -2 $\times$ S/N.}
    \label{fig:AGC219369_chan}
\end{figure}

\begin{figure}
    \centering
    \includegraphics[width=0.52\textwidth]{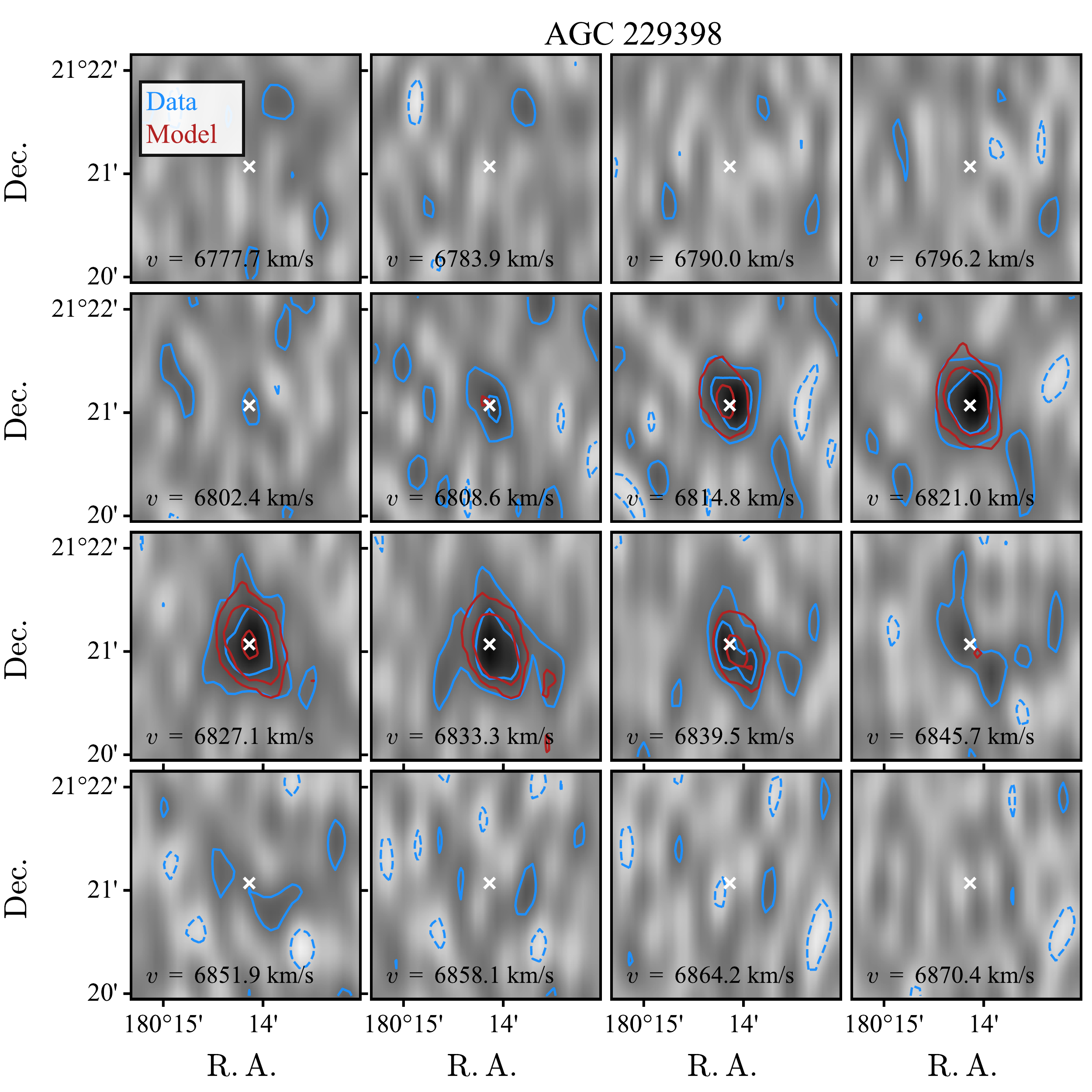}
    \caption{HI emission-line channel maps for AGC~229398 from WSRT data.  The white cross marks the center of the galaxy and the velocity for each channel map is listed in the lower left corner.  The emission of the galaxy is displayed by gray background and blue contour lines.  The best-fitting model is shown by the red contour lines.  Solid contours are at 2, 4, and 8 times the S/N of the data, and dashed contours represent -2 $\times$ S/N.}
    \label{fig:AGC229398_chan}
\end{figure}

\begin{figure}
    \centering
    \includegraphics[width=0.52\textwidth]{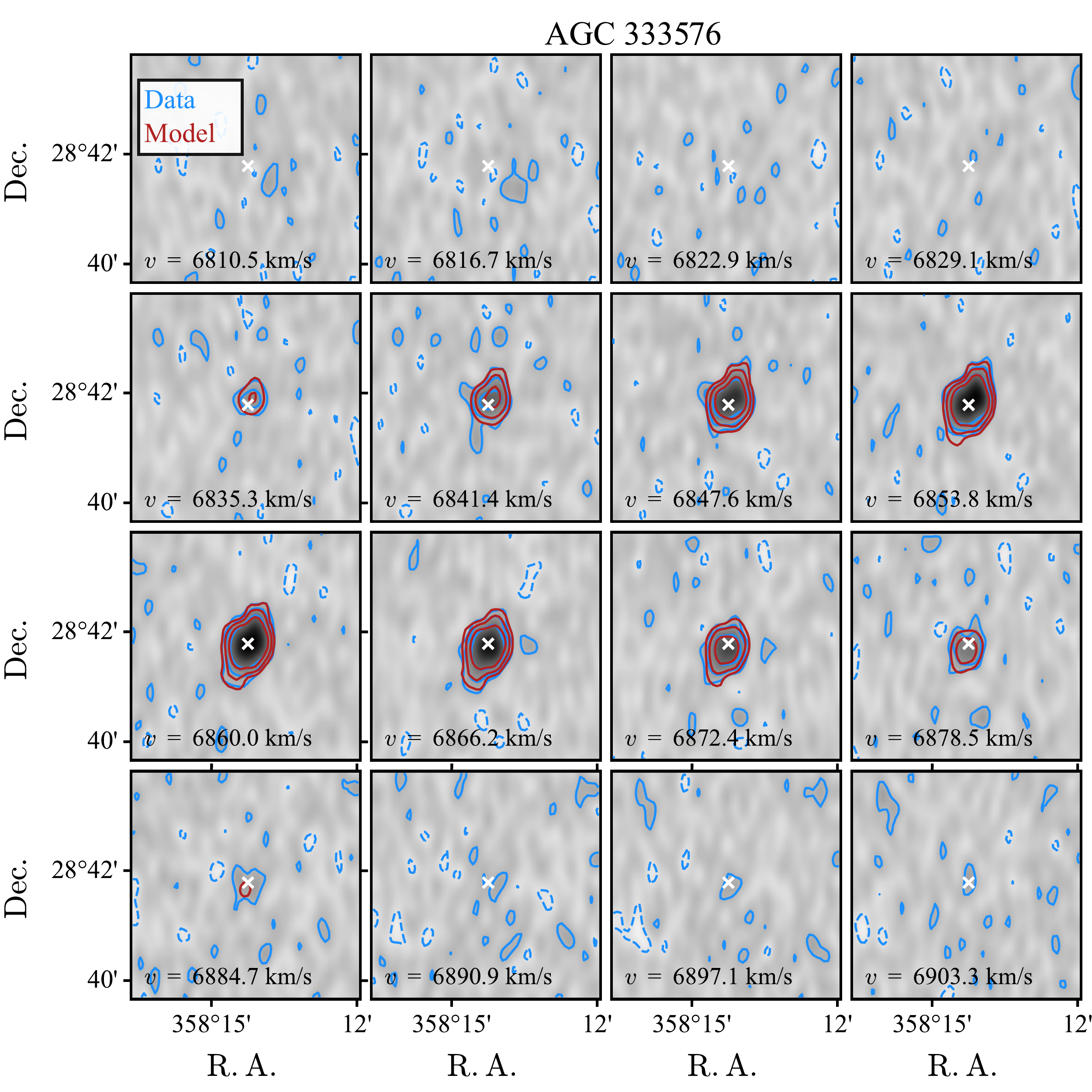}
    \caption{HI emission-line channel maps for AGC~333576 from WSRT data.  The white cross marks the center of the galaxy and the velocity for each channel map is listed in the lower left corner.  The emission of the galaxy is displayed by gray background and blue contour lines.  The best-fitting model is shown by the red contour lines.  Solid contours are at 2, 4, and 8 times the S/N of the data, and dashed contours represent -2 $\times$ S/N.}
    \label{fig:AGC333576_chan}
\end{figure}

Here we present the WSRT HI emission-line channel maps and the corresponding best-fitting kinematic models for AGC~123216 (Figure \ref{fig:AGC123216_chan}), AGC~219369 (Figure \ref{fig:AGC219369_chan}), AGC~229398 (Figure \ref{fig:AGC229398_chan}), and AGC~333576 (Figure \ref{fig:AGC333576_chan}).  AGC~123216 and AGC~333576 show the clearest signs of rotation and have models that most closely resemble the data, followed by AGC~219369.  The data for AGC~229398 do not demonstrate a clear rotation gradient, and the model is a poor fit for the data, which indicates it may be more pressure supported than rotation supported.

\bibliography{TDGbib}{}
\bibliographystyle{aasjournal}



\end{document}